\documentclass[aps,prb,amsmat,amssymb,superscriptaddress, twocolumn]{revtex4}
\usepackage[english]{babel}

\usepackage{graphicx}
\usepackage{transparent}
\usepackage{amsmath}
\usepackage{amssymb}
\usepackage{epstopdf}
\usepackage{amsfonts}
\usepackage{bm}
\usepackage{times}
\usepackage{pdfpages}
\usepackage{mathtools}
\usepackage{hyperref}
\usepackage{tabularx}
\usepackage{hhline}
\usepackage[normalem]{ulem}
\usepackage{grffile} 
\usepackage{verbatim}

\usepackage[section]{placeins}
\newcommand{\astcycl}{\mathrlap{\kern0.085em{\circlearrowright}}\ast}
\newcommand{\taucycl}{\mathrlap{\kern0.42em{\bullet}}\circlearrowright}

\def\<{\left\langle}
\def\>{\right\rangle}

\newcommand{\tr}{\operatorname{Tr}}
\newcommand{\timeorder}{\mathcal{T}_\tau}
\newcommand{\etal}{\textit{et al.}}

\begin{document}

\title{Causal versus local $GW$+EDMFT scheme and application to the triangular-lattice extended Hubbard model}

\author{Jiyu Chen}
%\email{}
\affiliation{Department of Physics, University of Fribourg, 1700 Fribourg, Switzerland}
\author{Francesco Petocchi}
\affiliation{Department of Physics, University of Fribourg, 1700 Fribourg, Switzerland}
\author{Philipp Werner}
%\email{philipp.werner@unifr.ch}
\affiliation{Department of Physics, University of Fribourg, 1700 Fribourg, Switzerland}

\begin{abstract}
Using the triangular-lattice extended Hubbard model as a test system, we compare $GW$+EDMFT results for the recently proposed self-consistency scheme with causal auxiliary fields to those obtained from the standard implementation which identifies the impurity Green's functions with the corresponding local lattice Green's functions. Both for short-ranged and long-ranged interactions we find similar results, but the causal scheme yields slightly stronger correlation effects at half-filling. We use the two implementations of $GW$+EDMFT to compute spectral functions and dynamically screened interactions in the parameter regime relevant for 1$T$-TaS$_2$. We address the question whether or not the sample used in a recent photoemission study [Phys. Rev. Lett. 120, 166401 (2018)] was half-filled or hole-doped.  
\end{abstract}


\maketitle

\section{Introduction}

Strongly correlated electron materials exhibit remarkable phenomena such as quasi-particles with strongly renormalized masses and correlation-induced metal-insulator (Mott) transitions.\cite{mott1968,imada1998} The screening of the Coulomb interaction determines the effective electron-electron interaction in solids and thus plays an important role in the modeling of this class of materials.\cite{hedin1999,werner2016} The dynamically screened interaction has to be computed self-consistently, because the correlated electronic structure determines the relevant low-energy screening modes, while the interaction determines band renormalizations, quasi-particle lifetimes and Mott gaps.  A simple model which allows to study the interplay between correlations and screening is the single-band extended Hubbard model with local and nonlocal (possibly long-ranged \cite{hansmann2013}) interactions.\cite{ayral2012,huang2014} 

Here, we use a combination of the Extended Dynamical Mean-Field Theory (EDMFT) method \cite{sun2002} and the $GW$ method ($GW$+EDMFT) \cite{biermann2003,aryasetiawan2004,biermann2004} to study the phase diagram and correlation functions of the single-band Extended Hubbard model on the two-dimensional triangular lattice. EDMFT provides a local self-energy and polarization, and allows to capture strong correlation effects and Mott physics, while $GW$ provides nonlocal self-energy and polarization components which capture nonlocal correlation and screening effects. The latter have been shown to be relevant for an accurate description of materials.\cite{petocchi2021} In order to treat long-ranged interactions, and compare the results for models with short-ranged and long-ranged nonlocal interactions, we implement the Ewald lattice summation.\cite{ewald1921,harris1998,hansmann2013}

One purpose of this study is to systematically test the recently proposed manifestly causal $GW$+EDMFT scheme of Backes, Sim and Biermann.\cite{backes2020} The mapping of the lattice model to a single-site impurity model yields two dynamical mean fields: a fermionic mean field (the hybridization function), and a bosonic mean field (the effective local interaction).  In the conventional $GW$+EDMFT implementation, which accomplishes the mapping by identifying the local lattice Green's function and local screened interaction with the corresponding impurity quantities, these auxiliary fields are not necessarily causal. This is not in principle a problem, as long as all the physical quantities (e.g. the screened interaction or the Green's function) are causal,\cite{nilsson2017} but it can create numerical issues. Such problems are avoided by the modified self-consistency scheme of Ref.~\onlinecite{backes2020}, which by construction yields causal auxiliary fields. To clarify the influence of the implementation on the auxiliary fields and physical observables, we present a systematic comparison between the $GW$+EDMFT results obtained with the conventional (``local") and the new ``causal" self-consistency scheme for the extended Hubbard model. 

Our second purpose is to provide reference data for strongly correlated electron systems on a triangular lattice, such as 1$T$-TaS$_2$. We present the $GW$+EDMFT phase diagram of the triangular lattice model as a function of the onsite interaction $U$ and nonlocal interaction $V$, and determine the parameters appropriate for single-band extended Hubbard simulations of 1$T$-TaS$_2$. We then use these parameters to study the spectral functions and dynamically screened interactions of half-filled and chemically doped systems in the temperature region corresponding to the C phase ($T\lesssim 180$ K\cite{wang2020}), and compare to the photoemission spectrum from Ref.~\onlinecite{ligges2018}. The latter study concluded, based on the dynamics of photo-doped doublons, that their nominally half-filled sample may have been hole doped. Our comparison to the occupation functions for different electron fillings however demonstrates a good match between the experimental equilibrium spectrum and the result for the undoped model.  

The paper is organized as follows. In Sec.~\ref{sec:model} we describe the model and in Sec.~\ref{sec:gw+edmft} the $GW$+EDMFT method, as well as the two self-consistency schemes. In Sec.~\ref{sec:results}, after presenting the $GW$+EDMFT phase-diagram of the triangular lattice extended Hubbard model (\ref{sec:phasediagram}), we show the comparison of the two schemes in different relevant parameter regimes (\ref{sec:comparison}). We then present the results representative of 1$T$-TaS$_2$ in Sec.~\ref{sec:1TTaS2} and compare our results to the equilibrium spectrum from Ref.~\onlinecite{ligges2018}. Section~\ref{sec:conclusions} contains the conclusions.

\section{Model and method}
\subsection{Model}\label{sec:model}
We consider the single-band Extended ($U$-$V$) Hubbard model on the two-dimensional triangular lattice. The grand-canonical Hamiltonian is given by
\begin{equation}
	\label{eq:hubbard}
\begin{aligned}
	H_{\text{Hubbard}}=& -\sum_{ij\sigma}t_{ij}c^\dagger_{i\sigma}c_{j\sigma}+U\sum_i n_{i\uparrow} n_{i\downarrow}\\&+\frac{1}{2}\sum_{ij}V_{ij}n_in_j-\mu \sum_{i} n_i,
\end{aligned}
\end{equation}
where the operators $\hat{c}_{i\sigma}$ and $\hat{c}^\dagger_{i\sigma}$ denote the annihilator and the creator of an electron with spin $\sigma\in\{\uparrow,\downarrow\}$ at the $i$-th lattice site. $n_{i\sigma}=c^\dagger_{i\sigma}c_{i\sigma}$ and $n_{i}=n_{i\uparrow}+n_{i\downarrow}$ are the particle number operators for site $i$. We will restrict the hoppings to nearest neighbor (NN) sites $\langle ij\rangle$, and write $t_{ij}=t\delta_{\langle ij\rangle}$. $U$ is the on-site repulsive interaction between two electrons with opposite spins on the same site, while $V_{ij}$ is the inter-site repulsive interaction, for which we choose the Coulomb form $V_{ij}\propto 1/r_{ij}$. Here, $r_{ij}$ denotes the physical distance between the sites $i$ and $j$. $\mu$ is the chemical potential for the grand-canonical ensemble.

We will use two approximations for the nonlocal interaction, to get some insights into the effects of the range of the interaction. On the one hand, we truncate the interaction to the NN components,  $V_{ij}=V_{01}\delta_{\langle ij \rangle}$.  On the other hand, we will treat the long-range Coulomb interaction via the Ewald lattice summation, \cite{hansmann2013} as explained below. This will define the second approximation, denoted as ``Ewald."

\subsection{Ewald lattice summation}

It is convenient to Fourier transform the Hamiltonian from real space to momentum space ($k$-space). For our two-dimensional homogeneous lattice model, this transformation yields $A(\mathbf{k})=\sum_i A_{0i}e^{i \mathbf{k}\cdot \mathbf{r}_{0i}}$, where $A_{ij}$ stands for the hopping $t_{ij}$ or the Coulomb repulsion $V_{ij}$. For the triangular lattice, in the translationally invariant NN case, using $A_{ij}\equiv A_{01}\delta_{\langle ij\rangle}$, the band dispersion becomes $\epsilon_\mathbf{k}=-2t(\cos k_x+\cos(-\frac{1}{2}k_x+\frac{\sqrt{3}}{2}k_y)+\cos(\frac{1}{2}k_x+\frac{\sqrt{3}}{2}k_y))$ and the bare interaction $v_\mathbf{k}= U+V_\mathbf{k}$ with $V_\mathbf{k}=2V_{1}(\cos k_x+\cos(-\frac{1}{2}k_x+\frac{\sqrt{3}}{2}k_y)+\cos(\frac{1}{2}k_x+\frac{\sqrt{3}}{2}k_y))$.  In addition to this, we also consider long-range interactions using the Ewald lattice summation, as was previously done in Ref.~\onlinecite{hansmann2013}. More specifically, we will assume an infinite-range Coulomb repulsion, $V_{ij}=V_{1}/r_{ij}$, which scales as one over the distance between the sites, and which is parametrized by the nearest-neighbor interaction $V_{1}$. The Ewald lattice summation method \cite{ewald1921,harris1998} provides a way of regularizing the interaction energy and avoiding problems of conditional convergence. To this end, the long-range Coulomb repulsion is represented as the sum of two parts, corresponding to a real-space summation and a reciprocal-space summation.\cite{harris1998}  First, we re-express the Coulomb potential in terms of a Gaussian integral.  Second, we divide the Gaussian integral into two parts, by means of a parameter $\eta$ which fixes the partitioning between short-range and long-range components.  Third, the summation in the two parts is performed in real space and reciprocal space, respectively. Details are presented in Appendix~\ref{app_ewald}. The final expression for the Ewald long-range interaction becomes
\begin{equation}\label{eq:ewald_potential}
	V_{ij}=V_1\Bigg\{\frac{1}{r_{ij}}\operatorname{erfc}\left( \frac{r_{ij}}{2\eta}\right)+\frac{1}{N_\mathbf{k}}\sum_{\mathbf{k}\neq 0} e^{i \mathbf{k}\cdot \mathbf{r}_{ij}}\frac{2\pi}{k}\operatorname{erfc}\left( k\eta \right) \Bigg\}.
\end{equation}

\begin{figure}[t]
	\centering
	\includegraphics[width=0.9\linewidth]{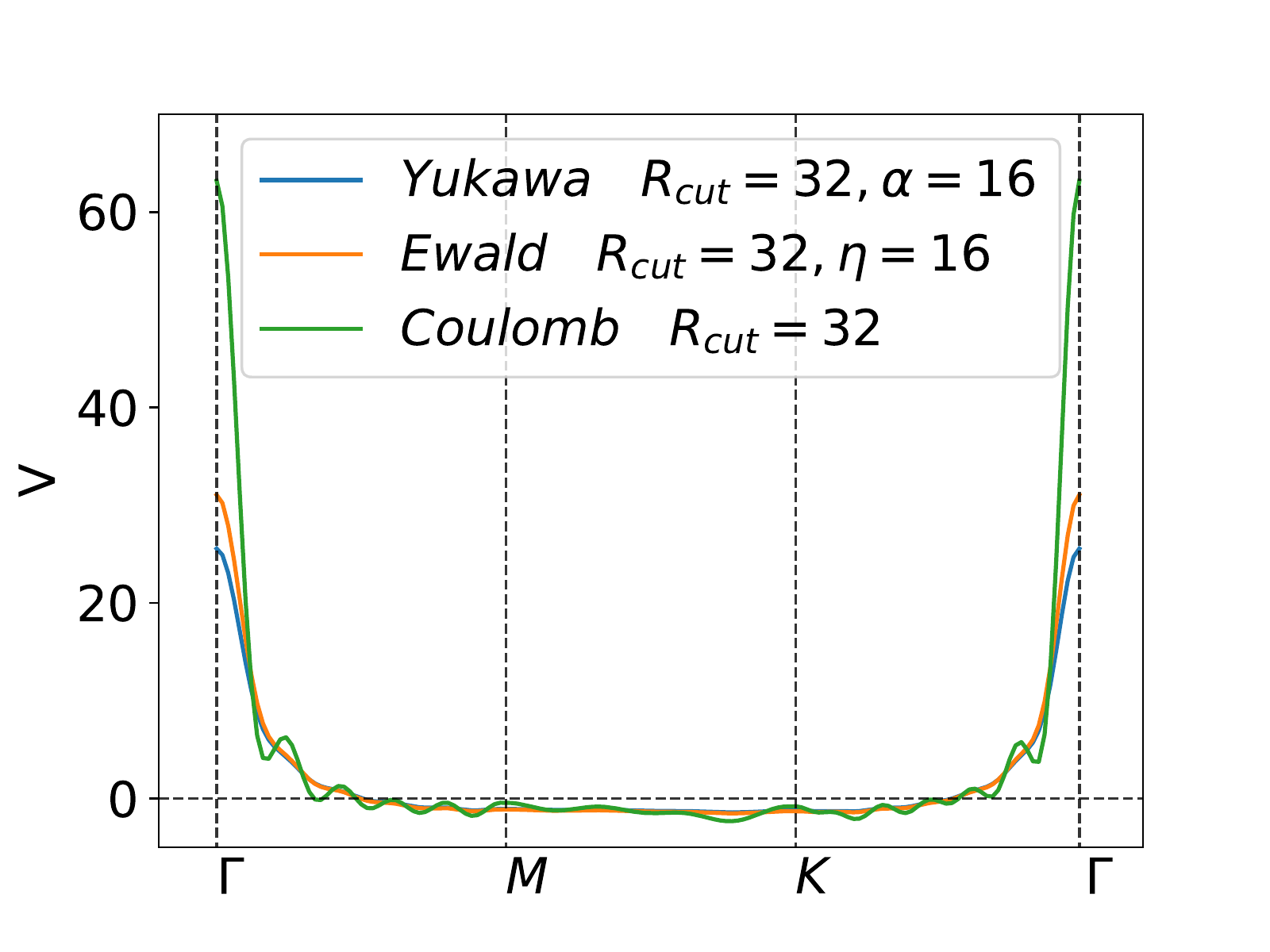}
	\caption{Comparison of the interaction $V(q)$ in the first Brillouin zone for three different potentials, with the same NN interaction $V_1=1$ and lattice spacing $a=1$. ($\Gamma$, M, K denote the centre, the midpoint of edge and the corner of the hexagonal first Brillouin zone, respectively.)
	}\label{fig:Vk}
\end{figure}

In Fig.~\ref{fig:Vk} we plot $V(k)$ along the indicated path in the Brillouin zone (BZ). The green curve shows the interaction obtained by truncating the Coulomb interaction at the 32nd neighbors. The orange line plots the Ewald interaction (\ref{eq:ewald_potential}) for $\eta=16$. In our implementation, we also use a cutoff at the 32nd neighbors in the Ewald case, which explains the negative values in some regions of $k$ space. For comparison, we also show a Yukawa potential $V_\text{Yukawa}(r)=\frac{V_1}{r} e^{-r/\alpha}$ for $\alpha=16$. One sees that the regularization with the $\eta$ parameter in the Ewald summation has a similar effect as a cutoff parameter $\alpha=\eta$ in the Yukawa case, smoothing the potential and regularizing the divergence at the $\Gamma$ point.

\subsection{Action}

In order to compute the Green's function and other physical quantities, it is convenient to reformulate the model in terms of the action using the coherent-state path-integral representation.\cite{negele1988} For an arbitrary Hamiltonian $H(c^\dagger_\alpha,c_\alpha)$, where $c^\dagger_\alpha$ and $c_\alpha$ are fermionic creators and annihilators indexed by $\alpha$, the grand-canonical partition function  $Z=\tr e^{-\beta H}$ can be expressed in the path-integral form as $Z=\int_{\phi(\beta)=-\phi(0)} \mathcal{D}[\phi_{\alpha}^{*}(\tau), \phi_{\alpha}(\tau)]e^{-S[\phi^{*},\phi]}$ with action
\begin{equation}
\begin{aligned}
	S[\phi^*,\phi]\equiv&\int_{0}^{\beta} d \tau\bigg\{\sum_{\alpha} \phi_{\alpha}^{*}(\tau)(\partial_\tau-\mu) \phi_{\alpha}(\tau) \\
	& +  H(\phi_{\alpha}^{*}(\tau), \phi_{\alpha}(\tau))\bigg\}.
\end{aligned}
\end{equation}
Here, $\phi^*_\alpha$ and $\phi_\alpha$ denote the corresponding anti-periodic Grassmann numbers satisfying $\phi_\alpha(\tau+\beta)=-\phi_\alpha(\tau)$. Specifically, for the Hamiltonian of the extended Hubbard model defined in Eq.~\eqref{eq:hubbard}, the action is
$S[c^{*}, c]$ $=\int_{0}^{\beta} d \tau \{\sum_{i, \sigma} c_{i \sigma}^{*}(\tau)(\partial_{\tau}-\mu) c_{i \sigma}(\tau) -\sum_{ij\sigma}t_{ij}c_{i \sigma}^{*}(\tau)c_{j \sigma}(\tau) $ $+U \sum_{i} n_{i \uparrow}(\tau) n_{i \downarrow}(\tau)+\sum_{i <j} V_{ij} n_{i}(\tau) n_{j}(\tau)\}.$
By applying the identity $n_in_i=n_i+2n_{i\uparrow}n_{i\downarrow}$ and introducing $v_{ij}=U\delta_{ij}+V_{ij}$ and $\tilde{\mu}=\mu+U/2$, we can write it in the more compact form
\begin{align}\label{eq:hubbard_action2}
	S\left[c^{*}, c\right]= &\int_{0}^{\beta} d \tau\bigg\{\sum_{i j \sigma} c_{i \sigma}^{*}(\tau)\left[\left(\partial_{\tau}-\tilde{\mu}\right) \delta_{i j}-t_{i j}\right] c_{j \sigma}(\tau) \nonumber\\
	& +\frac{1}{2} \sum_{i j} v_{i j} n_{i}(\tau) n_{j}(\tau)\bigg\}.
\end{align}

\subsection{$GW$+EDMFT}
\label{sec:gw+edmft}

\subsubsection{General remarks}

Within EDMFT the lattice model is mapped to an auxiliary single-site impurity model with two dynamical mean fields: the fermionic bath Green's function $\mathcal{G}^{-1}=i\omega_n+\mu-\Delta$,  which contains information on how electrons are connected to the bath through the hybridization function $\Delta$, and the bosonic field $\mathcal U$, which is a frequency-dependent effective impurity interaction.\cite{sun2002,ayral2013} This scheme identifies the local projection of the lattice self-energy $\Sigma$ and polarization $\Pi$ with the corresponding fields provided by an effective impurity model, which is solved in a self-consistent manner. 

By adding the nonlocal self-energy and polarization components obtained with the $GW$ method\cite{hedin1965} to the EDMFT local solution, the $GW$+EDMFT\cite{biermann2003}  scheme is able to simultaneously treat screening effects induced by local and nonlocal charge fluctuations and strong local correlations. Even if short-range correlations and frustration effects produced by the lattice geometry are 
not fully captured, 
$GW$+EDMFT in realistic materials contexts has been shown to yield accurate results for the effective interaction strengths in solids.\cite{petocchi2020,petocchi2021}

The standard (``local") self-consistency scheme of $GW$+DMFT fixes the auxiliary fields by identifying the local self-energy and local polarization with the impurity counterparts.\cite{biermann2003,ayral2013}  This can lead to noncausal dynamical mean fields, which is by itself not a problem (since the physical observables such as the Green's function and screened interaction are causal),\cite{nilsson2017} but it can potentially lead to numerical issues, such as sign problems in Monte-Carlo based impurity solvers, or incompatibilities with standard analytical continuation procedures that assume causality.  

In Ref.~\onlinecite{backes2020}, a modified $GW$+EDMFT self-consistency scheme has been proposed, which by construction produces causal auxiliary fields. We sketch below the local and the new causal $GW$+EDMFT self-consistency loops, while a more detailed derivation of the causal self-consistent equations can be found in Appendix~\ref{app_causal}. \\
 
\subsubsection{EDMFT}

\begin{figure}[t]
	\centering
	\includegraphics[width=0.9\linewidth]{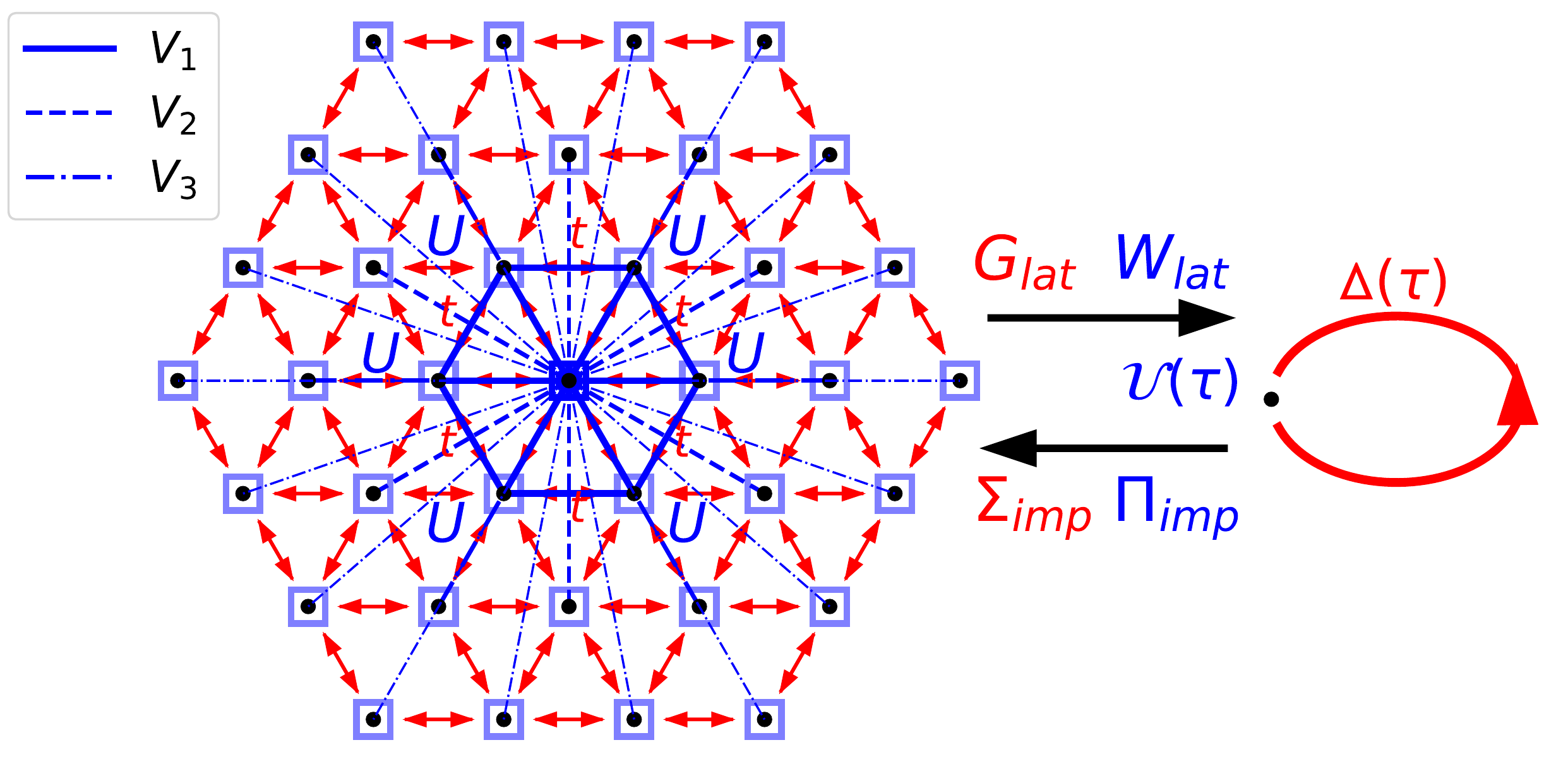}
	\caption{Left: The extended Hubbard model on the triangular lattice, with NN hopping $t$, on-site interaction $U$ and nonlocal interactions $V$. Right:  EDMFT maps the lattice model to an impurity model defined by a hybridization function $\Delta$ (or fermionic Weiss field $\mathcal{G}^{-1}=i\omega_n+\mu-\Delta$) and a bosonic field (retarded interaction) $\mathcal{U}$.}\label{fig:lattice}
\end{figure}

Even though the causal self-consistency loop differs from the local one only in the presence of nonlocal self-energy and polarization components, we will lay out the equations in this section in a way that is compatible with the new scheme of Backes \textit{et al.}\cite{backes2020} The EDMFT self-consistent equations map the lattice action defined in Eq.~\eqref{eq:hubbard_action2} to the impurity action
 \begin{align}
 	S^{\text{EDMFT}}=&-\int_0^\beta\sum_\sigma c^\dagger_\sigma(\tau)\mathcal{G}^{-1}(\tau-\tau')c_\sigma(\tau')\nonumber\\
	&+\frac{1}{2}\int_0^\beta d\tau d\tau' n(\tau)\mathcal{U}(\tau-\tau')n(\tau'),
\end{align} 
which contains the (fermionic) Weiss field $\mathcal{G}$ and the (bosonic) effective interaction $\mathcal{U}$ (see Fig.~\ref{fig:lattice}).

This mapping involves approximations in the case of finite-dimensional lattices,\cite{georges1996} but it becomes exact in the large-connectivity limit,\cite{si1996} i.e. for coordination number $z\to\infty$, if one properly renormalizes the hoppings ($t\to t/\sqrt{z}$) and interactions ($V\to V/z$). As noted in Ref.~\onlinecite{sun2002}, there are in principle two ways to formulate the self-consistency loop, either using the charge density $n$ and the density-density correlation function $\chi$, or (after a Hubbard-Stratonovich transformation of the lattice action Eq.~\eqref{eq:hubbard_action2}) using the auxiliary bosonic field $\phi$ and bosonic Green's function $W$. For the local EDMFT self-consistent equations the two formulations are equivalent,\cite{sun2002} while this equivalence may not hold any more for the causal EDMFT equations. In this work, following Ref.~\onlinecite{backes2020}, we use the formulation involving $n$ and $\chi$. The result of the generalized self-consistency equations (see Appendix~\ref{app_causal} for a detailed derivation) can be expressed in terms of corrections to the impurity Weiss fields:\cite{backes2020} 
\begin{equation}\label{eq:EDMFT_Generalised_sc}
	\begin{aligned}
		\mathcal{G}^{-1}\left(i\omega_n\right) &=\quad\langle G\rangle^{-1}_k+\langle\Sigma\rangle_k +\Delta_\text{cor},\\
		\mathcal{U}(i \nu_n)&=\left[\langle W\rangle^{-1}_k+\langle \Pi\rangle_k\right]^{-1}+\mathcal{U}_\text{cor},
	\end{aligned}
\end{equation}
where $G$ and $W$ denote the lattice Green's function and screened interaction, respectively, while $\Sigma$ is the self-energy and $\Pi$ the polarization. $\langle\rangle_k$ denotes an average over momentum in the BZ. The correction terms of the casual self-consistency scheme are given by\cite{backes2020} 
\begin{widetext}   
\begin{equation}\label{eq:EDMFT_Generalised_correction}
	\begin{aligned}
		\Delta_\text{cor} &= -\langle\Sigma G \Sigma\rangle_k+\langle\Sigma G\rangle_k\langle G\rangle_k^{-1}\langle G \Sigma\rangle_k-2\langle\Sigma\rangle_k+\langle\Sigma G\rangle_k\langle G\rangle_k^{-1}+\langle G\rangle_k^{-1}\langle G \Sigma\rangle_k, \\
		\mathcal{U}_\text{cor} &=-\langle \Pi W\rangle_k[\langle \Pi\rangle_k+\langle \Pi W \Pi\rangle_k]^{-1}\langle W \Pi\rangle_k+\langle \Pi\rangle_k\langle W\rangle_k[\langle \Pi\rangle_k+\langle \Pi\rangle_k\langle W\rangle_k\langle \Pi\rangle_k]^{-1}\langle W\rangle_k\langle \Pi\rangle_k.
	\end{aligned}
\end{equation}
\\
\end{widetext}

Note that momentum-independent $\Sigma$ and $\Pi$ yield by construction vanishing corrections $\Delta_\text{cor}=0$ and $\mathcal{U}_\text{cor}=0$. Therefore, the EDMFT self-consistency scheme, which assumes $\Sigma=\Sigma_\text{imp}$, $\Pi=\Pi_\text{imp}$ is not altered.\cite{ayral2013} The correction terms however have a nontrivial effect if $\Sigma$ and $\Pi$ are $k$-dependent.

\subsubsection{$GW$ method}

In this subsection, we discuss the $GW$ method,\cite{aryasetiawan1998,sun2002,biermann2003} which allows to construct a $k$-dependent self-energy $\Sigma^{GW}(k,i\omega)$ and polarization $\Pi^{GG}(k,i\omega)$ by truncating Hedin's vertex function to its first order.\cite{hedin1965,pavarini2011a} This is a weak-coupling approach that can be expected to work for weak or moderate interactions, or (in combination with EDMFT) for the less dominant nonlocal components in more strongly interacting systems. Hedin derived a set of exact coupled differential equations which link the Green's function $G$, self-energy $\Sigma$, vertex function $\Gamma$, screened interaction $W$, and polarization $\Pi$:\cite{hedin1965}
\begin{equation}
	\left\{
	\begin{aligned}
		\Sigma(1,2)&=-\int d(34) G\left(1,3^{+}\right) W(1,4) \Gamma(3,2,4), \\
		G(1,2)&=G_{0}(1,2)+\int d(34) G_{0}(1,3) \Sigma(3,4) G(4,2), \\
		\Gamma(1,2,3)&=\delta(1-2) \delta(2-3) \\&+\int d(4567) \frac{\partial \Sigma(1,2)}{\partial G(4,5)} G(4,6) G(7,5) \Gamma(6,7,3), \\
		\Pi(1,2)&=\int d(34) G(1,3) \Gamma(3,4,2) G\left(4,1^{+}\right), \\
		W(1,2)&=v(1,2)+\int d(34) v(1,3) \Pi(3,4) W(4,2),
	\end{aligned}
	\right.
\end{equation}
Here, the numbers stand for $i\equiv (\tau_i,\mathbf{r}_i)$ and the spin indices are omitted. The second and the fifth equations are the fermionic and bosonic lattice Dyson equations. The $GW$ approximation neglects the vertex corrections in the third equation by setting $\Gamma(1,2,3)\approx\delta(1-2)\delta(2-3)$, even though this violates the Pauli principle.~\cite{pavarini2011a} With this approximation the first and the forth equations simplify to 
\begin{equation}\label{eq:GWA_1}
	\left\{
	\begin{aligned}
		\Sigma(1,2)&=- G(1,2^{+}) W(1,2), \\
		\Pi(1,2)&=G(1,2) G(2,1^{+}), 
	\end{aligned}
	\right.
\end{equation}
or in the notation with site indices and imaginary time,
\begin{equation}
	\left\{
	\begin{aligned} 
		\Sigma^{G W}_{ij}(\tau)&=-G_{ij}(\tau)W_{ij}(\tau),\\
		\Pi^{GG}_{ij}(\tau)&=2G_{ij}(\tau) G_{ji}(-\tau).
	\end{aligned}
	\right.
\end{equation}
Here the factor of two comes from the spin summation. In momentum space, we have
\begin{equation}\label{eq:GW_sigma_pi}
	\begin{aligned}
		\Sigma^{G W}(k,\tau)&=-\frac{1}{S_\text{BZ}}\int d p G(p,\tau)W(k-p ,\tau)=-G*W,\\
		\Pi^{G G}(k,\tau)&=\frac{2}{S_\text{BZ}}\int d{p}G({p},\tau) G({p}-{k},-\tau)=2G*G,
	\end{aligned}
\end{equation}
where ``$*$" denotes a momentum convolution, and $S_\text{BZ}$ the integral over the first BZ. 

%\begin{comment}
%In practical calculations, we separate $W(k,\tau)=v_k\delta(\tau)+W^{c}(k,\tau)$, so that the self-energy becomes
%\begin{equation}
%	\begin{aligned}
%	\Sigma^{G W}({k},\tau)=&-\frac{1}{S_{BZ}}\int d{p} %2 
%	\left [G({p},0^+)v({k}-{p})\right.\\&\left.+G({p},\tau)W^c({k}-{p} ,\tau)\right ],
%	\end{aligned}
%\end{equation}
%where $v(k)\equiv v_k$ is the bare interaction (the non-interacting $W_0$ in the lattice Dyson equation). In the Hartree-Fock approximation, the self-energy is $\Sigma^{HF}=- G*v$, corresponding to the instantaneous term. Thus, the $GW$ approximation can be viewed as a generalization of the Hartree-Fock approximation, where in the retarded contribution, the bare Coulomb interaction $v$ is replaced by the dynamically screened Coulomb interaction $W$, which yields $-G*W^c$.\cite{ayral2012}
%The $GW$ approximation may be viewed as a correction to the random phase approximation (RPA),\cite{nilsson2017} which gives the polarization $\Pi_\text{RPA}=2 G_0*G_0$ and the screened interaction $W_{\operatorname{RPA}}=v/(1-v\Pi_\text{RPA})$. In particular, the $GW$ approximation replaces $\Pi_\text{RPA}$ by the bubble of interacting Green's functions, $G_0*G_0 \rightarrow G*G$.
%\end{comment}

In 1960 Luttinger and Ward \cite{luttinger1960} proposed an explicit rule to construct a functional $\Psi[G]$, from which the self-energy $\Sigma$ can be obtained by a functional derivative. Later, Almbladh\cite{almbladh1999} extended this ``LW functional" to $\Psi[G,W]$ in a formalism which includes the polarization $\Pi$ and the self-energy, and which yields the following variational relations,
\begin{equation}
	\left(\frac{\delta\Psi[G,W]}{\delta G}\right)_W=\Sigma,\quad \left(\frac{\delta\Psi[G,W]}{\delta W}\right)_G=-\frac{1}{2}\Pi.
\end{equation}
Looking at Eq.~\eqref{eq:GWA_1}, one finds that the $GW$ approximation corresponds to the functional $\Psi^{GW}[G,W]=-\frac{1}{2}\tr[GWG]$, which is the lowest order approximation in $W$.\cite{nilsson2017}

\subsubsection{$GW$+EDMFT schemes}

As mentioned before, EDMFT and $GW$ are complementary methods, in the sense that the former allows to capture strong local correlations and the latter weak nonlocal correlations. It is thus natural to split the functional  $\Psi[G,W]$ into two terms, such that EDMFT contributes to its local part and the $GW$ scheme contributes to its non-local part,
\begin{equation}
	\Psi \approx \Psi^{\mathrm{EDMFT}}\left[G_{i i}, W_{i i}\right]+\Psi_{\text {nonloc }}^{G W}\left[G_{i j}, W_{i j}\right].
\end{equation}
The local contribution of the $GW$ approximation is subtracted to avoid double counting,
\begin{equation}
	\Psi_{\text {nonloc }}^{G W}\left[G_{i j}, W_{i j}\right]=\Psi^{G W}\left[G_{i j}, W_{i j}\right]-\Psi^{G W}\left[G_{i i}, W_{i i}\right].
\end{equation}
Applying the relations $\Sigma=\frac{\delta \Psi}{\delta G}$ and $  \Pi=-2 \frac{\delta \Psi}{\delta W}$, we then obtain
\begin{equation}
	\left\{
	\begin{aligned}
		\Sigma_{ij}=\Sigma^{\operatorname{EDMFT}}_{ii}\delta_{ij}+\Sigma^{GW}_{ij}(1-\delta_{ij}),\\
		\Pi_{ij}=\Pi^{\operatorname{EDMFT}}_{ii}\delta_{ij}+\Pi^{GG}_{ij}(1-\delta_{ij}).
	\end{aligned}
	\right. 
\end{equation}
Substituting this non-local self-energy and polarization into the self-consistency equations \eqref{eq:EDMFT_Generalised_sc} and \eqref{eq:EDMFT_Generalised_correction}, we can solve the lattice model in a self-consistent way. Ignoring the correction terms $\Delta_\text{cor}$ and $\mathcal{U}_\text{cor}$ in Eq.~\eqref{eq:EDMFT_Generalised_sc} yields the standard (local) $GW$+EDMFT scheme, while including these terms corresponds to the causal $GW$+EDMFT scheme proposed in Ref.~\onlinecite{backes2020}.

In the following, we summarize the two $GW$+EDMFT self-consistency loops, see also Fig.~\ref{fig:EDMFT+GW}. The terms in brackets are added in the causal variant.  

\begin{enumerate}
\item Start from some initial guess for the lattice self-energy $\Sigma(k,i\omega_n)$ and the lattice polarization $\Pi(k,i\nu_n)$.

\item With the lattice Dyson equation, calculate the lattice Green's function $G$ and screened interaction $W$,
\begin{equation}
	\begin{aligned}
		& G^{-1}(k,i\omega_n) = G_0^{-1}(k,i\omega_n)-\Sigma(k,i\omega_n), \\
		& W^{-1}(k,i\nu_n) = W_0^{-1}(k,i\nu_n)-\Pi(k,i\nu_n),
	\end{aligned}
\end{equation}
where the non-interacting lattice Green's function $G_0$ and screened interaction $W_0$ are given by
\begin{equation}
	\begin{aligned}
		& G_0^{-1}(k,i\omega_n) = i\omega_n+\mu-\epsilon_k,\\
		& W_0^{-1}(k,i\nu_n) = v^{-1}_k.
	\end{aligned}
\end{equation}

\item Use the (causal) EDMFT self-consistent equations in Eq.~\eqref{eq:EDMFT_Generalised_sc} 
to calculate the dynamical mean fields $\mathcal{G}(i\omega_n)$ and $\mathcal{U}(i\nu_n)$ of the impurity model,
\begin{equation*}
	\begin{aligned}
		\mathcal{G}^{-1}\left(i\omega_n\right) &=\quad\langle G\rangle^{-1}+\langle\Sigma\rangle \,\,\,\, (+\,\,\Delta_\text{cor}),\\
		\mathcal{U}(i \nu_n)&=\left[\langle W\rangle^{-1}+\langle \Pi\rangle\right]^{-1} \,\, (+\,\,\mathcal{U}_\text{cor}).
	\end{aligned}
\end{equation*}

\item Solve the impurity model defined by $\mathcal{G}(i\omega_n),\mathcal{U}(i,\nu_n)$ numerically with the continuous-time quantum Monte Carlo (CTQMC) method.\cite{werner2006,werner2010} The solver produces the impurity Green's function $G_{\text{imp}}(i\omega_n)$ and density-density correlation function $\chi_{\text{imp}}(i\nu_n)$. The bosonic Green's function can be calculated with the relation
\begin{equation}
	W_{\text{imp}}(i\nu)=\mathcal{U}(i\nu_n)-\mathcal{U}(i\nu_n)\chi_{\text{imp}}(i\nu_n)\mathcal{U}(i\nu_n).
\end{equation}

\begin{figure}[t]
	\centering
	\includegraphics[width=\linewidth]{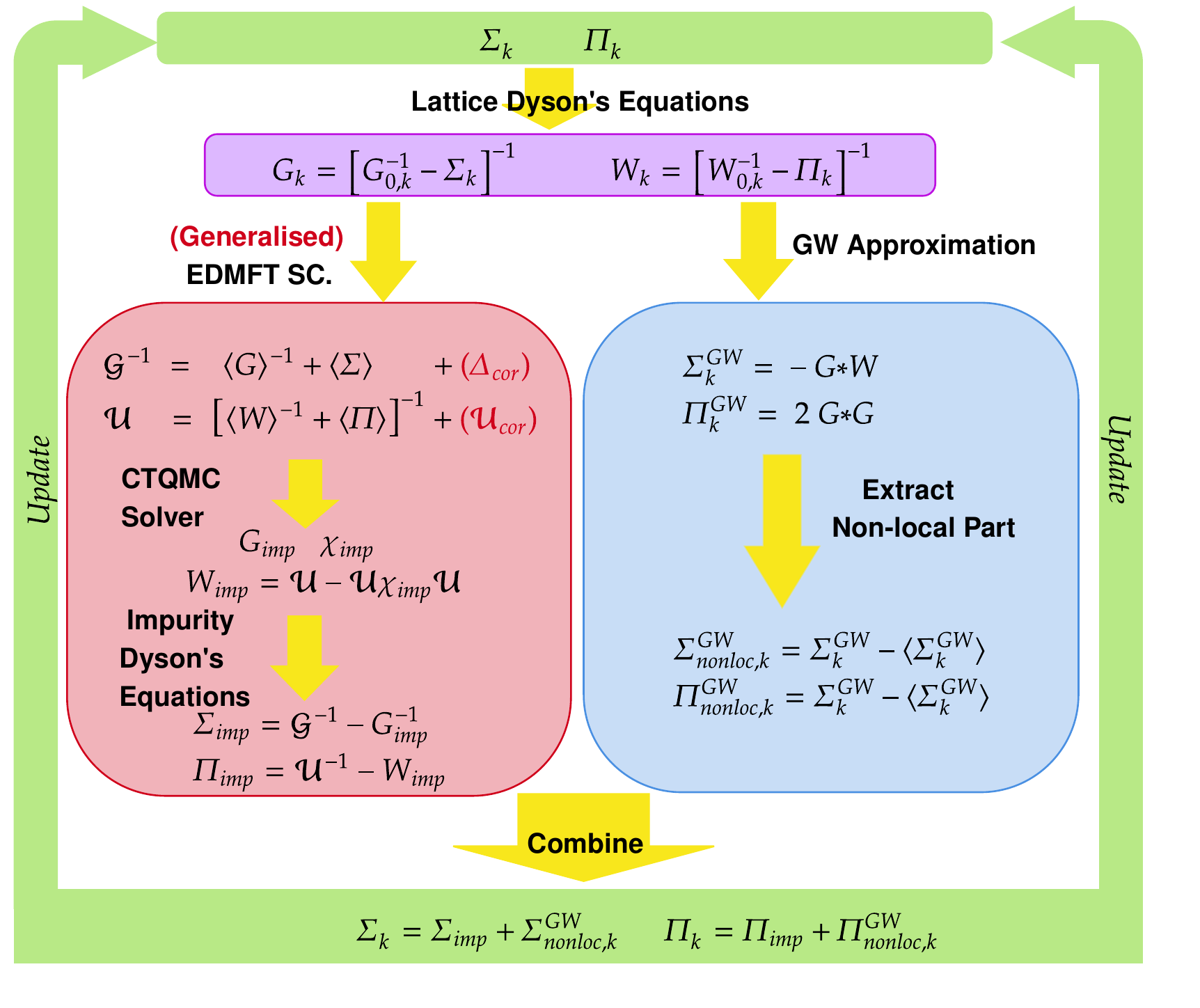}
	\caption{Flow-chart of the $GW$+EDMFT schemes. The red parts in brackets are added in the causal version of $GW$+EDMFT, while in the absence of these terms, the scheme corresponds to the standard local version of $GW$+EDMFT.}\label{fig:EDMFT+GW}
\end{figure}

\item Extract the impurity self-energy and polarization from the impurity Dyson equations,
\begin{equation}
	\begin{aligned}
		& \Sigma_{\text{imp}}(i\omega_n) = \mathcal{G}^{-1}(i\omega_n)-G_{\text{imp}}^{-1}(i\omega_n),\\
		& \Pi_{\text{imp}}(i\nu_n) = \mathcal{U}^{-1}(i\nu_n) - W_{\text{imp}}^{-1}(i\nu_n).
	\end{aligned}    
\end{equation}

\item Calculate the $GW$ self-energy and polarization,
\begin{equation}
	\begin{aligned}
		\Sigma^{GW}(k,\tau)=&-\frac{1}{N_p}\sum_p G(p,\tau)W^c(k-p,\tau)\\&-\frac{1}{N_p}\sum_p G(p,0^+)v(k-p),\\
		\Pi^{GG}(k,\tau)=&2\frac{1}{N_p}\sum_p G(p,\tau)G(p-k,-\tau),
	\end{aligned}
\end{equation}
and extract their non-local parts,
\begin{equation}
	\begin{aligned}
		\Sigma^{GW}_\text{nonloc}(k,\tau)&=\Sigma^{GW}(k,\tau)-\frac{1}{N_k}\sum_{k}\Sigma^{GW}(k,\tau),\\
		\Pi^{GG}_\text{nonloc}(k,\tau)&=\Pi^{GW}(k,\tau)-\frac{1}{N_k}\sum_k\Pi^{GG}(k,\tau).
	\end{aligned}
\end{equation}
Here, $\frac{1}{N_k}\sum_k\ldots$ denotes the numerical average over the first BZ and $W^{c}(k,\tau)=W(k,\tau)-v_k\delta(\tau).$
 
\item Obtain the $GW$+EDMFT self-energy and polarization by combining the non-local $GW$ contribution and the local EDMFT contribution,
\begin{equation}
	\begin{aligned}
		\Sigma^{GW+\text{EDMFT}}(k,i\omega_n)&=\Sigma_\text{imp}(i\omega_n)+\Sigma^{GW}_\text{nonloc}(k,i\omega_n),\\
		\Pi^{GW+\text{EDMFT}}(k,i\nu_n)&=\Pi_\text{imp}(i\omega_n)+\Pi^{GG}_\text{nonloc}(k,i\omega_n).
	\end{aligned}
\end{equation}
\end{enumerate}

With this, we have updated the momentum-dependent lattice self-energy $\Sigma(k,i\omega)$ and polarization $\Pi(k,i\nu_n)$ and can go back to step 2 for the next loop. Upon convergence, we obtain the fermionic and bosonic Green's functions in the imaginary frequency (or time) domain. The complete $GW$+EDMFT scheme is illustrated in Fig.~\ref{fig:EDMFT+GW}.

\section{Results}
\label{sec:results}

\subsection{General remarks}

In this section, we use the two self-consistency schemes to study the $GW$+EDMFT solution of the triangular-lattice extended Hubbard model with repulsive interactions $U$, $V>0$. First, we map out the phase diagram for hopping $t=-0.02$ eV, corresponding to a bandwidth $W=|{9t}|=0.18$ eV. This choice is motivated by the width of the Hubbard bands in the C phase of 1$T$-TaS$_2$.\cite{ligges2018,wu2021} Next, we compare the results from the causal and local self-consistency scheme in different regions of the phase diagram. Finally, we identify the interaction parameters appropriate for 1$T$-TaS$_2$ and study the doping dependence of the screening properties and the electronic structure. Unless otherwise stated, we run the simulations for inverse temperature $\beta=500$ eV$^{-1}$, corresponding to $T=23$~K. In the following, we measure energies in units of eV and omit the units for convenience. The lattice constant is set to $a=1$. 

\subsection{Phase diagram}\label{sec:phasediagram}

\begin{figure}[!]
	\centering
	\includegraphics[width=0.99\linewidth]{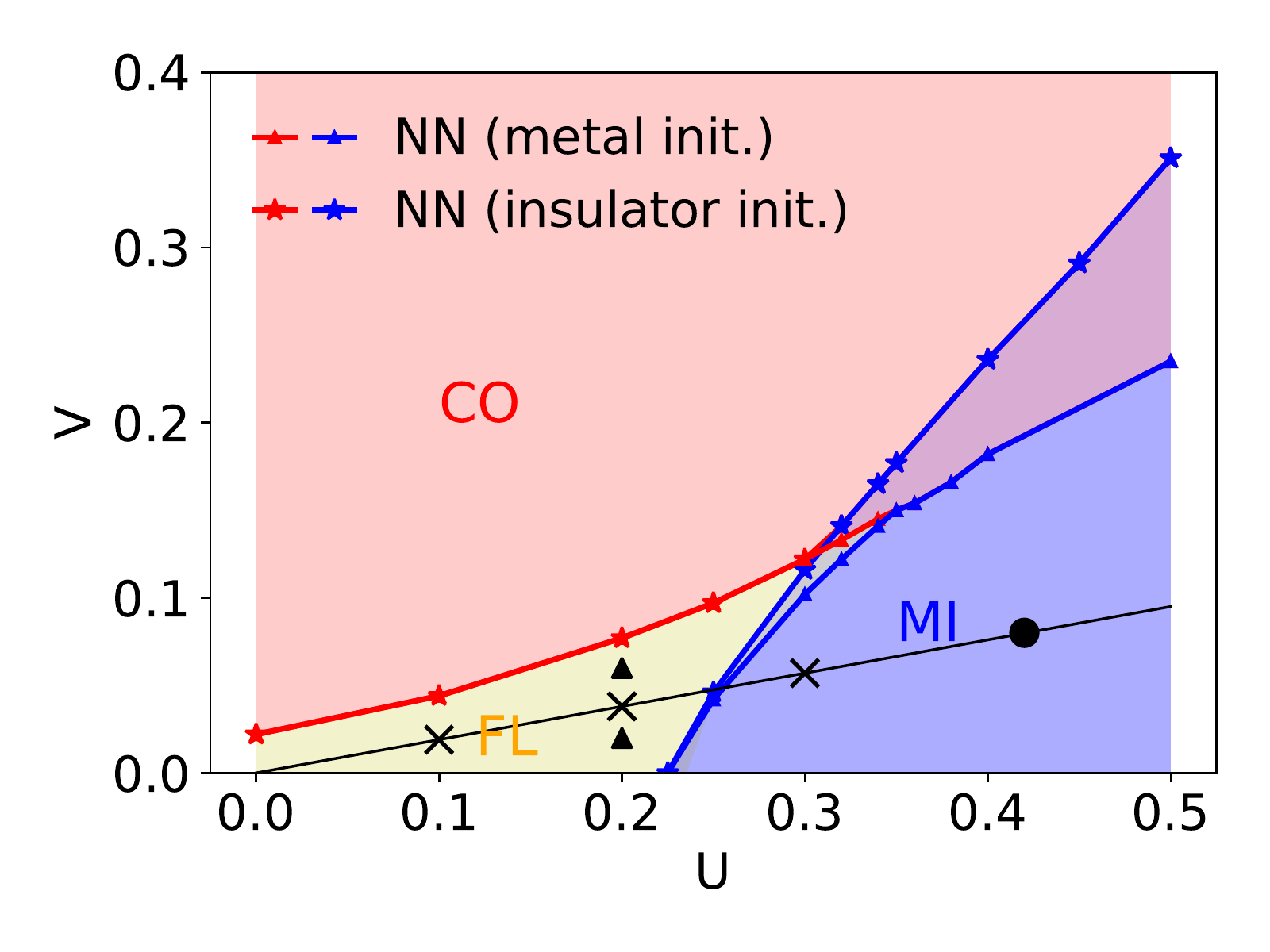}
	\includegraphics[width=0.99\linewidth]{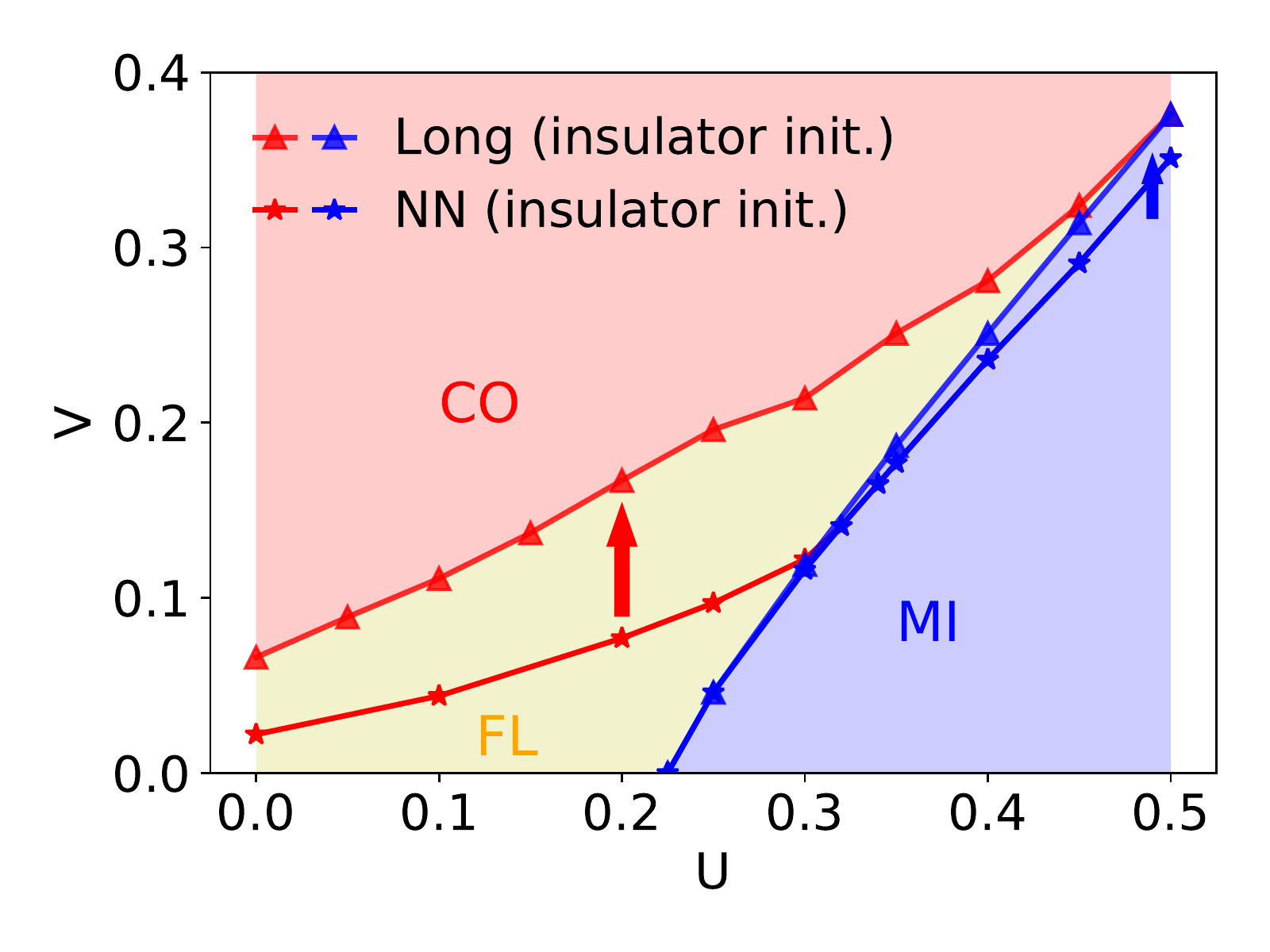}
	\caption{$GW$+DMFT phase diagrams of the half-filled single-band extended Hubbard model on the triangular lattice. Top: Phase diagram with only local and NN interactions, for an initial metallic (triangles) and Mott insulating (stars) solution in the self-consistency loop. The black marks indicate the $U$ and $V$ parameters which will be further studied below. The black line corresponds to the fixed ratio $V/U=0.19$. Bottom: Phase diagram for the model with long-range interaction treated by Ewald summation (triangles, starting from a Mott insulating solution), and for comparison the result for the NN interaction case (stars). The red and blue arrows indicate the shift of the phase boundary as a result of the longer-ranged interaction. (Local self-consistency scheme, $\beta=500$, $t=-0.02$.)
	}\label{fig:phase_diagram}
\end{figure}

We first apply the local $GW$+EDMFT scheme to map out the phase diagram of the half-filled extended Hubbard model on the triangular lattice.

In the weak coupling regime, $V/t \ll 1$ and $U/t\ll 1$, electrons can hop easily between neighboring sites, which results in a metallic phase (M). For dominant local interactions, $U/t \gg 1$ and $V/U\ll 1$, a Mott insulating state (MI) induced by strong local correlation is expected. This state, which minimizes the density of doublons, and thus the potential energy, is unstable to antiferromagnetic order on a bipartite lattice, if spin symmetry breaking is allowed. Similarly, in the $V/U\gg 1$ limit, the electrons on a bipartite lattice would arrange themselves in an ordered pattern, with one sublattice almost doubly occupied and the other sublattice almost empty, corresponding to a commensurate charge-ordered (CO) phase. Since the triangular lattice of our study is not bipartite, a suppression of the aforementioned ordering instabilities due to lattice frustration can be expected, although this effect is not fully captured by $GW$+EDMFT, which yields a result representative of the bipartite situation.
In the following, we will restrict our calculations to paramagnetic solutions, without imposing any constraint on the charge ordering tendencies for $V/U\gg 1$.  

An effect which is captured by $GW$+EDMFT regardless of the lattice geometry is that nonzero inter-site interactions $V$ produce a screening of the on-site $U$, through charge density fluctuations on other sites. We thus expect that the metallic phase is stabilized by increasing $V$.

\begin{figure*}[ht!]
	\centering
	\includegraphics[width=0.33\linewidth]{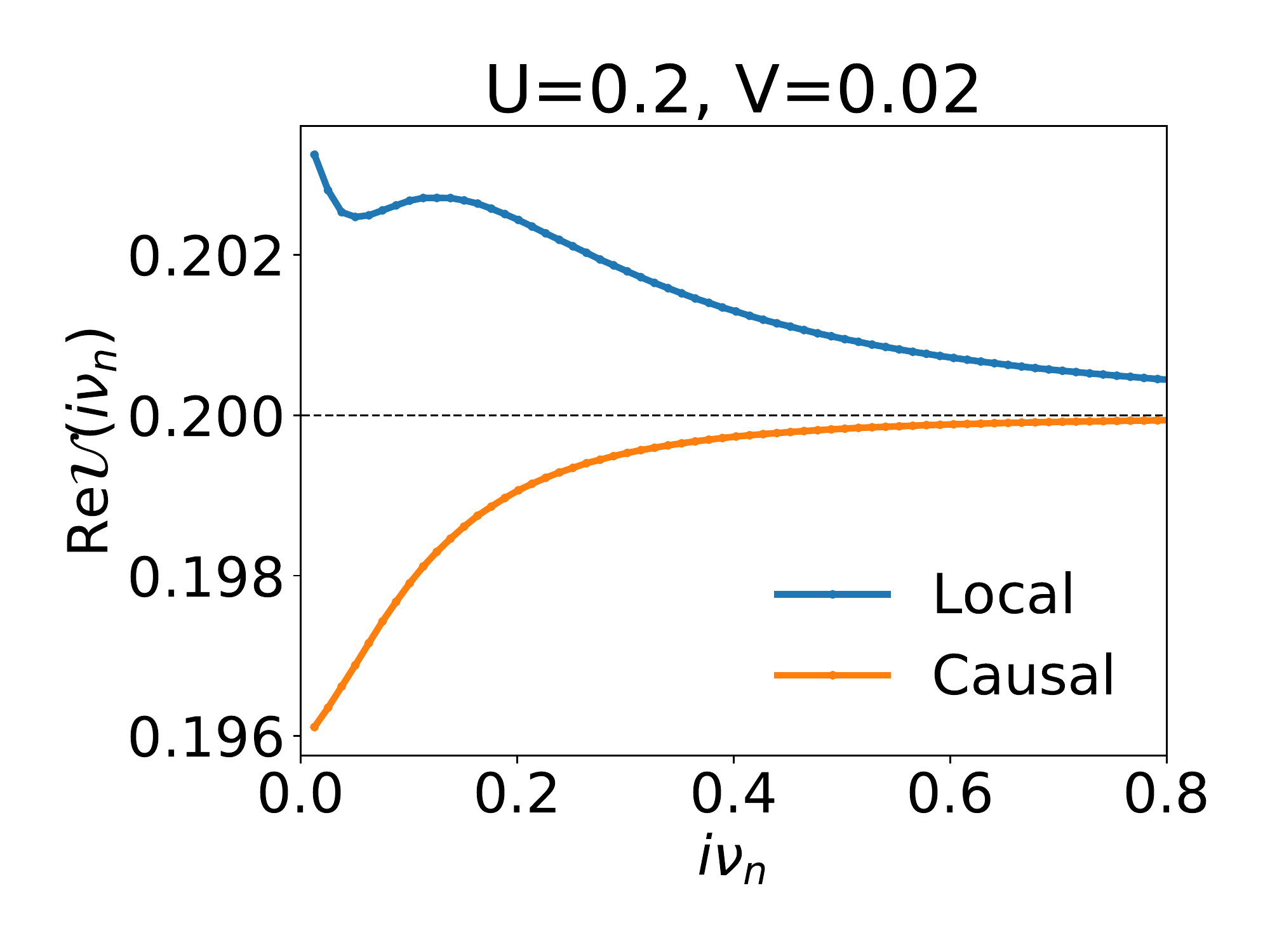}
	\includegraphics[width=0.33\linewidth]{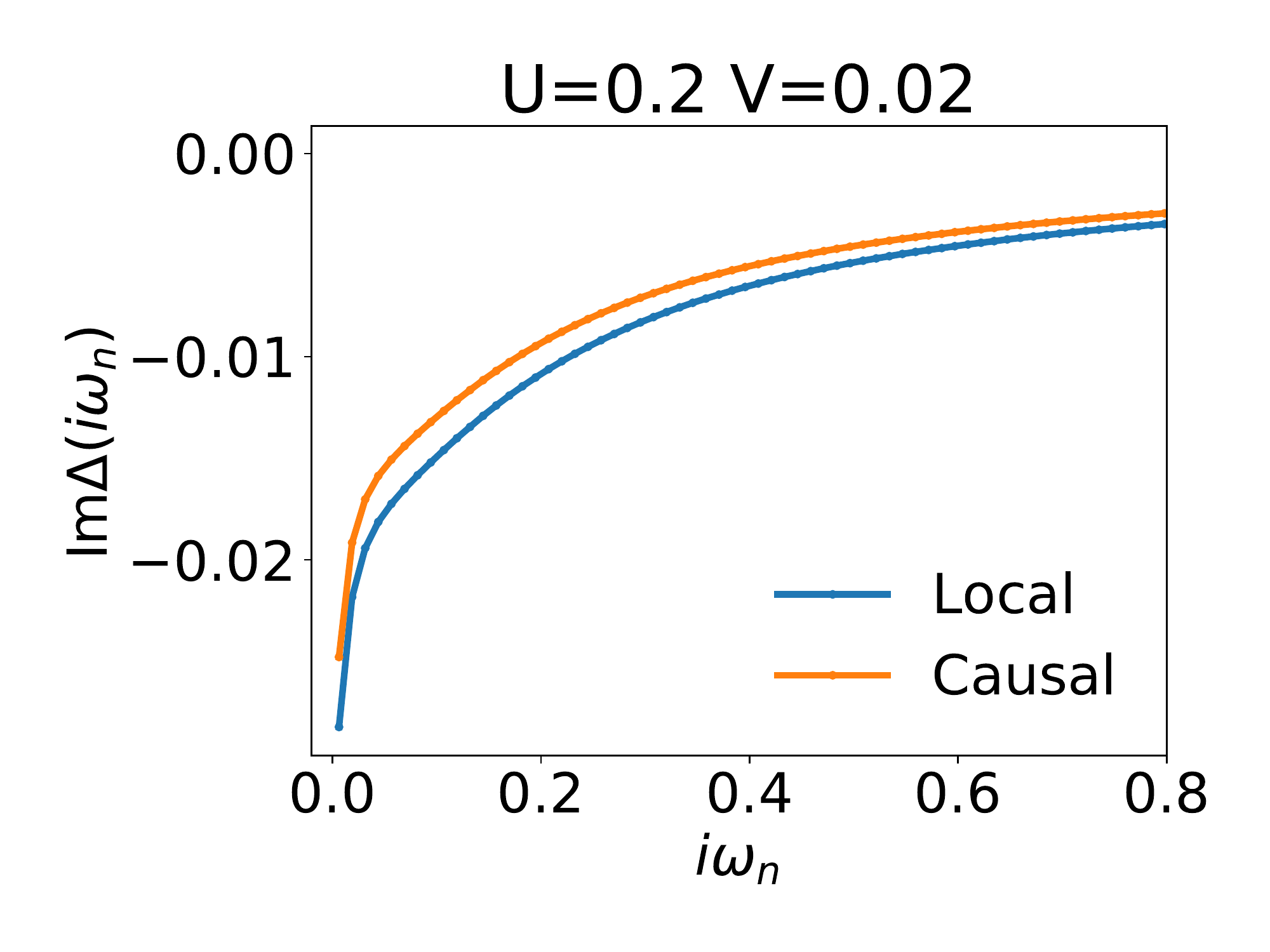}
	\includegraphics[width=0.33\linewidth]{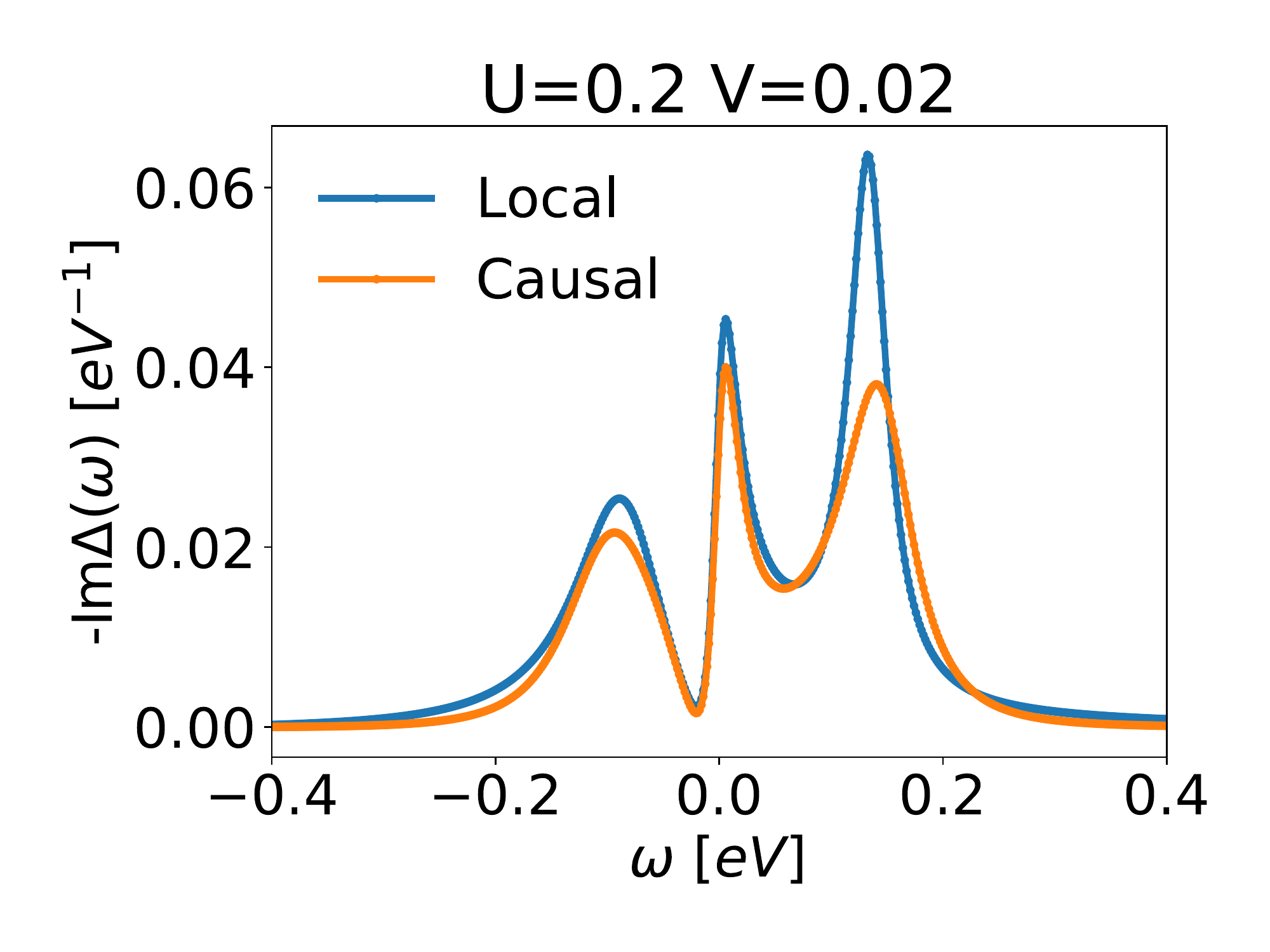}
	\includegraphics[width=0.33\linewidth]{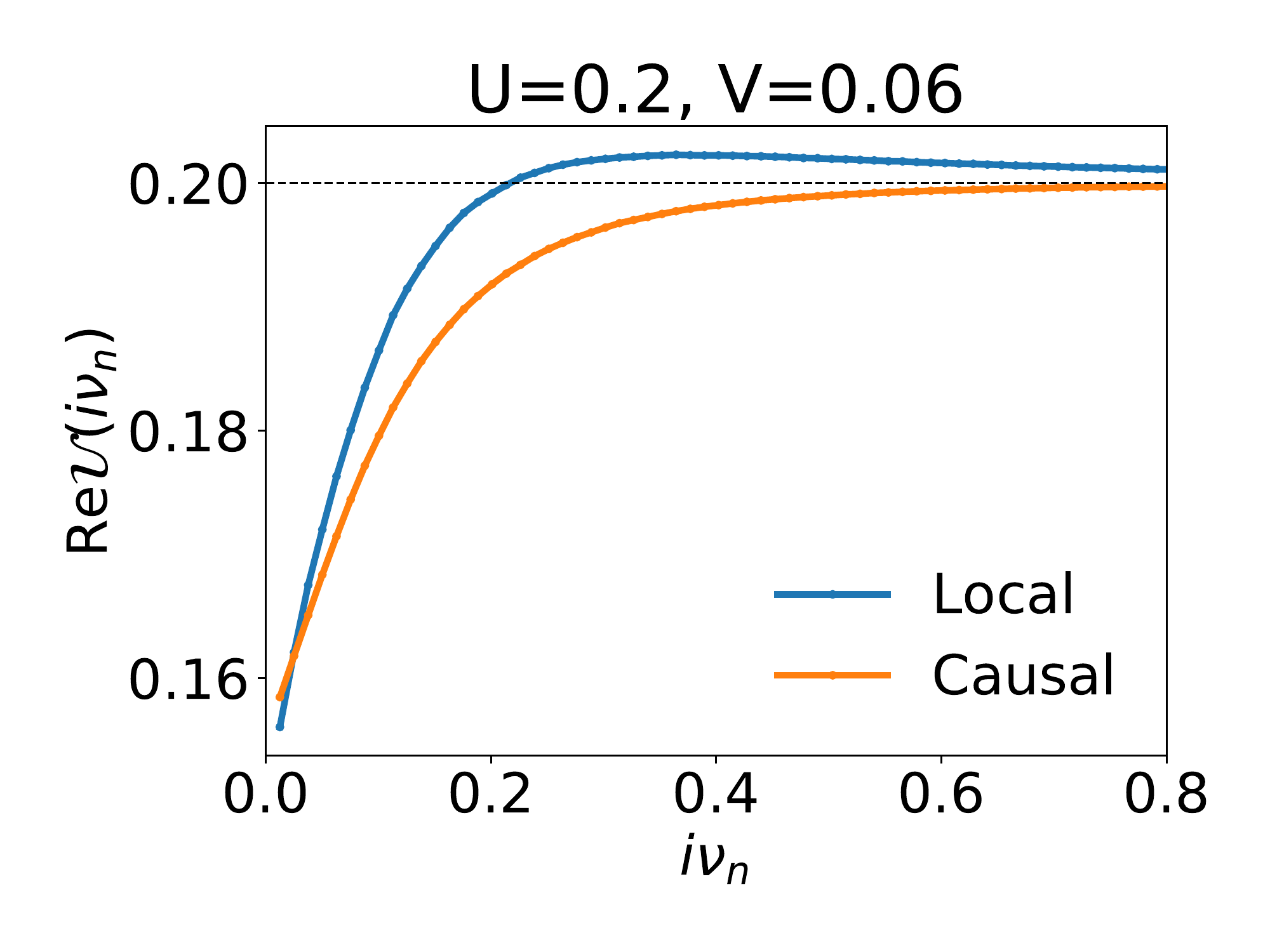}
	\includegraphics[width=0.33\linewidth]{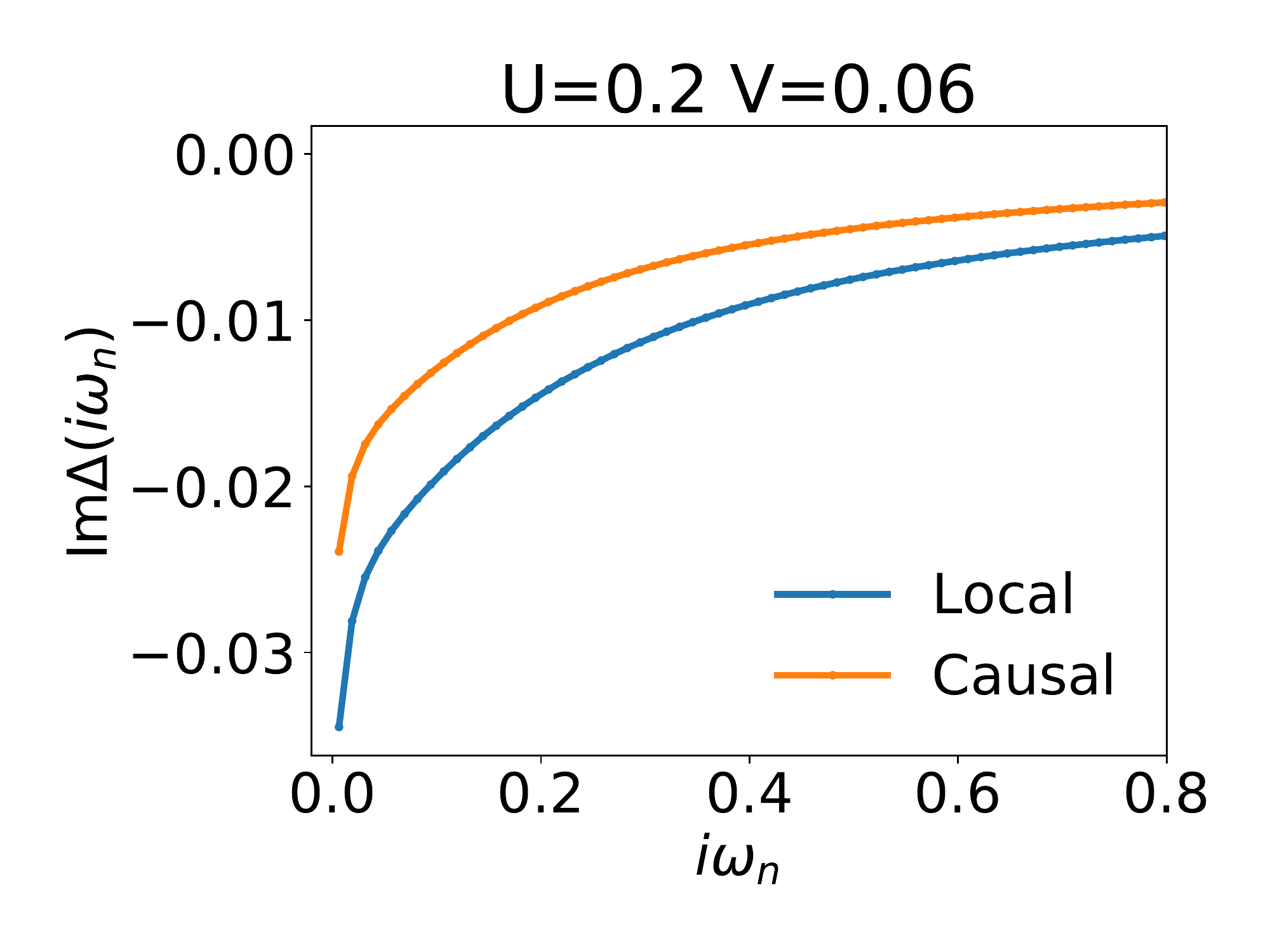}
	\includegraphics[width=0.33\linewidth]{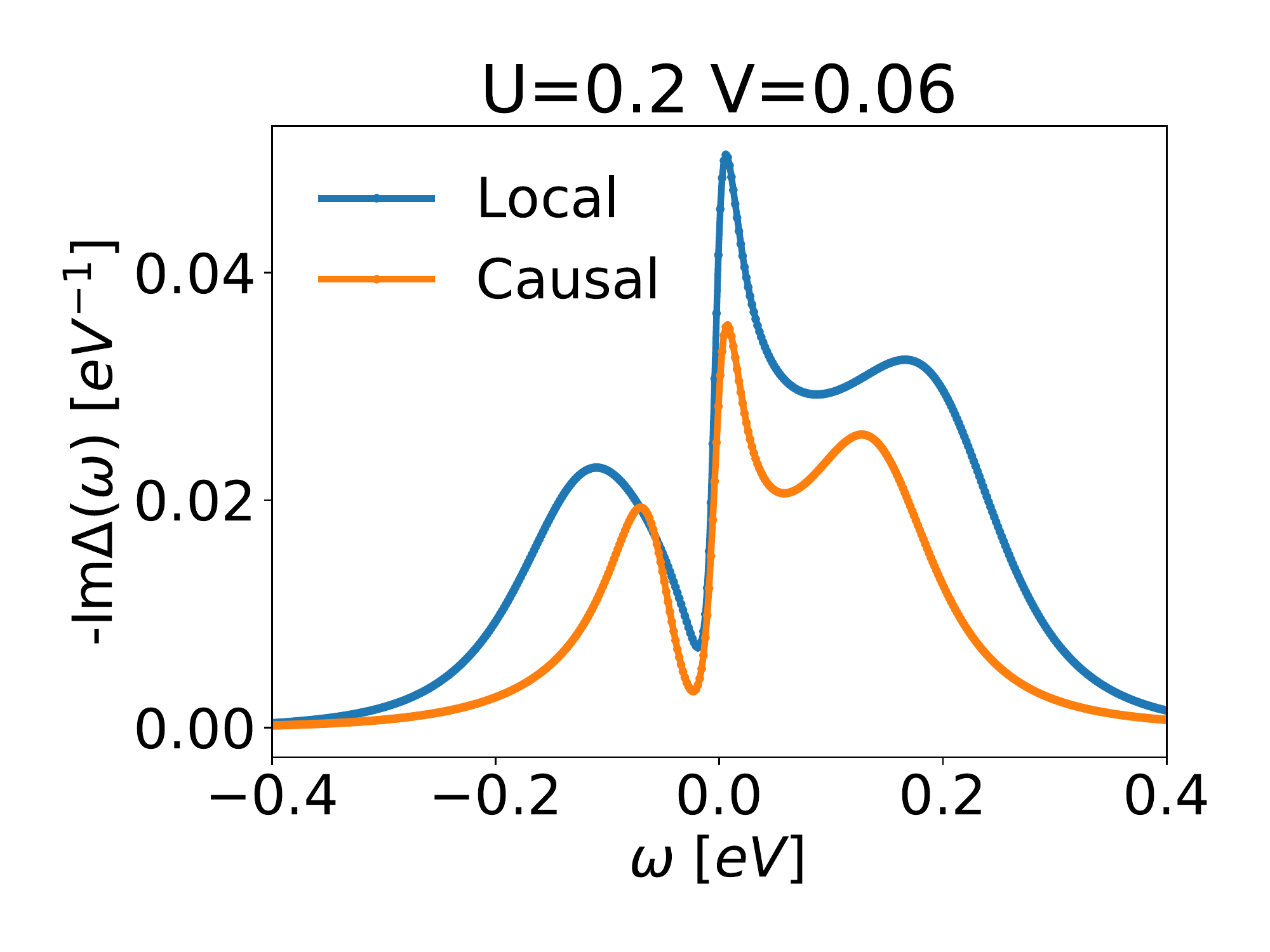}
	\caption{From left to right column: real part of the impurity interaction Re$\mathcal{U}(i\nu_n)$, imaginary part of the hybridization function Im$\Delta(i\omega_n)$ and spectral function $-\text{Im}\Delta(\omega)$ obtained by Pad\'e analytical continuation. The interaction parameters (triangles in the top panel of  Fig.~\ref{fig:phase_diagram}) are chosen so as to highlight the effect of an increase in the non-local interaction in the metallic phase. The data are for the model with NN nonlocal interactions, treated with the local and causal self-consistency scheme.}
	\label{fig:Weiss_V}
\end{figure*} 

In Fig.~\ref{fig:phase_diagram}, we present the phase diagrams of the triangular lattice model as a function of the local and non-local interaction strength in the $U$-$V$ domain for two kinds of non-local interactions: The top panel shows the result for the model with $V$ truncated to the nearest  neighbor (NN) lattice sites. Here we indicate two boundaries for the MI phase, obtained by starting the self-consistency loop from a metallic (line with triangles) and Mott insulating (line with stars) solution, respectively. The bottom panel shows the phase diagram for the model with long-range interactions treated with the Ewald lattice summation (proportional to the result shown in Fig.~\ref{fig:Vk}). The boundary between the Mott insulator and metal is determined from the low-frequency behavior of Im$G(i\omega_n)$,\cite{sun2002} while the boundary to the charge-ordered (CO) phase is deduced from the divergence of the local charge susceptibility $\chi_\text{loc}(i\nu=0)$.\cite{huang2014} Beyond this phase boundary we do not attempt to stabilize the CO phase, which might anyhow be an artifact of the DMFT-based treatment.  In the $V=0$ limit, i.e. the conventional Hubbard model, the Mott transition occurs at $U/W=1.25$. As $V$ is increased, the metallic phase is indeed stabilized and the phase boundaries for the two models start to deviate. In the bottom panel of  Fig.~\ref{fig:phase_diagram}, looking at the phase boundary to the CO phase, we note that including the long-range interactions suppresses the CO phase a lot in the large-$V$ and small/intermediate-$U$ region (indicated by the red arrow) and less so in the large-$U$ region (indicated by the blue arrow). This shows that within the $GW$+EDMFT description, the long-range interaction introduces additional frustration, which makes it difficult to stabilize the ordered phase. A similar effect was also observed in the case of the bipartite square lattice.\cite{huang2014}

\subsection{Comparison of the two self-consistency schemes}
\label{sec:comparison}

To compare the auxiliary fields and physical quantities obtained with the two self-consistency schemes, we simulate the half-filled extended Hubbard with different $U$ and $V$ values representative of the 
metal and Mott insulator phase,
see the black markers (crosses and triangles) 
in the top panel of Fig.~\ref{fig:phase_diagram}. The black line indicates the ratio $U/V=0.19$ which, as explained later, corresponds to the ratio estimated for 1$T$-TaS$_2$. Here, we focus on the results obtained with the NN nonlocal interaction. 

\begin{figure*}[ht!]
	\centering
	\includegraphics[width=0.33\linewidth]{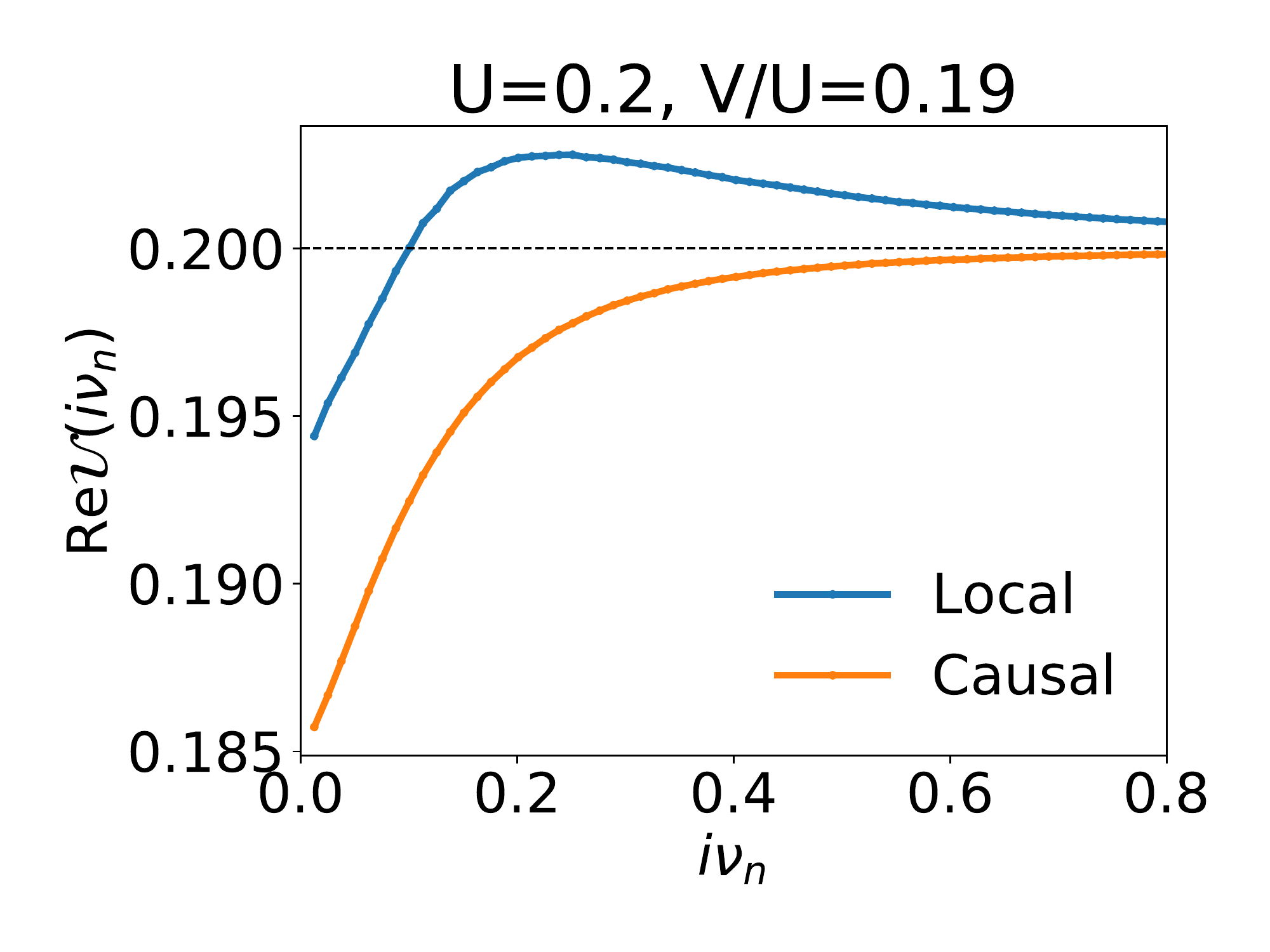}
	\includegraphics[width=0.33\linewidth]{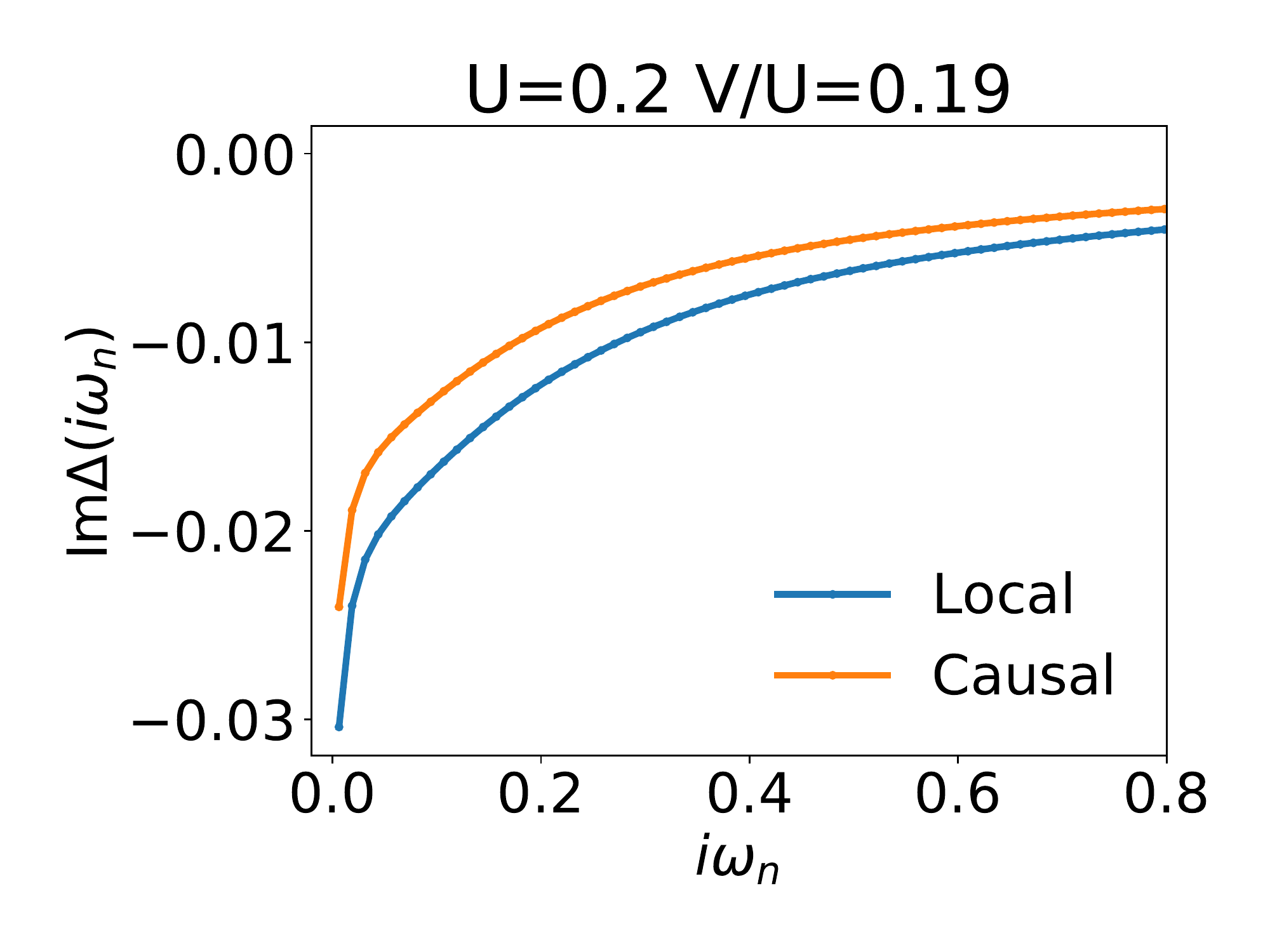}
	\includegraphics[width=0.33\linewidth]{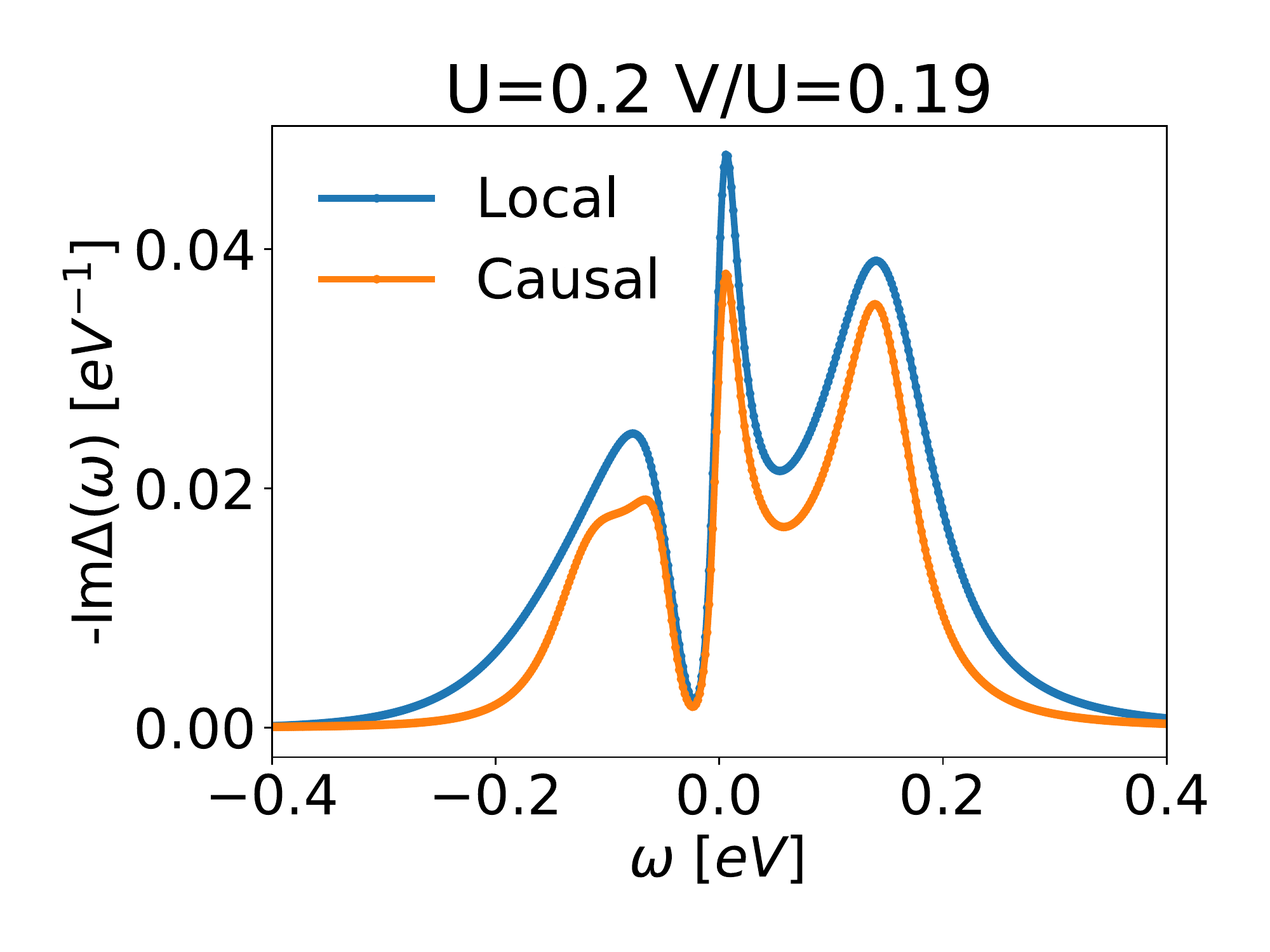}
	\includegraphics[width=0.33\linewidth]{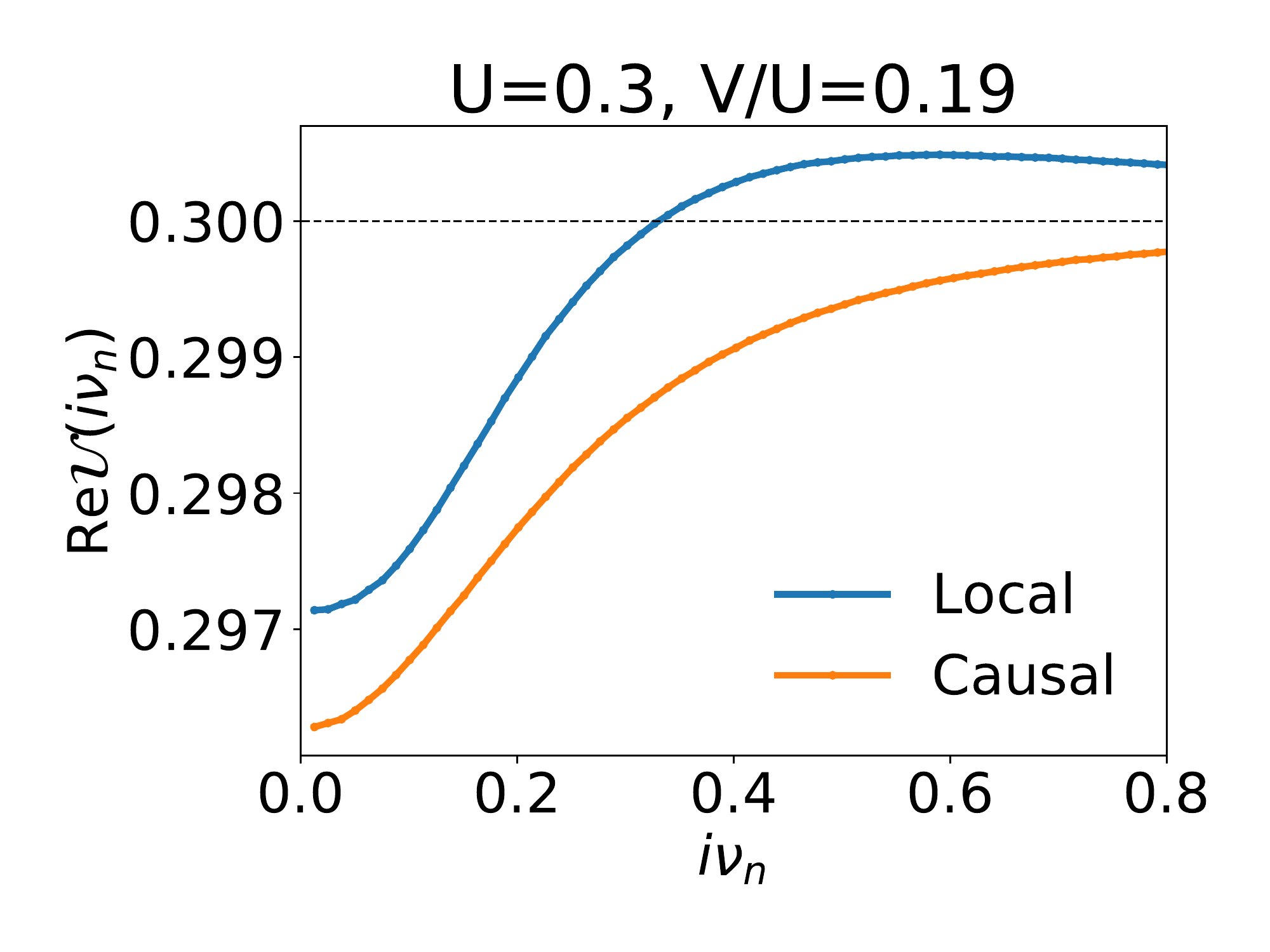}
	\includegraphics[width=0.33\linewidth]{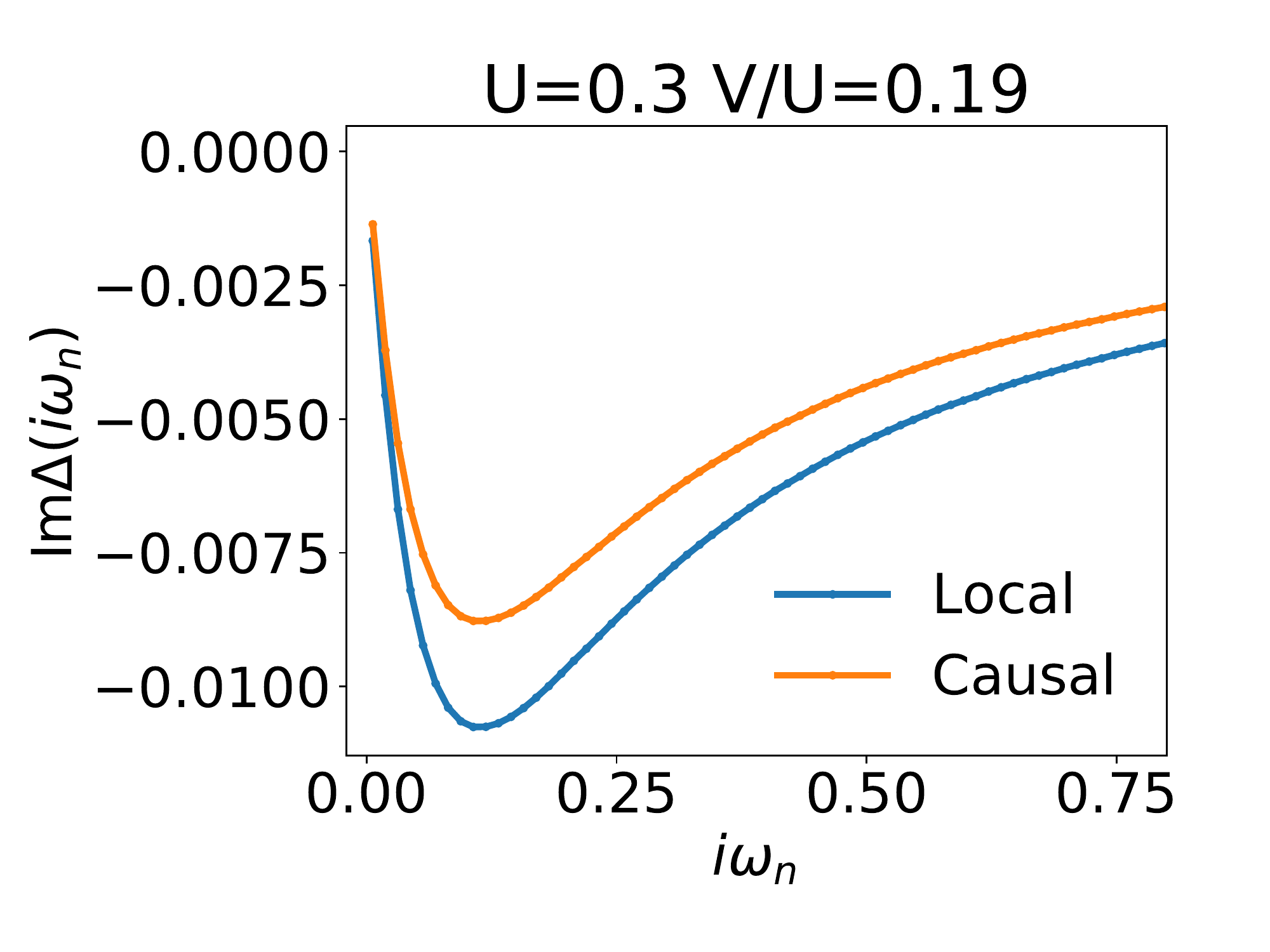}
	\includegraphics[width=0.33\linewidth]{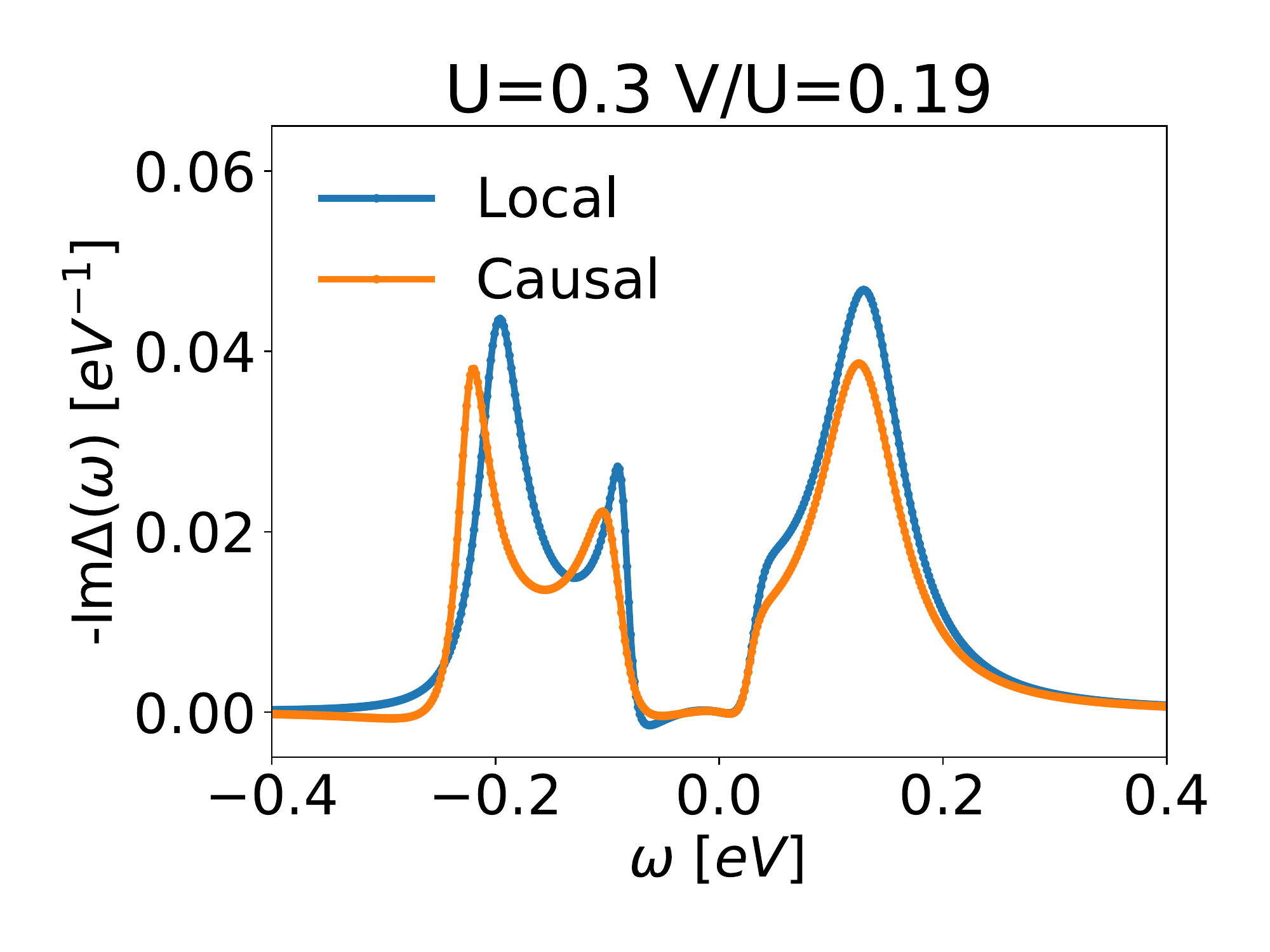}
	\caption{From left to right column: real part of the impurity interaction Re$\mathcal{U}(i\nu_n)$, imaginary part of hybridization function Im$\Delta(i\omega_n)$ and spectral function $-\text{Im}\Delta(\omega)$ for increasing onsite interaction $U$ and fixed $V/U$ ratio (crosses in the top panel of  Fig.~\ref{fig:phase_diagram}) with NN nonlocal interactions.
	}
	\label{fig:Weiss_U}
\end{figure*} 

In Fig.~\ref{fig:Weiss_V} and  Fig.~\ref{fig:Weiss_U}, we show the evolution of the bosonic and fermionic Weiss fields as a function of the interaction parameters of the model. 
The corresponding points in the phase diagram in the top panel of Fig.~\ref{fig:phase_diagram} are indicated by triangles for Fig.~\ref{fig:Weiss_V} and by crosses for Fig.~\ref{fig:Weiss_U}. In the latter case the first two solutions are in the metallic region, and the last one is Mott insulating. While the quantities defined on the Matsubara axis are a direct output of the $GW$+EDMFT simulation, the hybridization function on the real axis, $-\text{Im}\Delta(\omega)$ has been obtained by Pad\'e analytical continuation.\cite{vidberg1977} 

A noncausal impurity interaction $\mathcal{U}(i\nu_n)$ is characterized by the appearance of a negative slope on the Matsubara-frequency axis.\cite{nilsson2017} As we can see in Fig.~\ref{fig:Weiss_V} and  Fig.~\ref{fig:Weiss_U}, all the local effective interactions obtained with the local self-consistency scheme are noncausal (blue lines). In contrast, the impurity interactions obtained with the causal scheme (orange lines) feature a positive slope at all Matsubara frequencies, and hence are causal. This confirms that the modified self-consistency scheme proposed in Ref.~\onlinecite{backes2020} indeed eliminates noncausal features in the impurity interaction.

\begin{figure*}[ht!]
	\centering
	\includegraphics[width=0.33\linewidth]{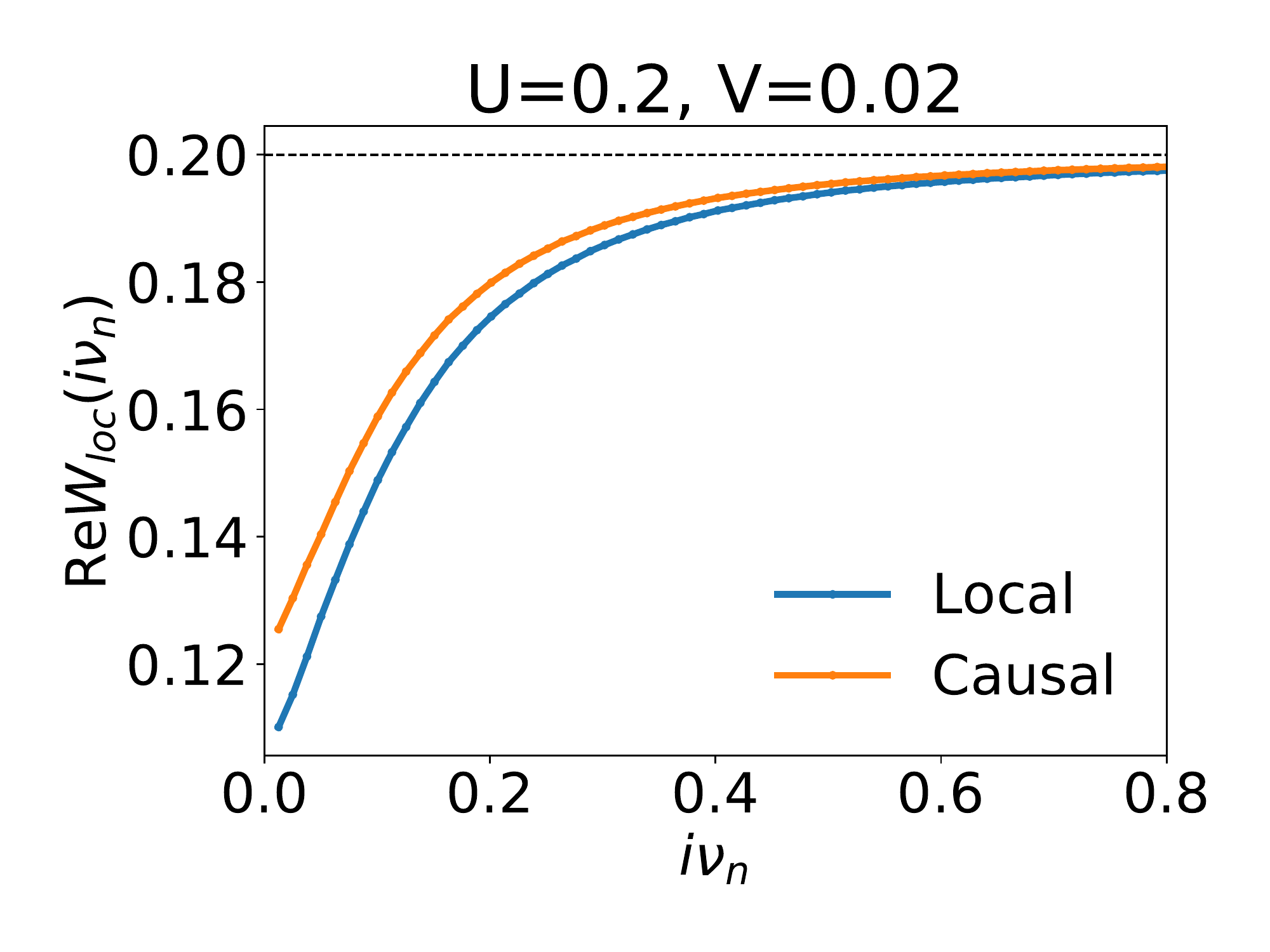}
	\includegraphics[width=0.33\linewidth]{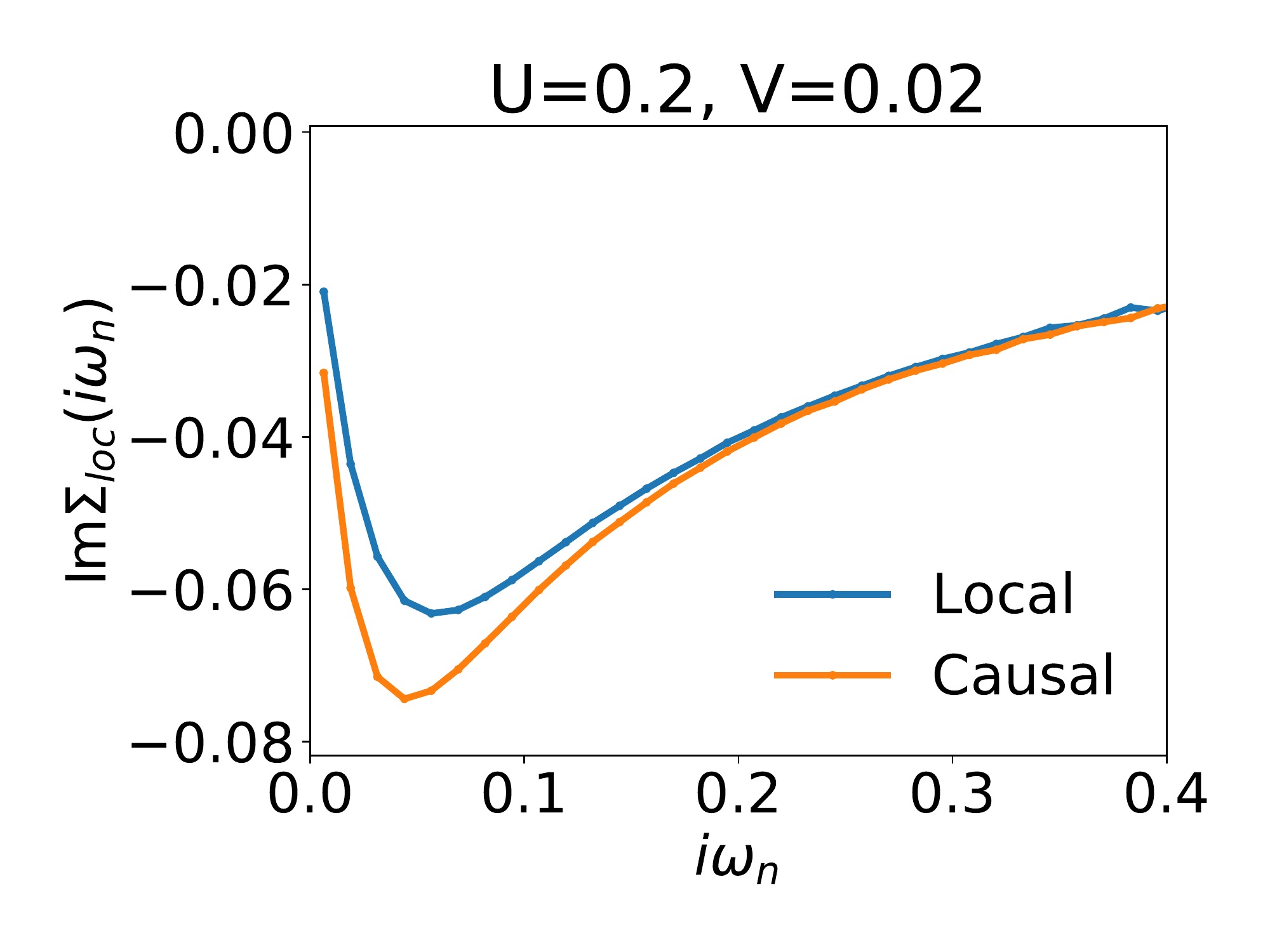}
	\includegraphics[width=0.33\linewidth]{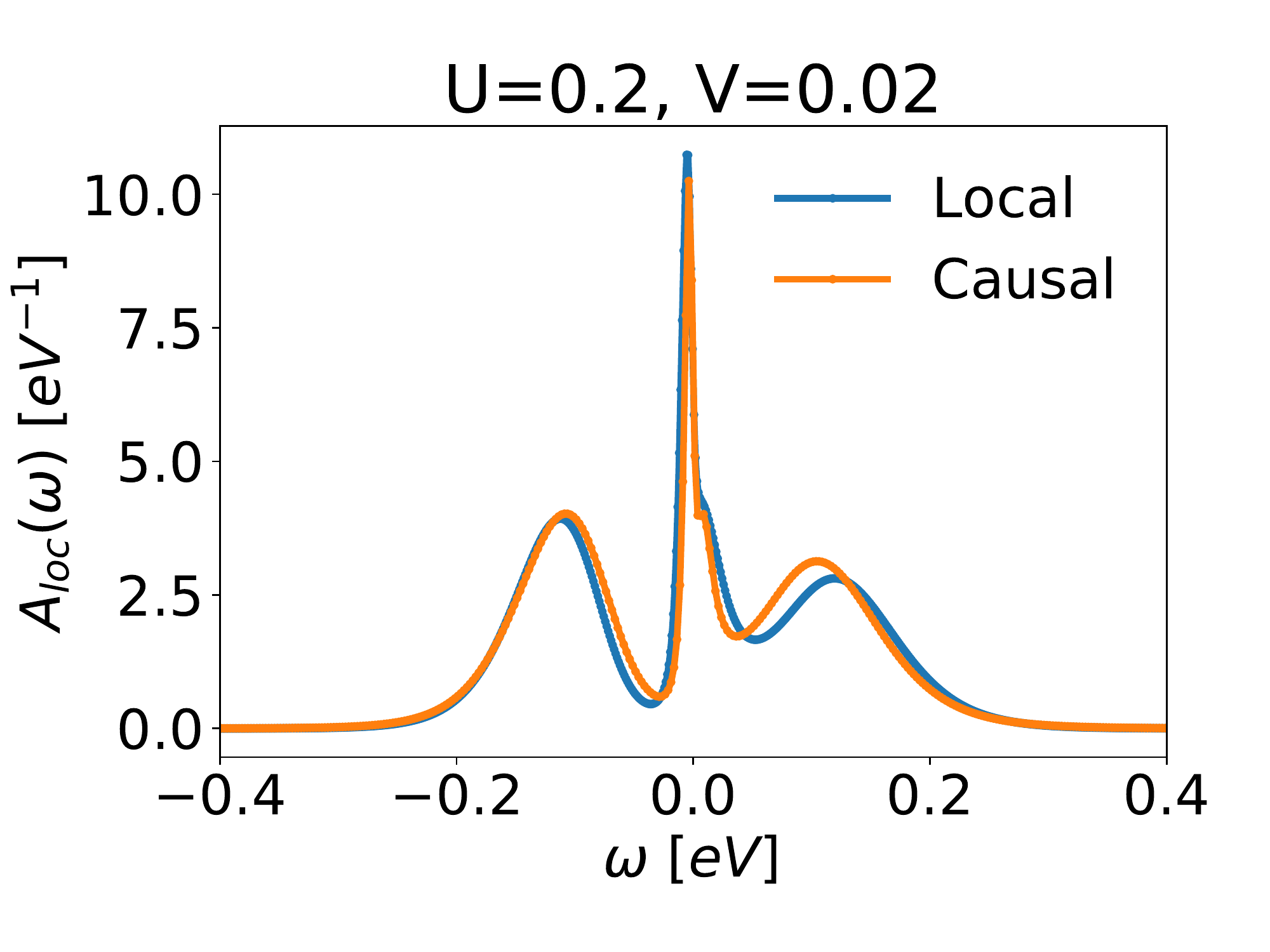}
	\includegraphics[width=0.33\linewidth]{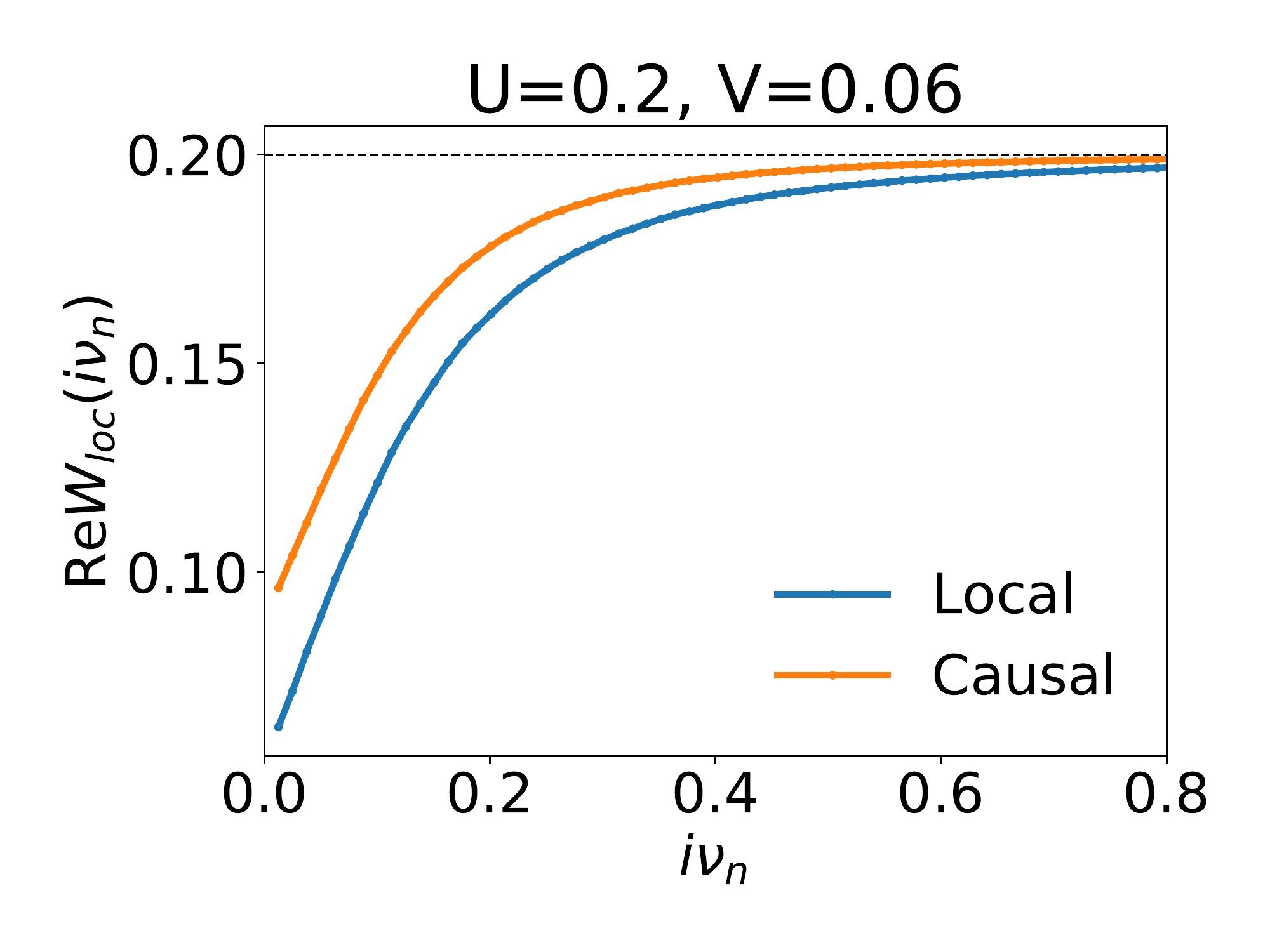}
	\includegraphics[width=0.33\linewidth]{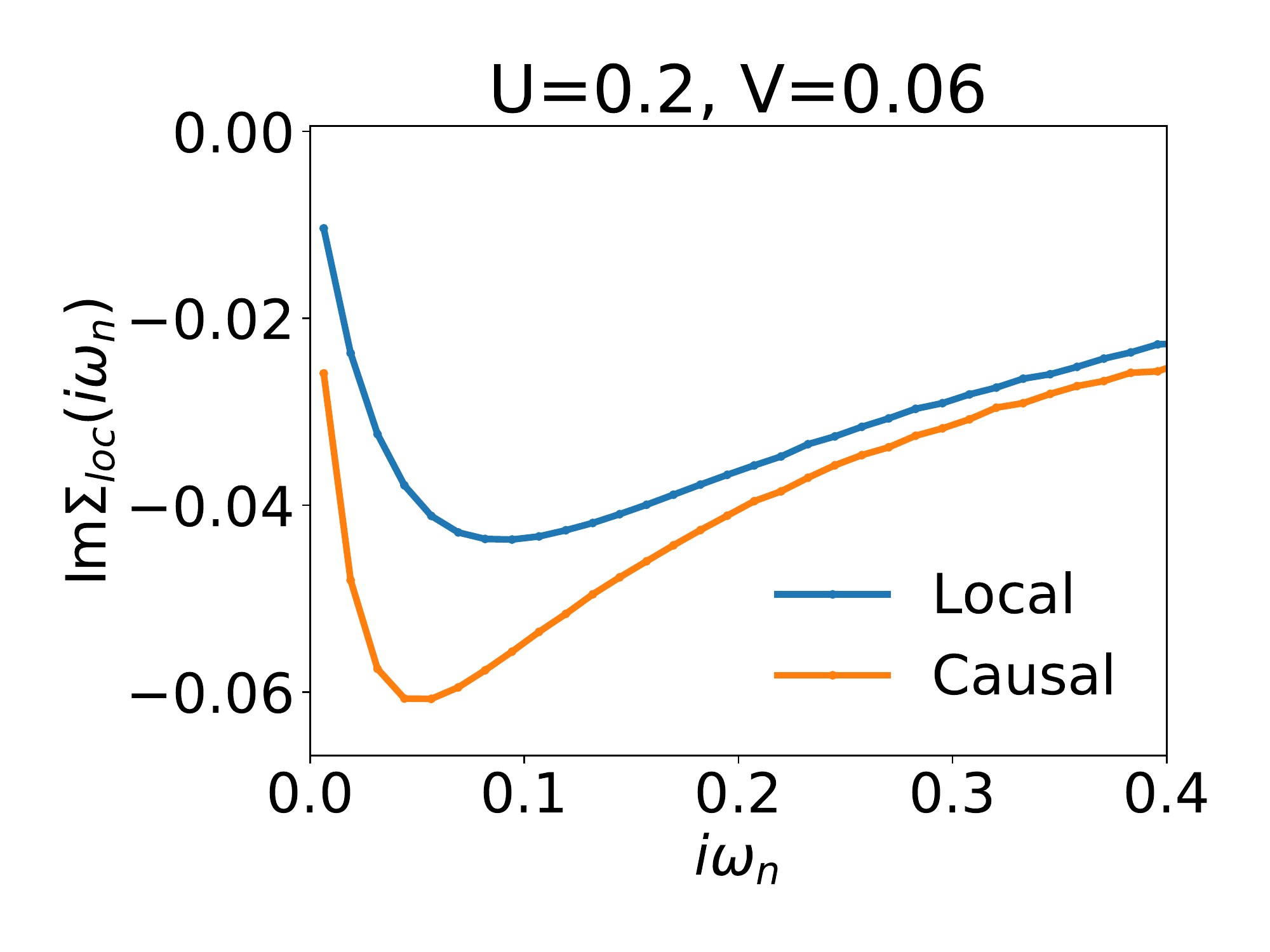}
	\includegraphics[width=0.33\linewidth]{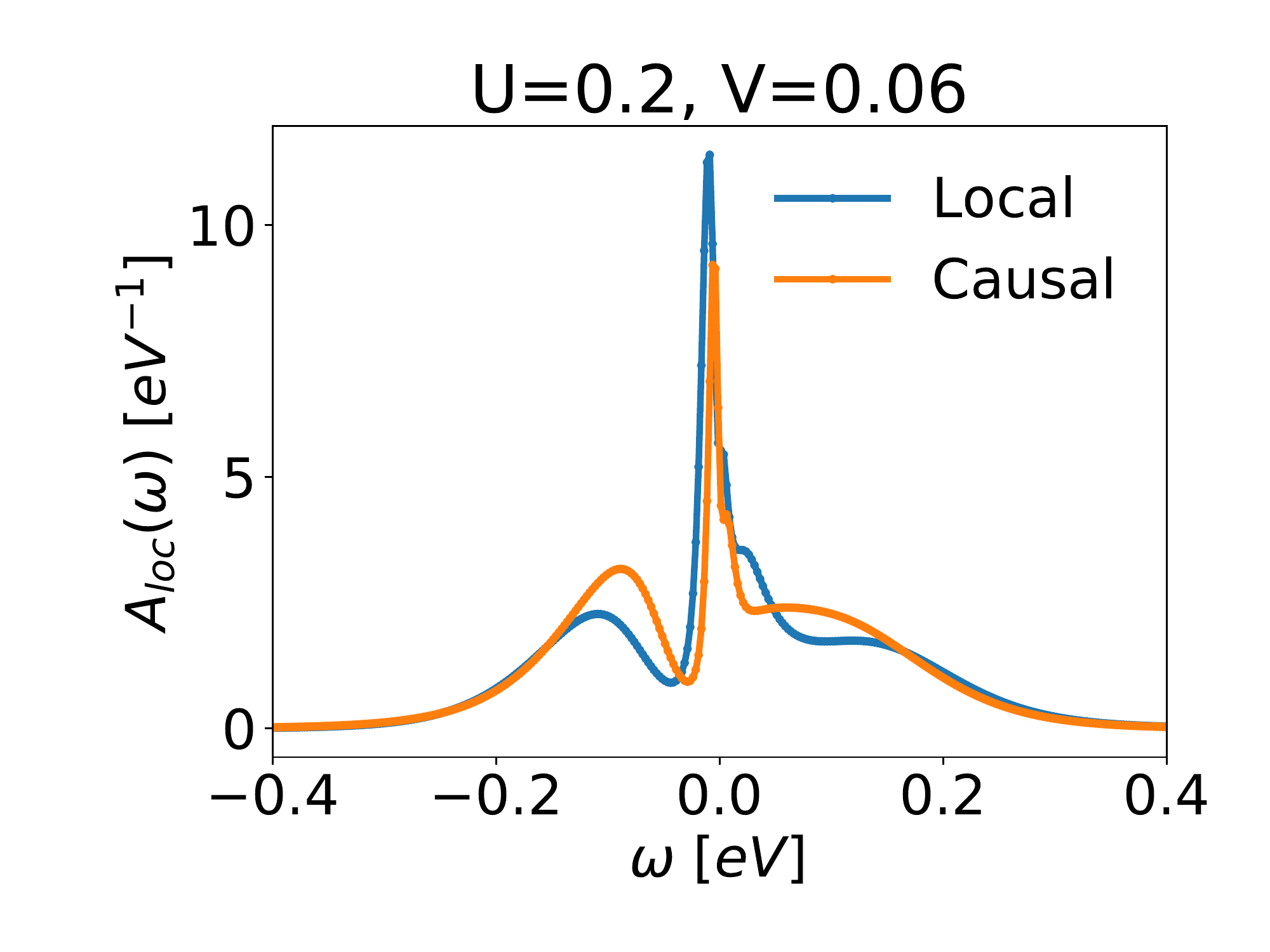}
	\caption{From left to right column: fully screened local interaction Re$W_\text{loc}(i \nu_n)$, local self-energy Im$\Sigma_\text{loc}(i\omega)$ and local fermionic spectral function $A_\text{loc}(\omega)$ obtained by the local and causal $GW$+EDMFT scheme for the model with fixed $U=0.2$ and increasing NN inter-site interactions. 
	 }
	\label{fig:Obs_V}
\end{figure*} 

\begin{figure*}[ht!]
	\centering
	\includegraphics[width=0.33\linewidth]{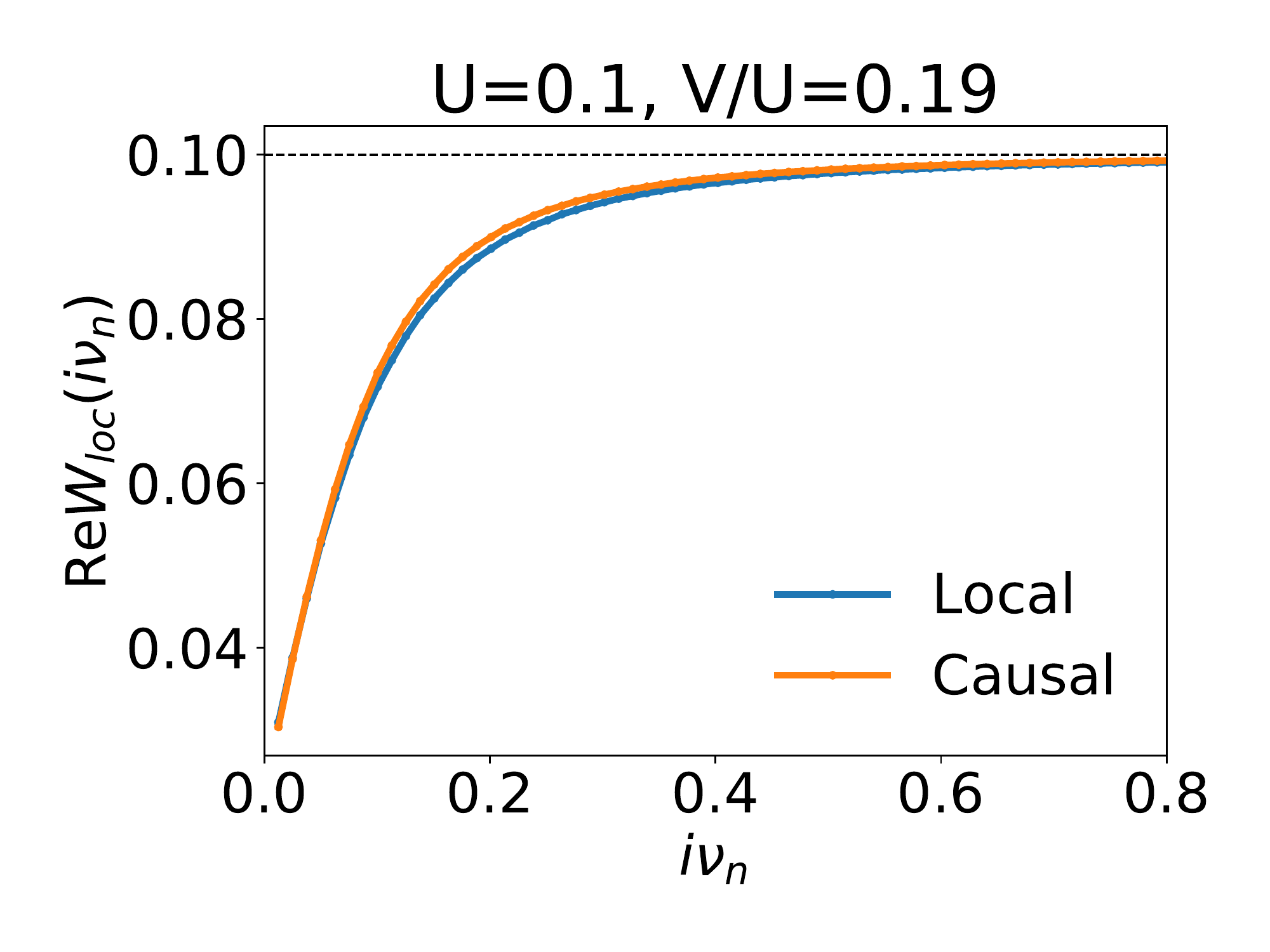}
	\includegraphics[width=0.33\linewidth]{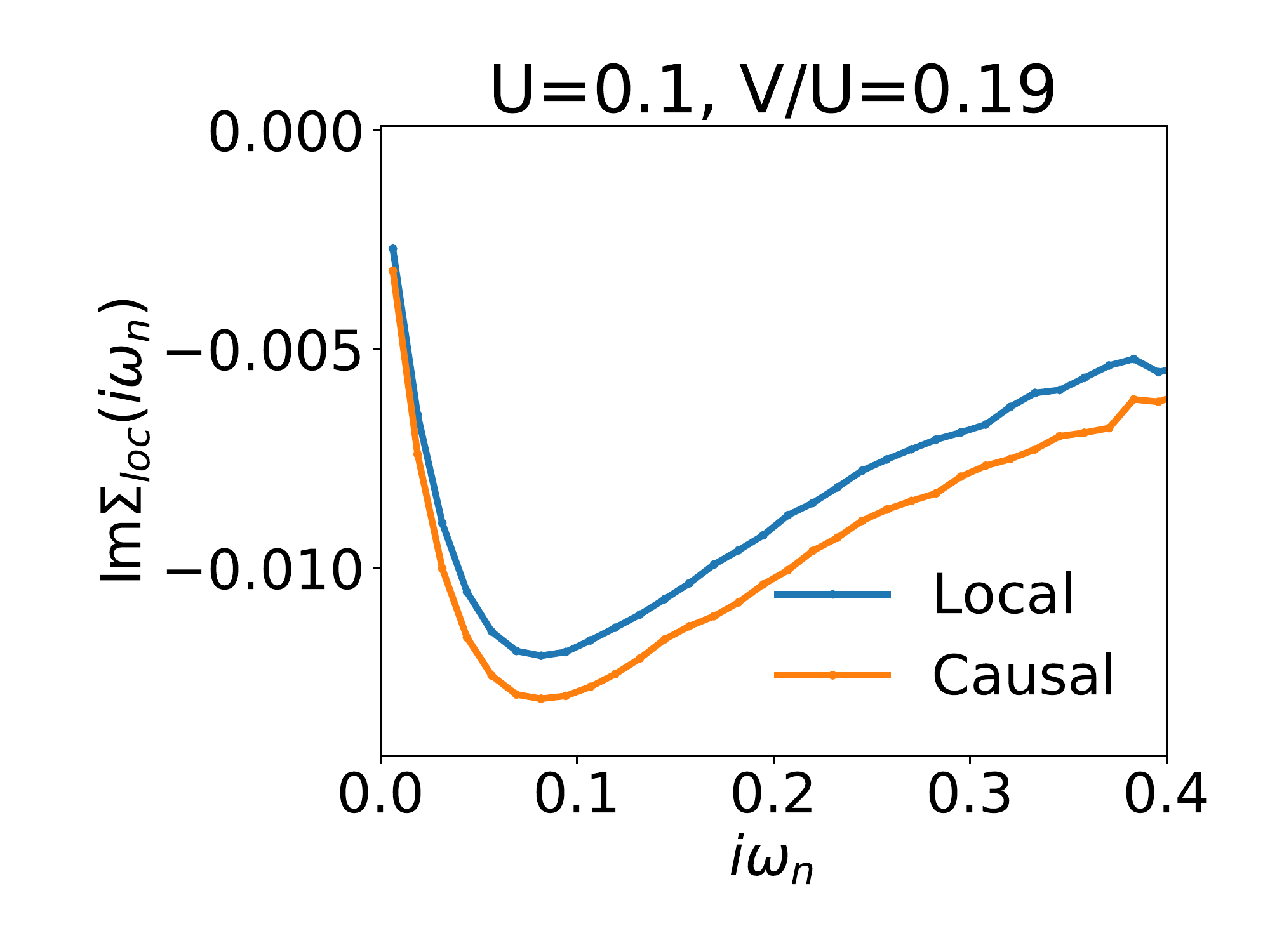}
	\includegraphics[width=0.33\linewidth]{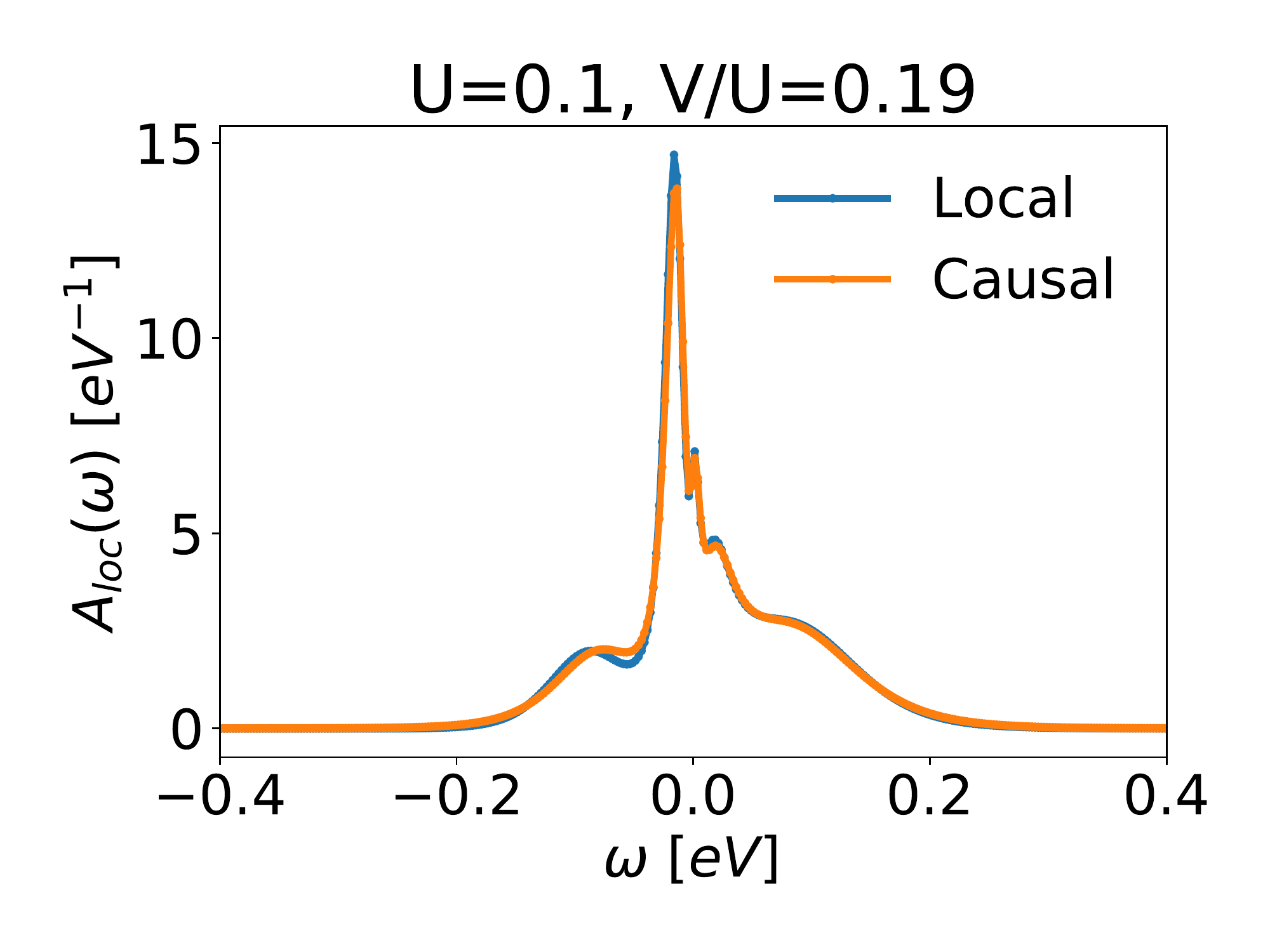}
	\includegraphics[width=0.33\linewidth]{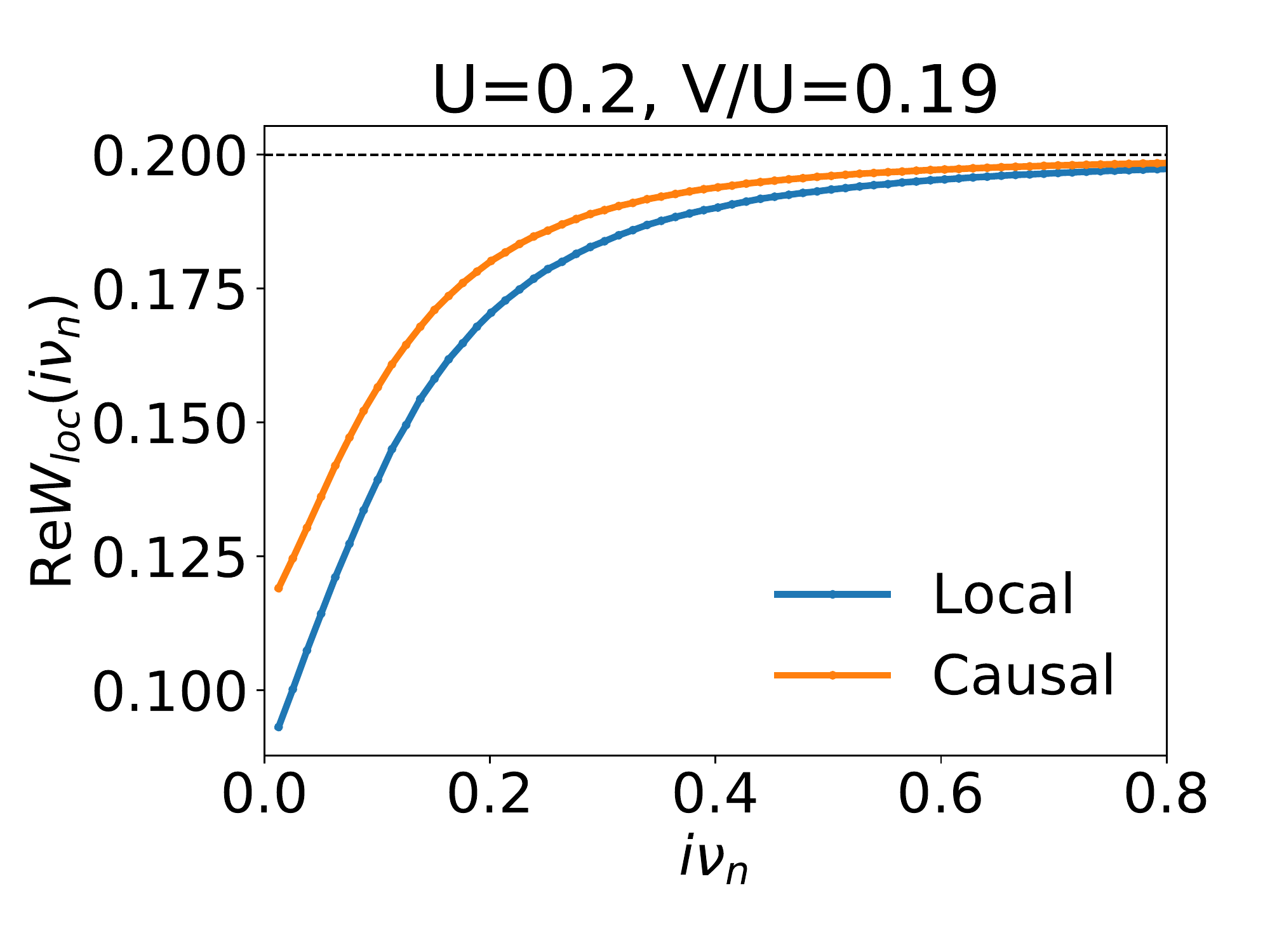}
	\includegraphics[width=0.33\linewidth]{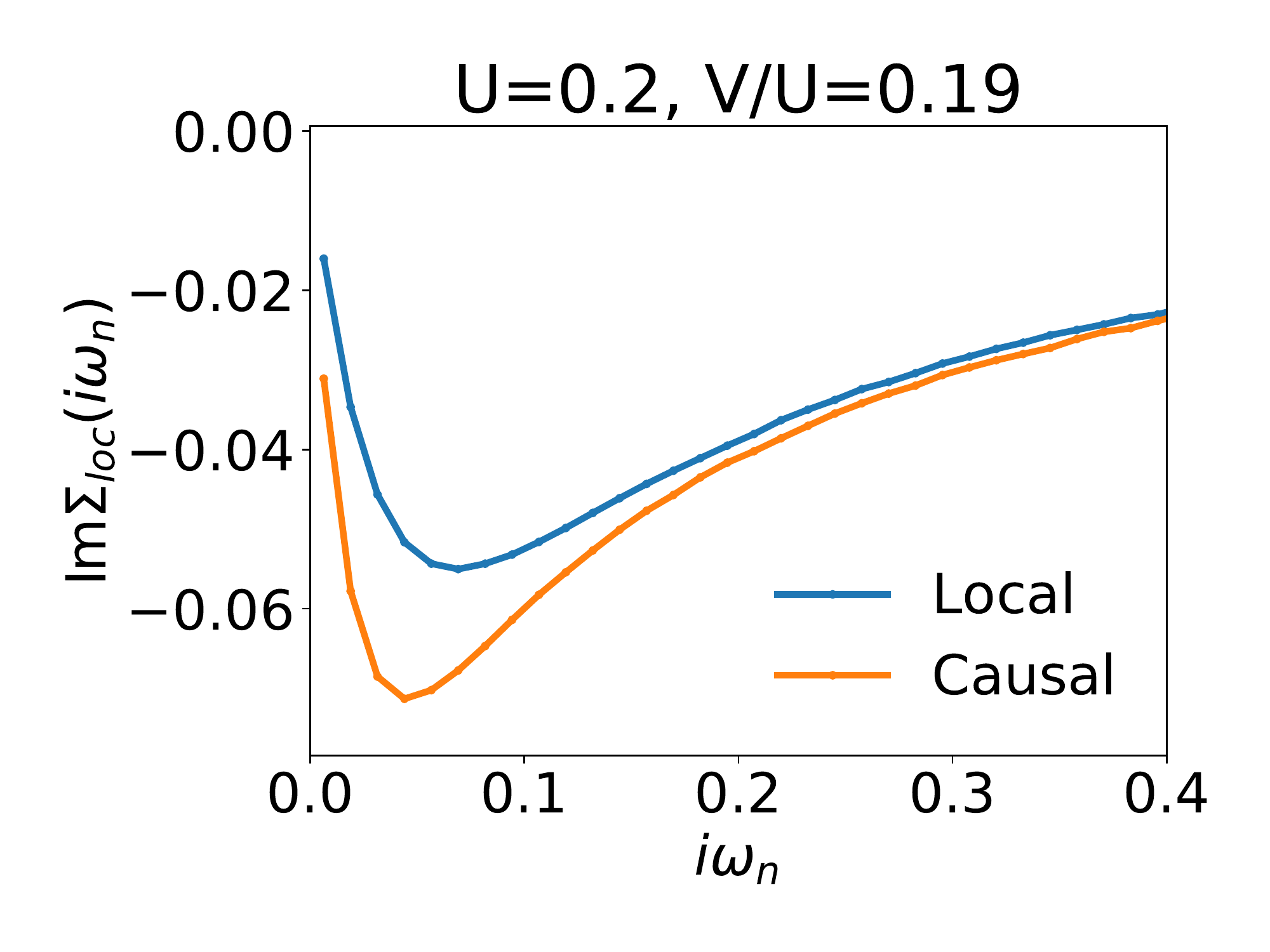}
	\includegraphics[width=0.33\linewidth]{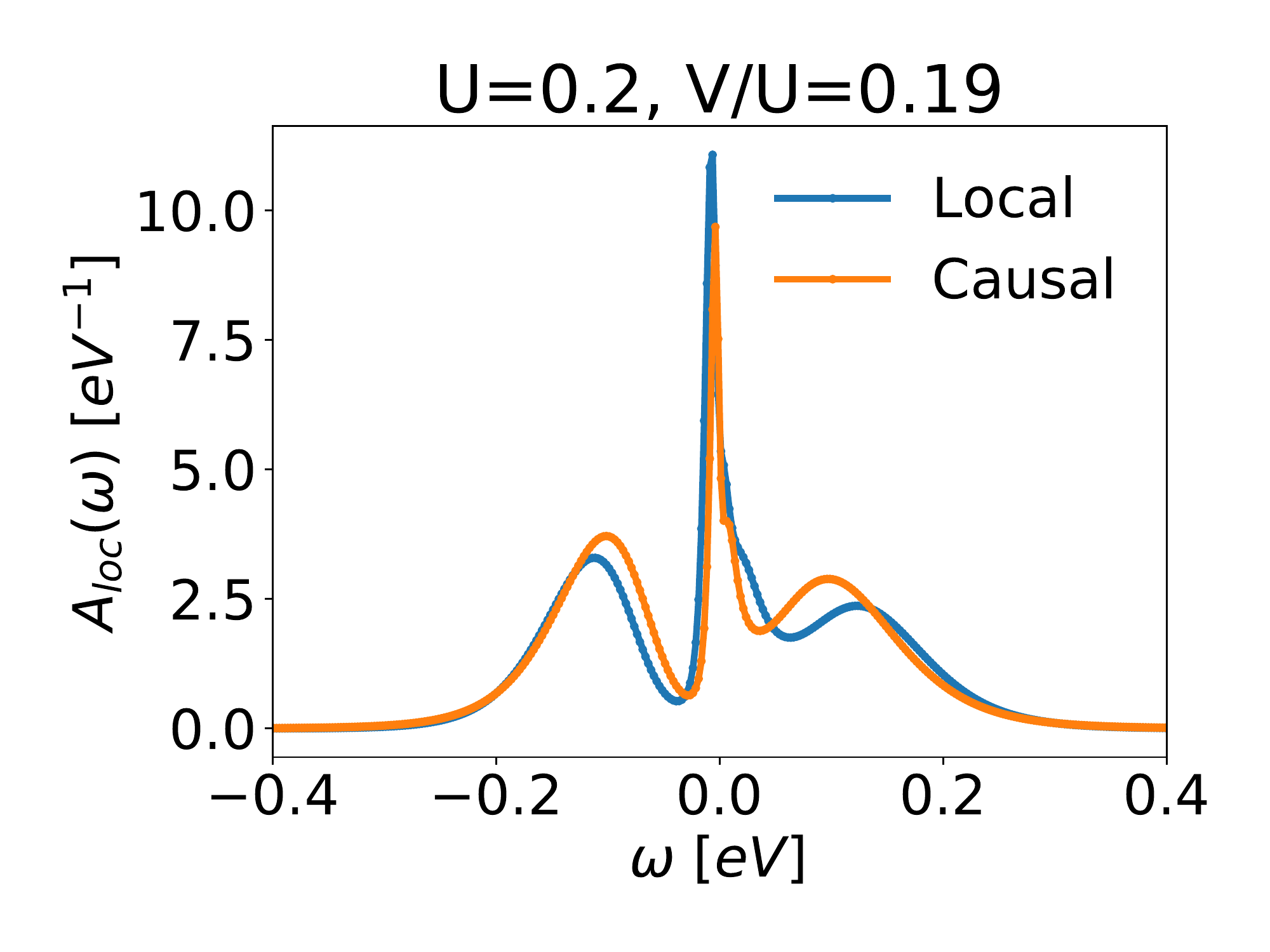}
	\includegraphics[width=0.33\linewidth]{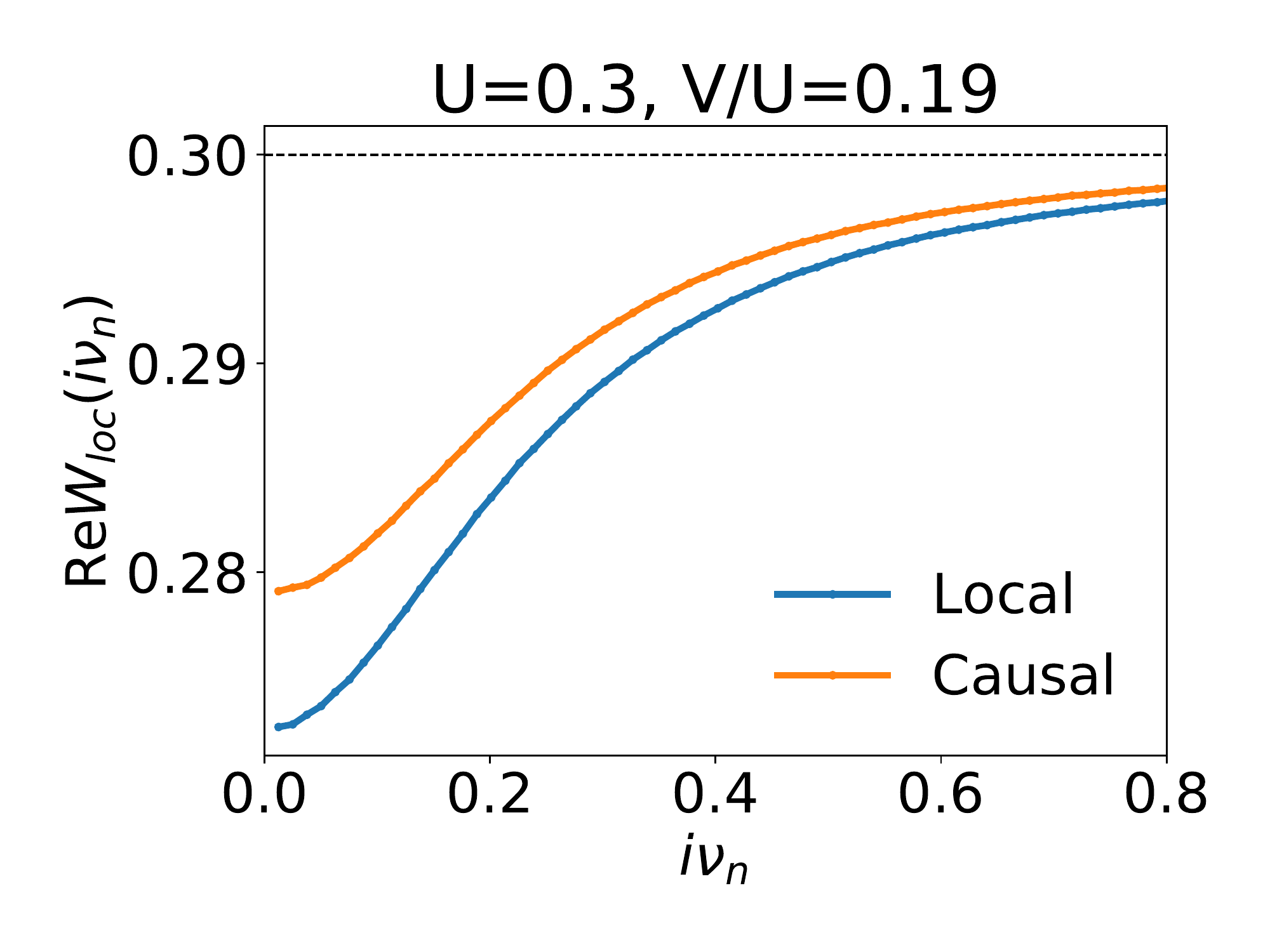}
	\includegraphics[width=0.33\linewidth]{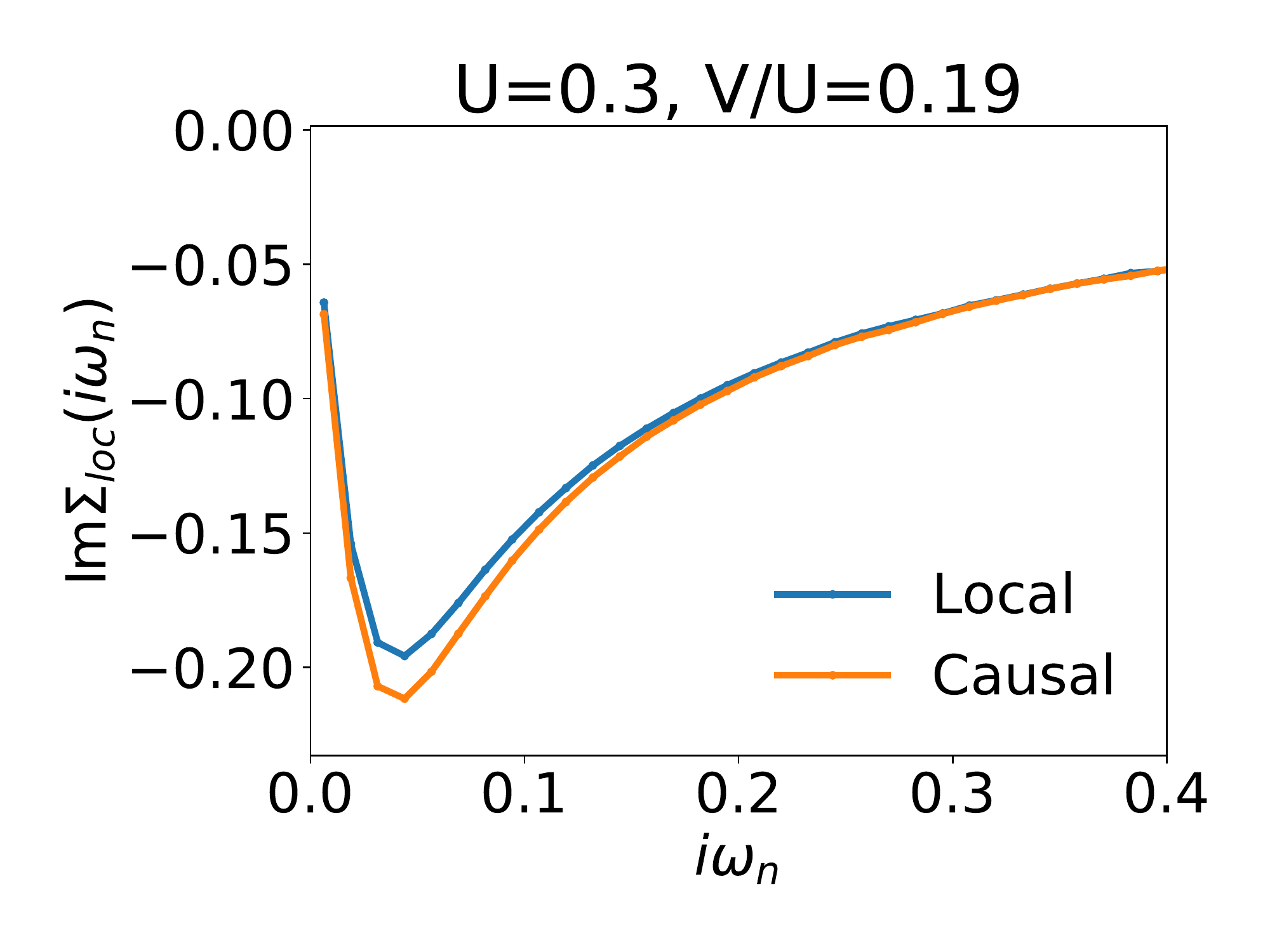}
	\includegraphics[width=0.33\linewidth]{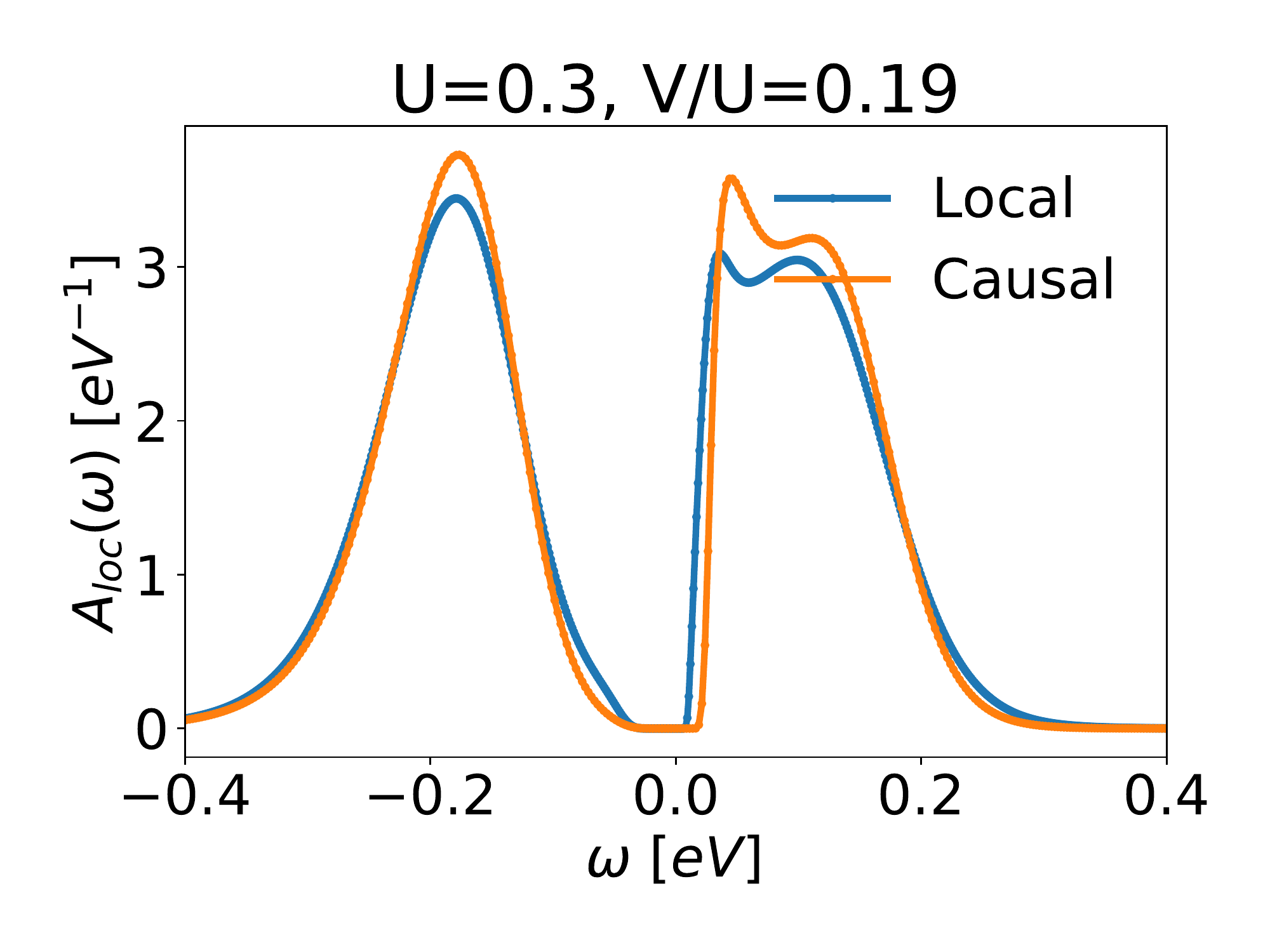}
	\caption{From left to right column: fully screened local interaction Re$W_\text{loc}(i \nu_n)$, local self-energy Im$\Sigma_\text{loc}(i\omega)$ and local fermionic spectral function $A_\text{loc}(\omega)$ obtained for increasing onsite interaction $U$ and fixed $V/U$ ratio. }
	\label{fig:Obs_U}
\end{figure*} 

The effective interaction $\mathcal{U}$ from the causal scheme is consistently smaller than in the case of the local scheme, which suggests weaker correlation effects. On the other hand, the hybridization function from the causal scheme on the Matsubara-frequency axis is also smaller than the corresponding one obtained from the local scheme, which implies a reduced effective hopping and stronger correlations. The net effect of these two opposing trends is difficult to guess, but we expect that the interaction strength resulting from the two schemes will be similar. On the real axis $-\text{Im}\Delta(\omega)$ does not display any prominent noncausal structure in both schemes. This is consistent with our experience from numerous previous $GW$+EDMFT studies,\cite{boehnke2016,nilsson2017,petocchi2020,petocchi2021} which showed that non-causalities in $\Delta$ are generically not an issue in the application of the local scheme. What the spectral functions confirm is the systematically weaker hybridization in the solution obtained from the causal self-consistency loop. 

The corrections on the Weiss fields in Eqs.~\eqref{eq:EDMFT_Generalised_correction} do not affect the causality of the physical quantities $W_\text{loc}(i\nu_n)$ and $\Sigma_\text{loc}(i\omega_n)$ that we report in Fig.~\ref{fig:Obs_V} and Fig.~\ref{fig:Obs_U} for the same points in the phase diagram as considered above. The application of the causal self-consistency scheme results, for all the $U$-$V$ points we studied at half-filling, in larger screened interactions and self-energies, indicating an enhanced correlation strength. The same effect can be noticed in the local spectral functions $A_\text{loc}(\omega)=-\frac{1}{\pi}\text{Im}G_\text{loc}(\omega)$, where one finds that the quasi-particle peak near $\omega\approx 0$ in the metallic solutions is narrower for the causal scheme than for the local scheme, while the gap in the Mott insulating solution is larger. 

By comparing the first column of Fig.~\ref{fig:Obs_V} and the middle row of Fig.~\ref{fig:Obs_U} (with the same on-site interaction $U=0.2$) one finds that an increase of $V$ in the metallic phase systematically enhances the screening of the local interaction $W$, and the same holds for the impurity interaction $\mathcal{U}$.  Considering the rows in Fig.~\ref{fig:Obs_U}, where we increase $U$ while fixing the $V/U$ ratio, one can see that the relative change between the bare and static screened interaction is reduced with increasing correlation strength (again the same holds for $\mathcal{U}$), and that the screening effect is strongly reduced after the transition from the metallic to the Mott insulating phase (note the different scales).

These findings are consistent with the results reported for a dimer by Backes \etal,\cite{backes2020} although less pronounced. 
As noted above, the net increase in the interaction strength, which we observe in the causal scheme at half-filling, is the result of a decrease in the local effective interaction $\mathcal{U}(i\nu_n)$, which is over-compensated by an even stronger reduction of the hybridization strength. As will be shown below, the opposite trend can be found in doped systems, so that it is difficult to make a general statement about stronger or weaker interactions resulting from one or the other scheme. 
Overall, we find that the physical results ($G$, $W$, and corresponding spectra) are rather similar for the two schemes, which is comforting in the sense that the general approach appears to be robust. In fact, several recent applications of the (local) $GW$+EDMFT scheme to correlated materials provided meaningful estimates of the correlation strength.\cite{nilsson2017,petocchi2020,petocchi2020a,petocchi2021}

%%%%%%%%%%%%%%%%%%%%%%%%%%%%%%%%%%%%%%%%%%%%%%%%%%%

\subsection{Doping and temperature dependence in a model for  1$T$-TaS$_2$}
\label{sec:1TTaS2}

\subsubsection{Choice of parameters}

\begin{figure}[t]
	\centering
	\includegraphics[width=0.85\linewidth]{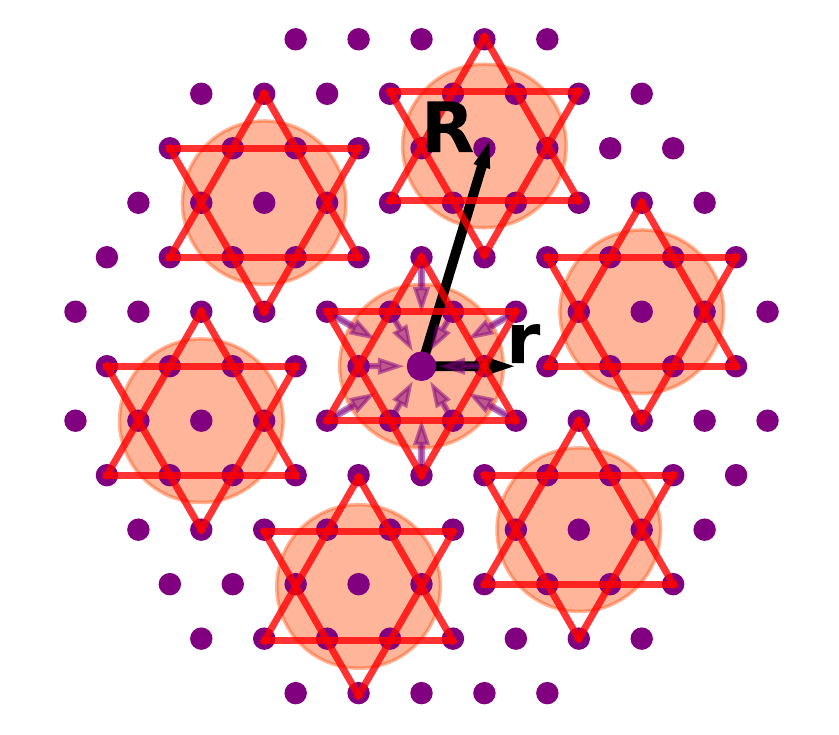}
	\caption{The C-CDW pattern in the low-temperature phase of 1$T$-TaS$_2$, sketched based on the experimental results in~Refs.~\onlinecite{cho2016,wang2020,wu2021}. Purple dots are the Ta atoms, distorted in the direction of the purple arrows, and the red lines indicate the David-star-like super-cells, each including 13 lattice sites. The red disk shows the area with high charge density seen by STM. $R=\sqrt{13}$ is the distance between two neighboring super-cells and $r$ is the size of the disk.}\label{fig:David_star}
\end{figure}

To connect our model calculations to experimental results for the layered transition-metal dichalcogenide $1T$-TaS$_2$,\cite{ cho2016,ligges2018,wu2021} we first estimate the realistic ratio between the on-site interaction $U$ and the NN interaction $V$ in the half-filled, Mott insulating system with poor screening. At low temperature ($T \leq 180$~K), $1T$-TaS$_2$  is in a commensurate charge density wave (C-CDW) state, characterized by a David-star-like lattice distortion,\cite{cho2016,wang2020,wu2021} which we illustrate in Fig.~\ref{fig:David_star}. This structural distortion modifies the electronic band structure, which at low energy can be represented by a single-band model on a triangular lattice. In this model, each lattice site corresponds to a 13-atom cluster. 

To determine the ratio $V/U$, we make the simple assumption that the electrons are distributed uniformly inside a disk representing the David-star-like molecular orbital. From the scanning tunneling microscope (STM) image,\cite{cho2016} one can roughly estimate the size $r\approx 0.4$ of this disk (in units of lattice spacing), while $R=\sqrt{13}$ is the distance between the molecular orbitals (in units of the lattice spacing of the undistorted lattice). The interactions $U$ and $V$ are then proportional to

\begin{equation}
	\begin{aligned}
		U\sim \int_{D_1}\int_{D_1} d \vec{r}_1 d \vec{r}_1' \frac{1}{|\vec{r}_1-\vec{r}_1'|}, \\
		V\sim\int_{D_1}\int_{D_2} d \vec{r}_1 d \vec{r}_2 \frac{1}{|\vec{r}_1-\vec{r}_2|},
	\end{aligned}
\end{equation}
where $\vec{r}_1$ and $\vec{r}'_1$ are in the same disk ($D_1$), while $\vec{r}_1$ and $\vec{r}_2$ are in neighboring disks ($D_1$ and $D_2$). The thus estimated ratio is $V/U\sim 0.19$. 

To estimate the values of $U$ and $t$, we compare the separation between the Hubbard bands and the width of the Hubbard bands in $A_\text{loc}(\omega)$ to the results from photoemission and STM studies. The on-site interaction $U$ essentially determines the gap between the upper Hubbard band (UHB) and the lower Hubbard band (LHB), which is about 0.4~eV, \cite{cho2016,wu2021} while the width of the Hubbard bands is approximately 0.2~eV.\cite{cho2016,wu2021} A good match with experiment is obtained for $U=0.42$ and $t=-0.02$ (the bandwidth is $W=9|t|$ for the triangular lattice). Given the onsite interaction of $U=0.42$ and the ratio $V/U=0.19$ we choose the NN interaction as $V=V_1=0.08$. This ``realistic" parameter set for 1$T$-TaS$_2$ is indicated by the black point in Fig.~\ref{fig:phase_diagram}. We note that it is not close to the CO phase boundary.

As in the previous sections, we set the inverse temperature to $\beta=500$ eV$^{-1}$ ($T=23$~K), unless otherwise stated, and use eV as the unit of energy.

\begin{figure*}[t!]
	\centering
	\includegraphics[width=0.33\linewidth]{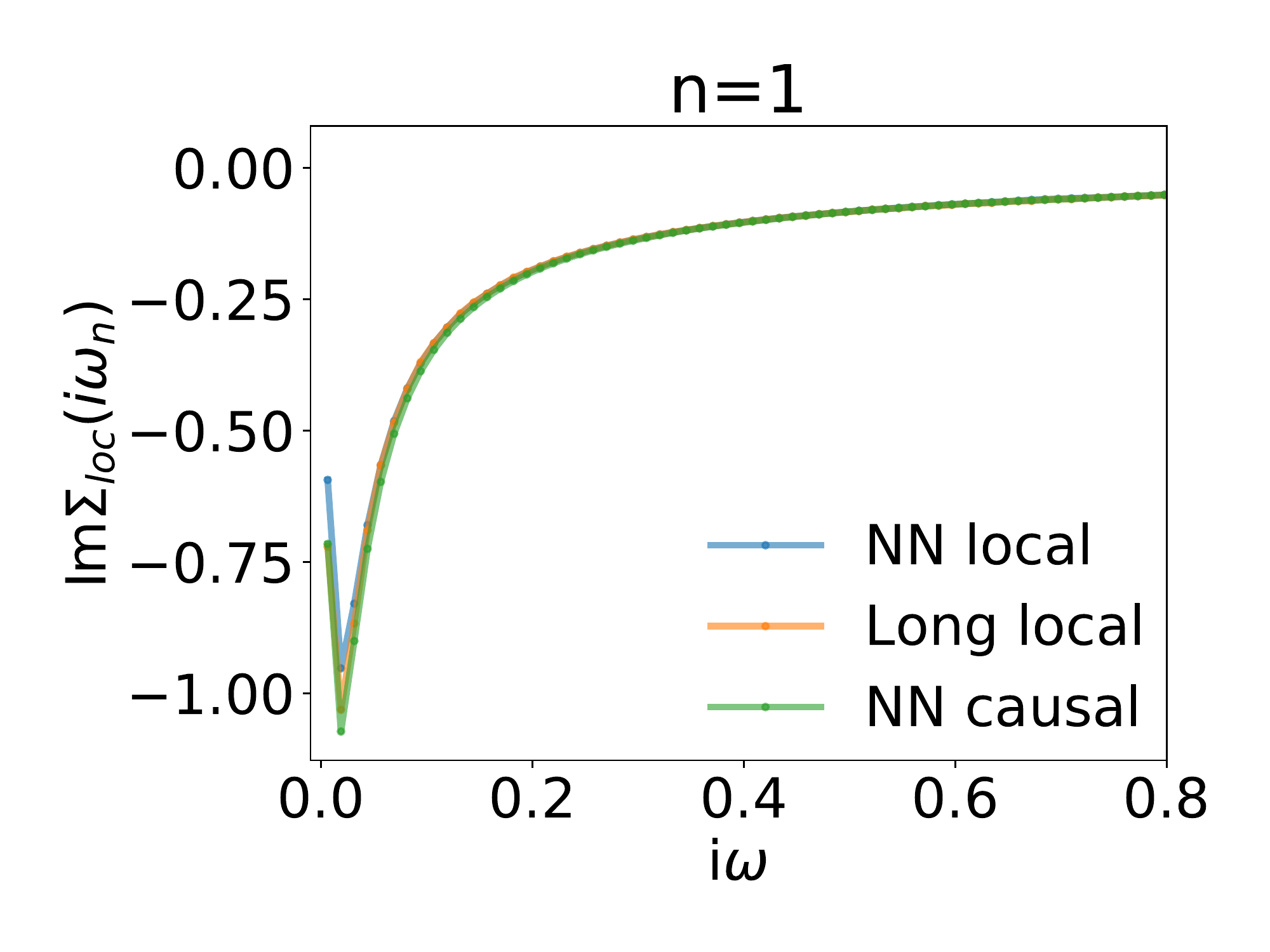}
	\includegraphics[width=0.33\linewidth]{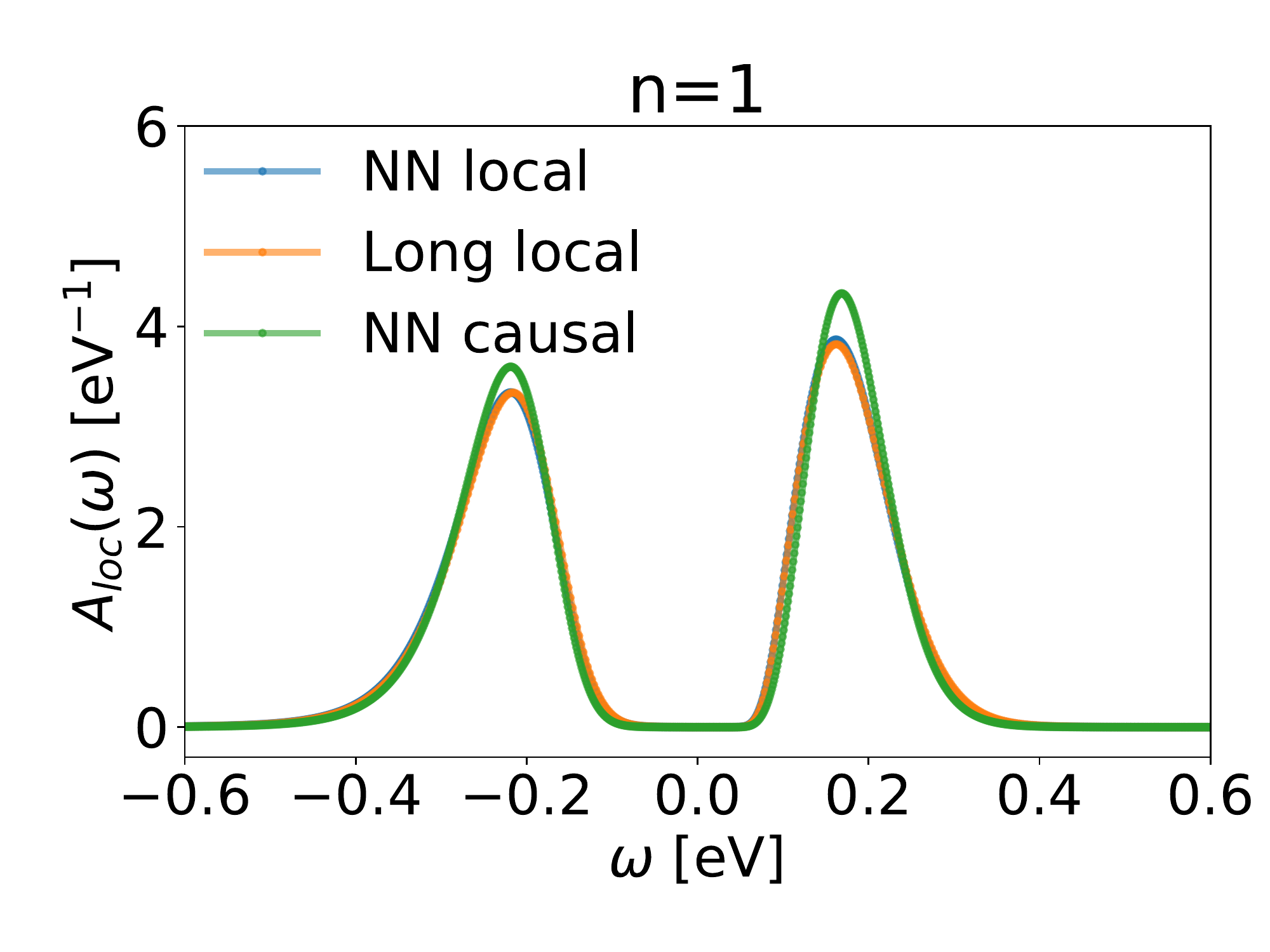}
	\includegraphics[width=0.33\linewidth]{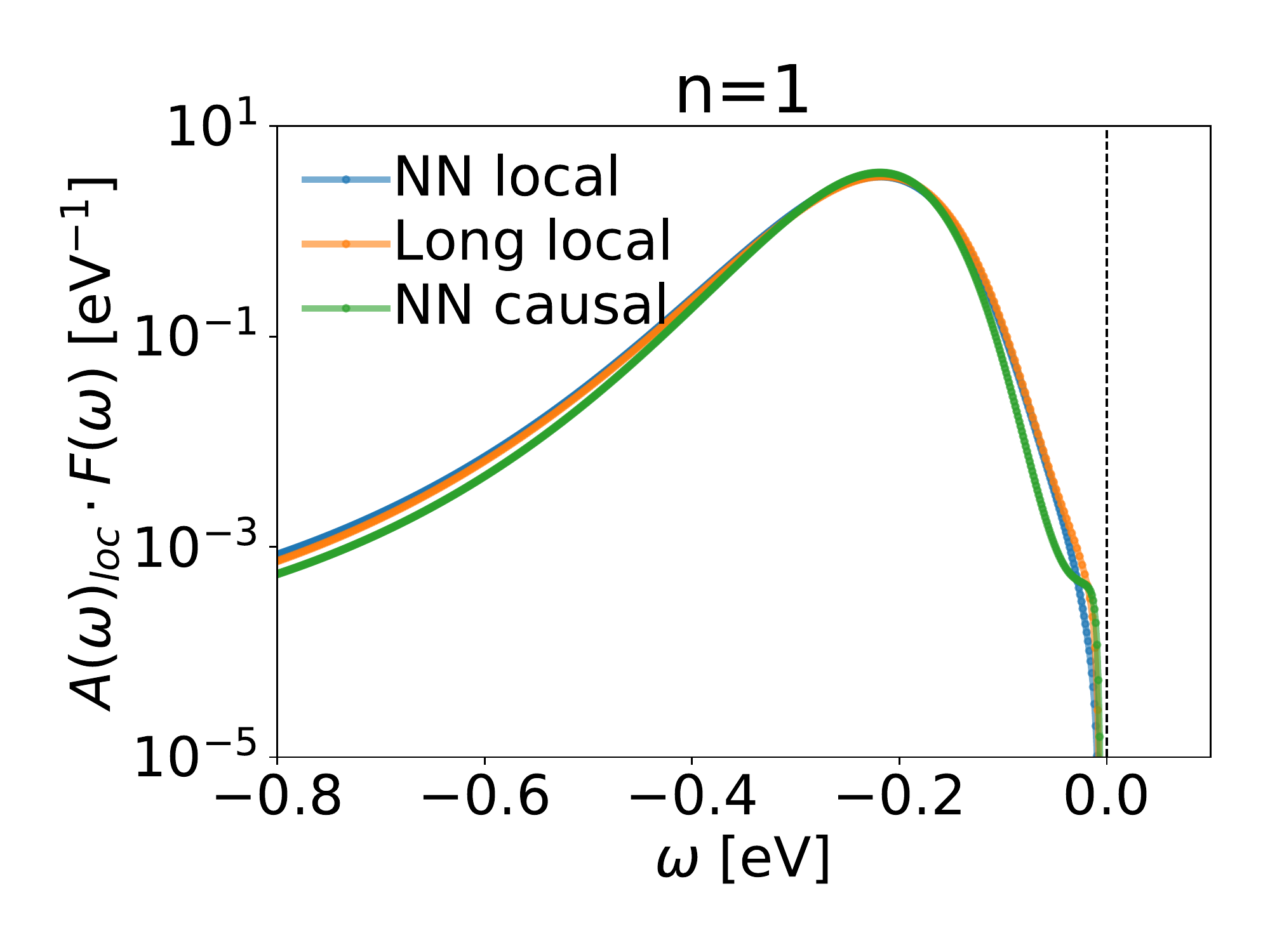}
	\includegraphics[width=0.33\linewidth]{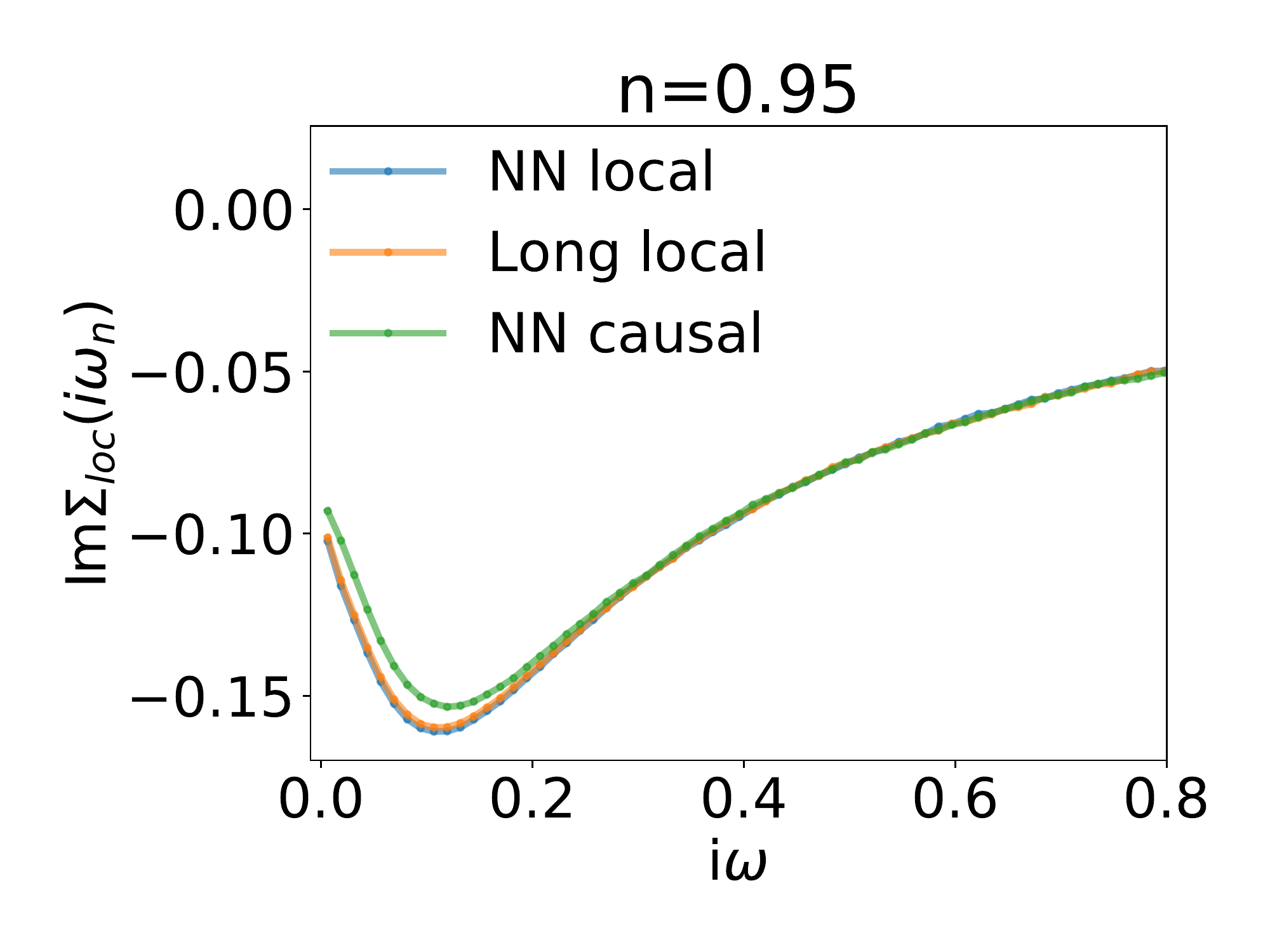}
	\includegraphics[width=0.33\linewidth]{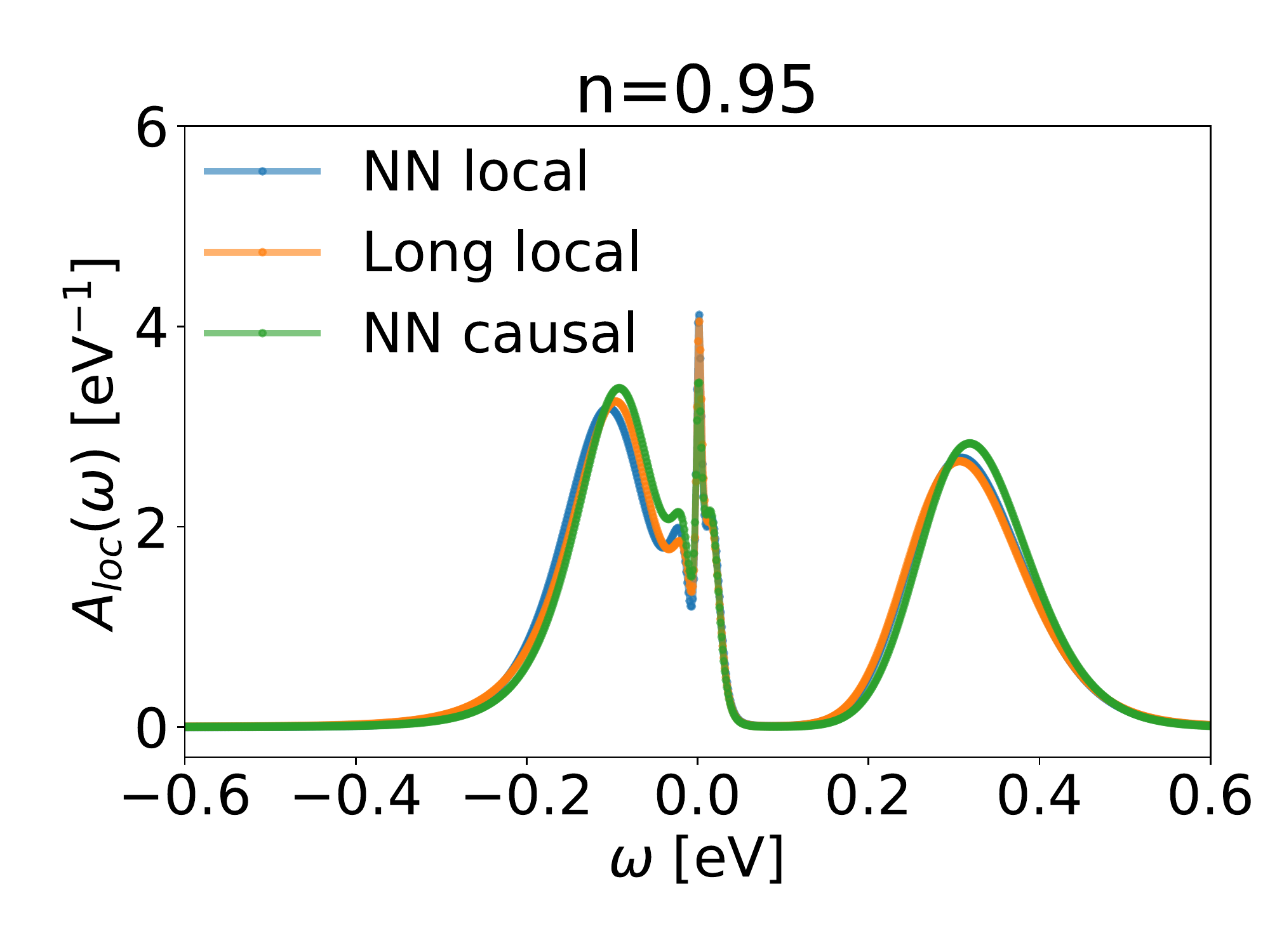}
	\includegraphics[width=0.33\linewidth]{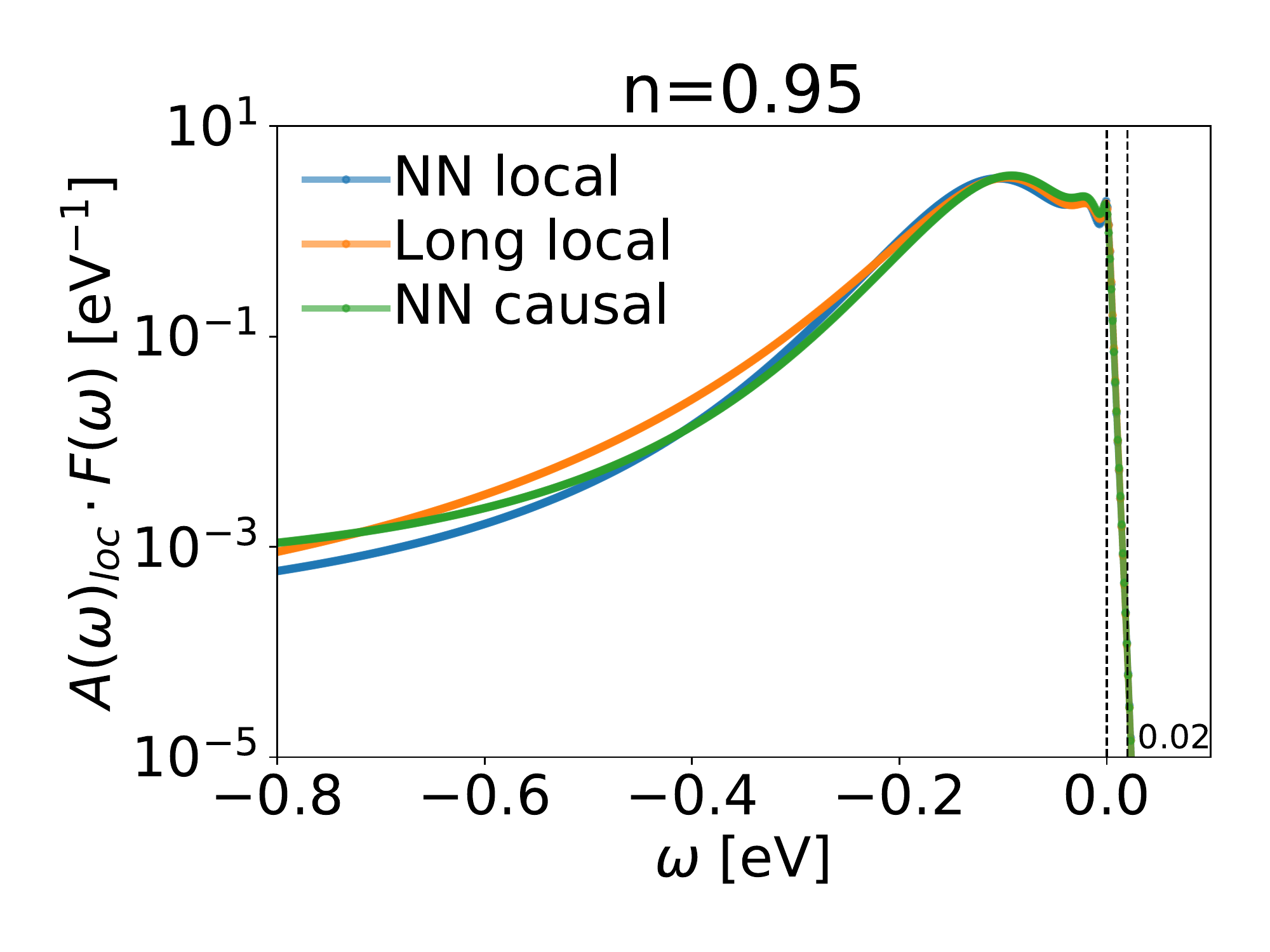}
	\includegraphics[width=0.33\linewidth]{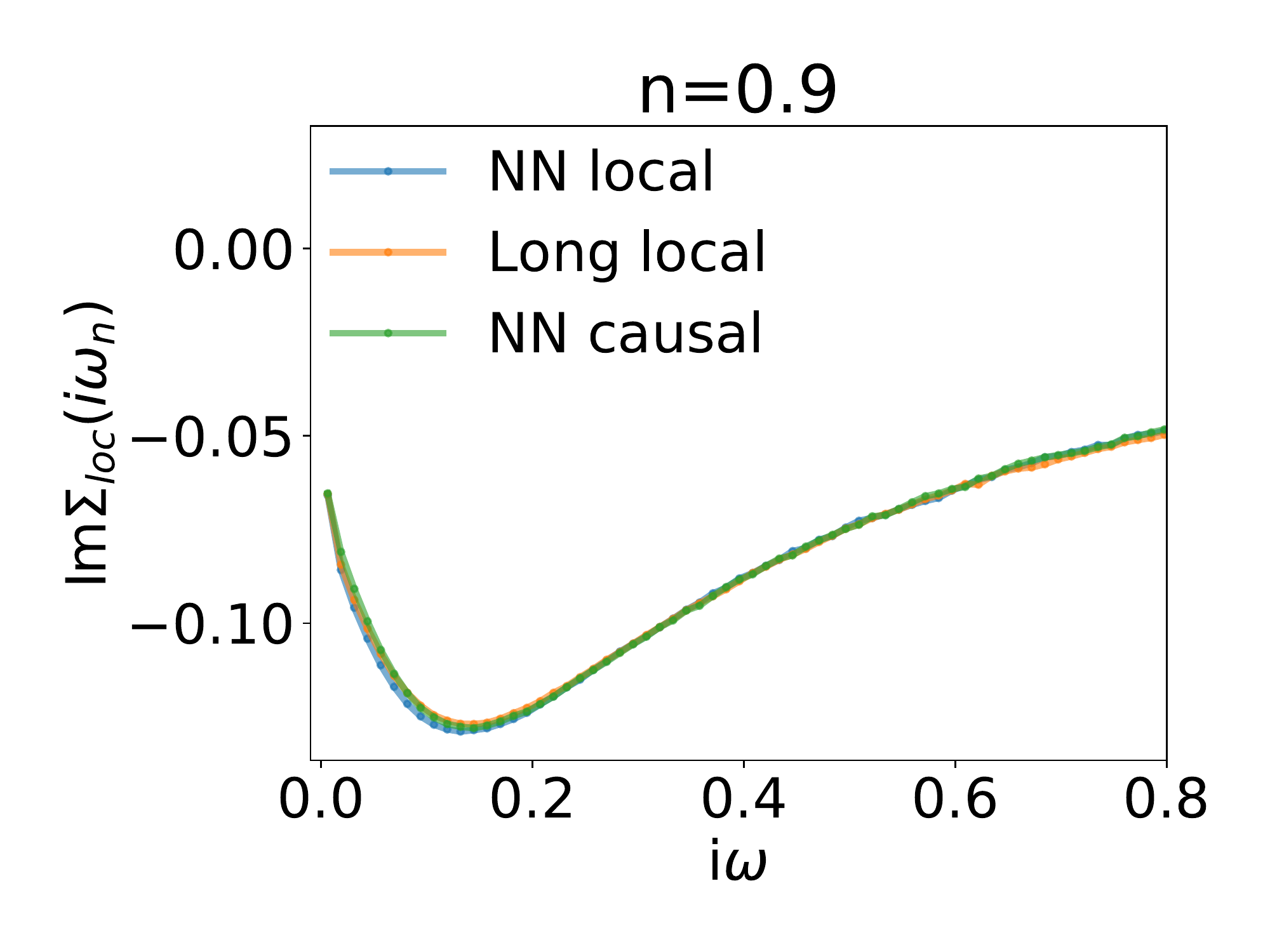}
	\includegraphics[width=0.33\linewidth]{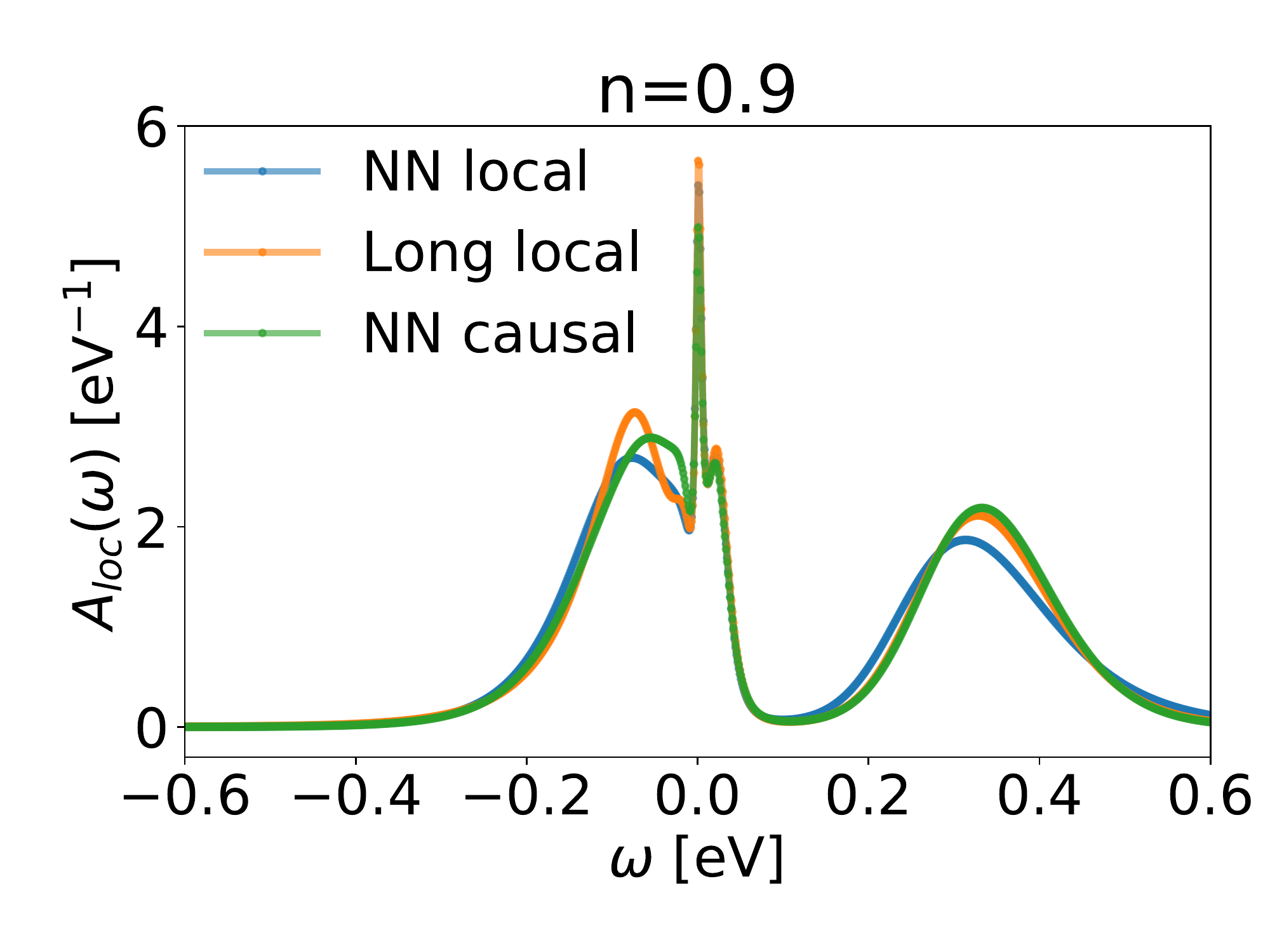}
	\includegraphics[width=0.33\linewidth]{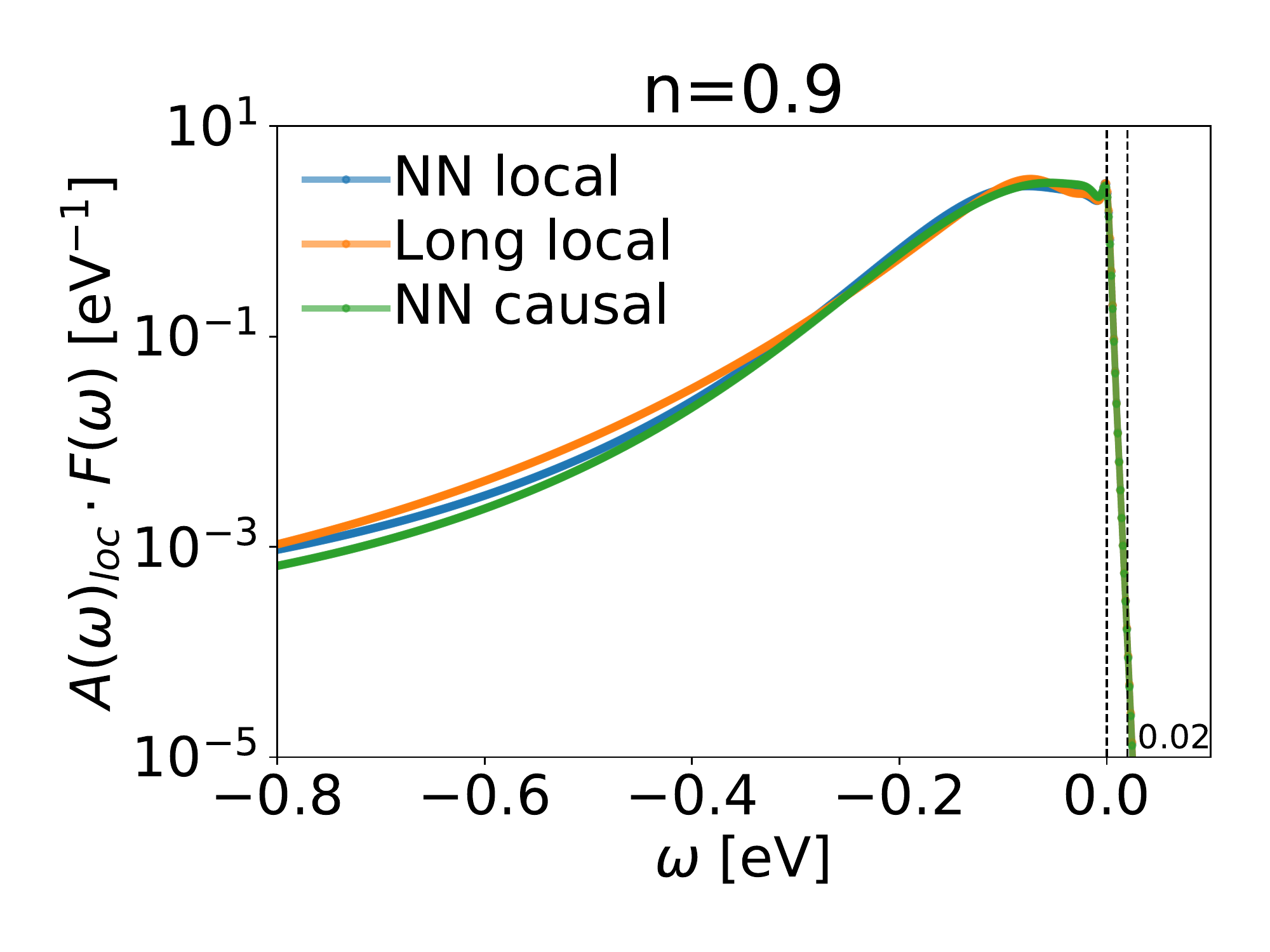}
	\caption{Local self-energy $\Sigma_\text{loc}$, fermionic spectral function $A_\text{loc}(\omega)$, and distribution function $A_\text{loc}(\omega)F(\omega)$ (from left to right). The doping is increased row-wise from the undoped configuration to $5\%$ and $10\%$ hole concentration. The interaction parameters are $U=0.42$ and $V=0.08$. 
	}\label{fig:doping_SGA}
\end{figure*}

\subsubsection{Doping dependence}

In a recent time-resolved photoemission study of 1$T$-TaS$_2$, the authors concluded that their nominally half-filled system may in fact be substantially hole-doped.\cite{ligges2018} The indirect evidence for this was the surprisingly fast recombination of photo-doped doublons and holons, which in a pure Mott state with the given ratio between gap size and bandwidth should be longer-lived.\cite{eckstein2011} This motivates us to compute the local spectral function for the realistic parameters and different hole dopings. In Figs.~\ref{fig:doping_SGA}, the three rows correspond to half-filling, 5\% hole doping ($n=0.95$) and 10\% hole doping ($n=0.9$). The half-filled system is Mott insulating, while the doped systems are metallic.

In Fig.~\ref{fig:doping_SGA}, the gap between the upper Hubbard band (UHB) and lower Hubbard band (LHB) is $0.4$ eV, and the width of the Hubbard bands is approximately $0.2$ eV, consistent with the experimental results. When the system is 5\% or 10\% hole doped, a prominent quasi-particle peak appears at the edge of the lower Hubbard band, which is shifted to $\omega=0$, while the peaks of the LHB and UHB are shifted to $-0.1$ eV and $0.3$ eV, respectively. 

A quasi-particle peak was not reported in Ref.~\onlinecite{ligges2018} in the initial equilibrium spectrum. 
The rightmost column of Fig.~\ref{fig:doping_SGA} shows the occupation function, i.e. the spectral function multiplied with the Fermi function $F(\omega)=1/(e^{\beta\omega}+1)$, on a logarithmic scale. This is the quantity measured in photo-emission experiments. The results for the undoped and doped systems differ substantially: near the Fermi level, the occupation function of the doped systems exhibit an almost linear decay (in the log plot) over three orders of magnitude, in an energy interval of width $\Delta \omega\approx 0.02$, which is determined by the temperature. In the undoped case, on the other hand, the first three orders of magnitude decrease from the peak value of the occupation is controlled by the shape of the lower Hubbard band, rather than by the Fermi distribution function, since in this case the chemical potential is in the gapped region. 

We plot in Fig.~\ref{fig:doping_SGA} the results from three different calculations: blue lines are for the local scheme with NN interactions, orange lines for the local scheme with long-ranged interactions, and green lines for the causal scheme with NN interactions. As far as the spectral function and occupation function are concerned, the differences between the models and schemes are minor and on a scale comparable to the intrinsic uncertainties of the MaxEnt approach.  

\begin{figure}[!]
	\centering
	\includegraphics[width=0.475\linewidth]{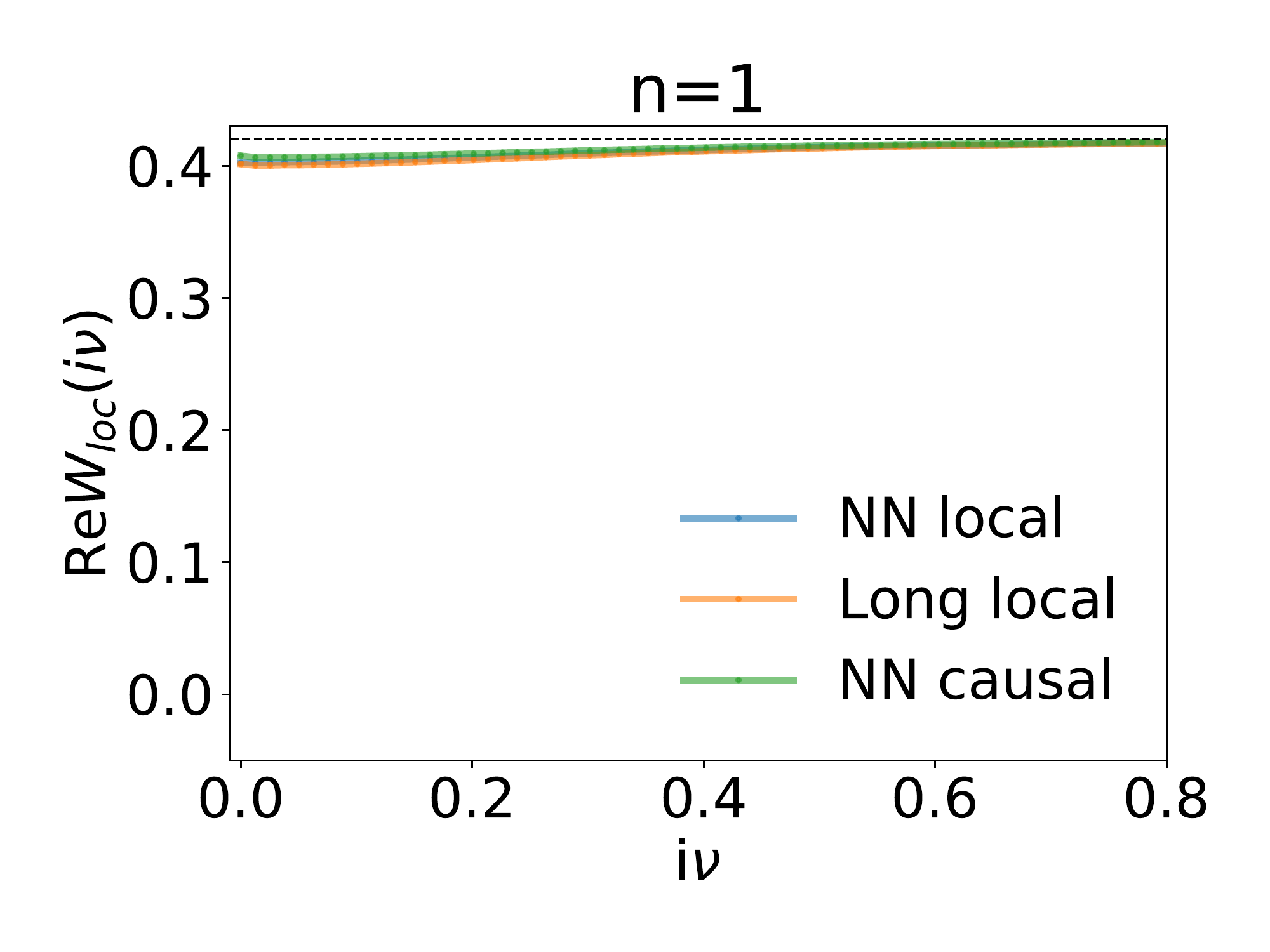}
	\includegraphics[width=0.475\linewidth]{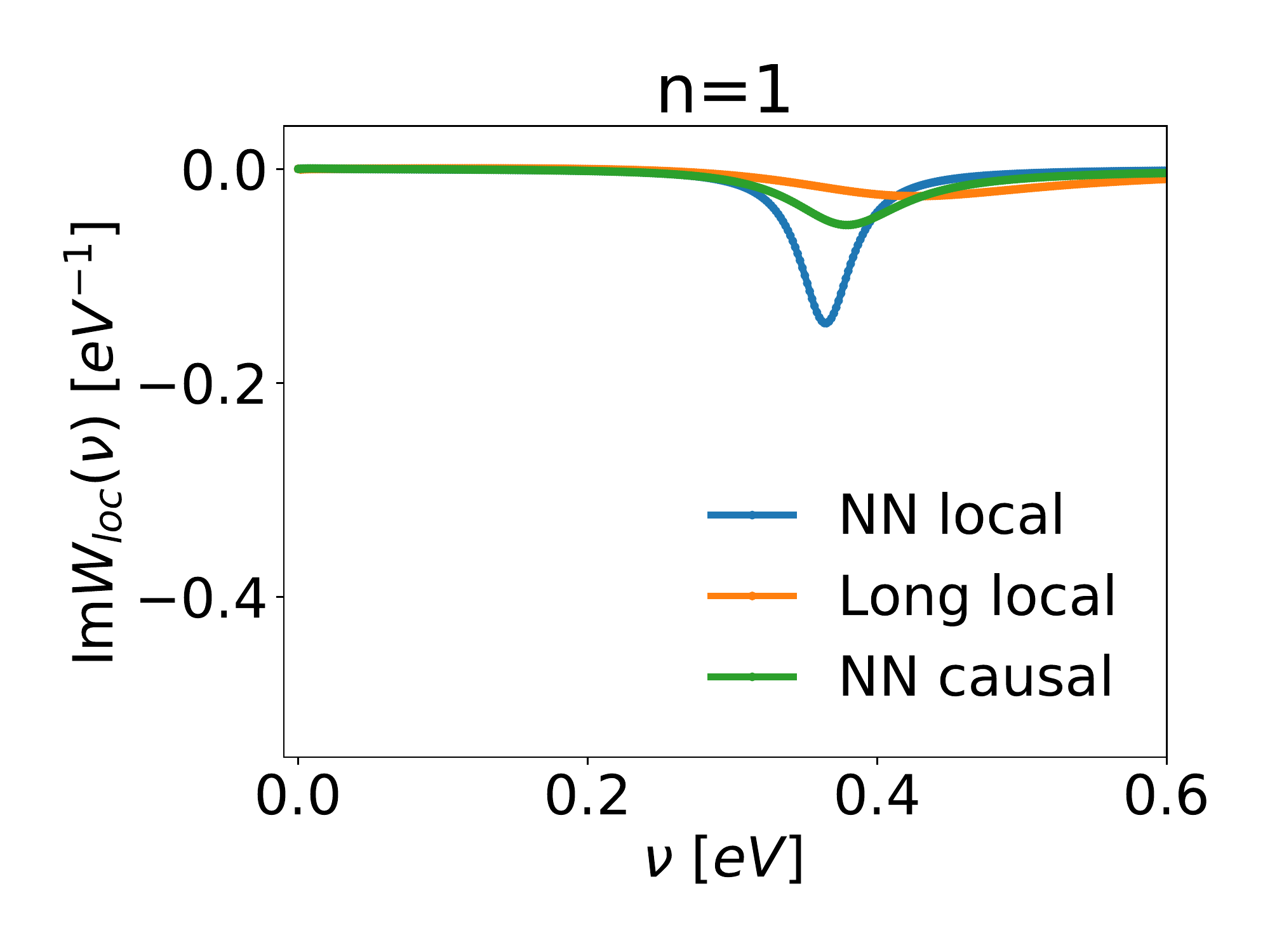}
	\includegraphics[width=0.475\linewidth]{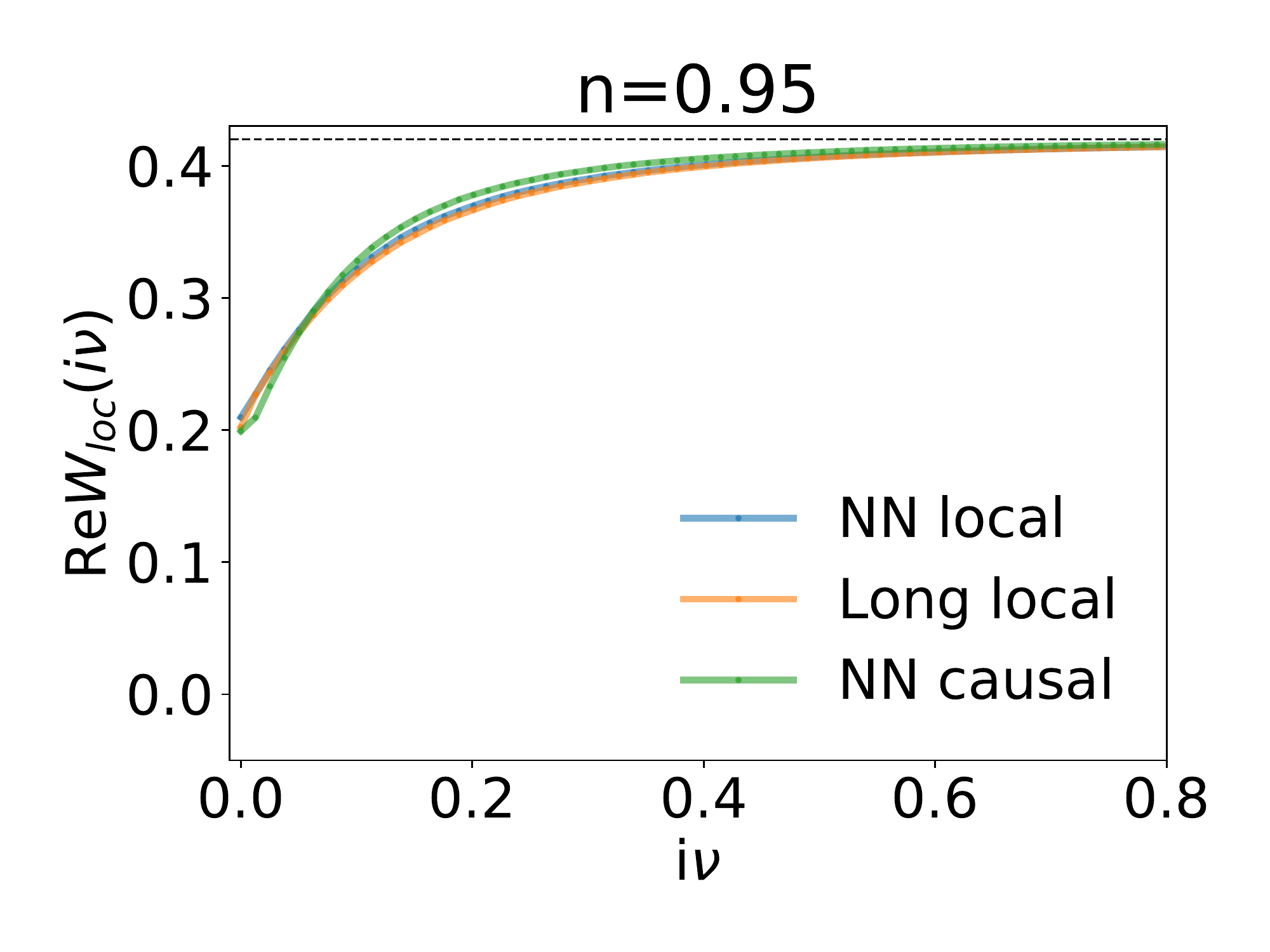}
	\includegraphics[width=0.475\linewidth]{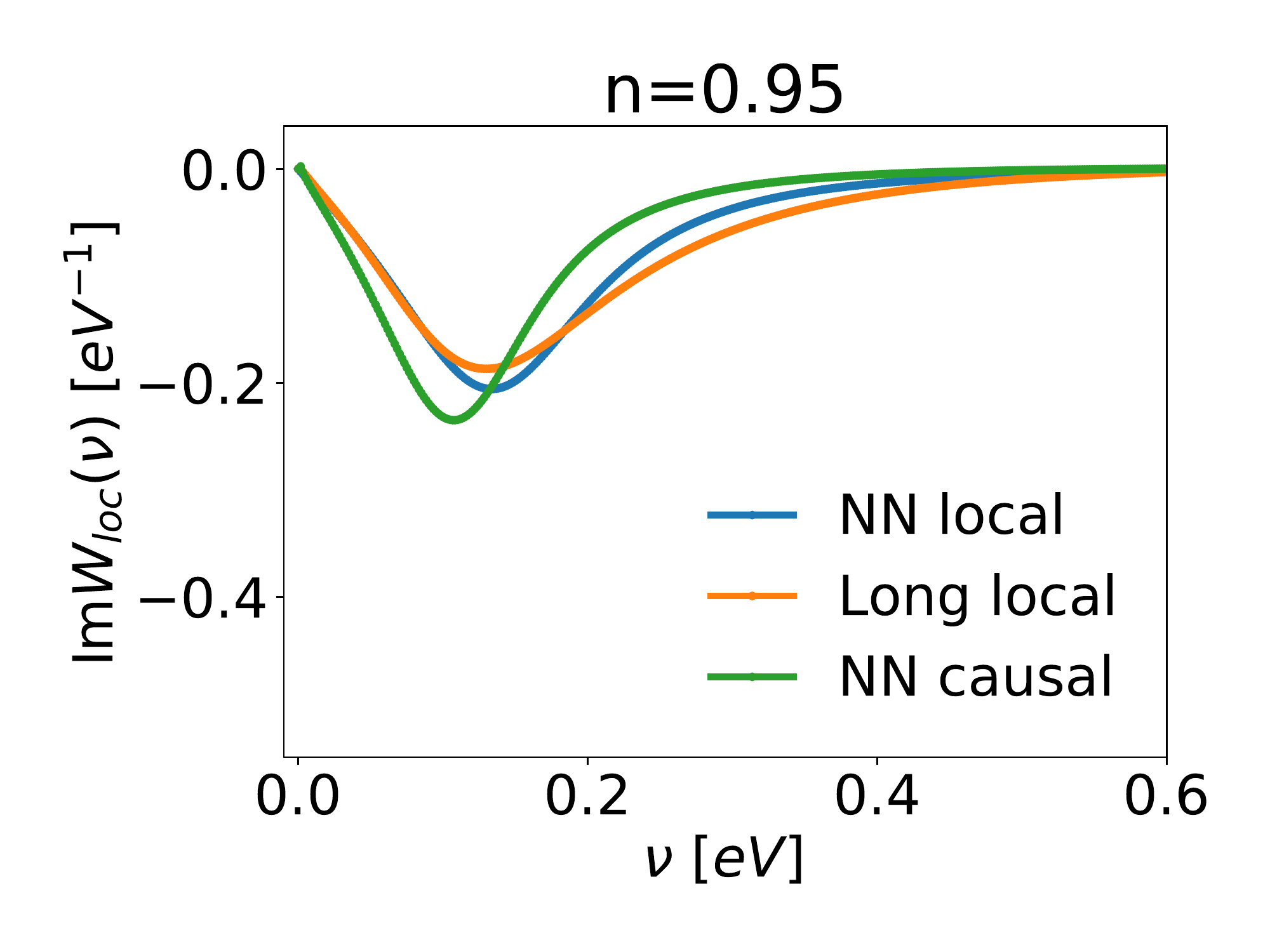}
	\includegraphics[width=0.475\linewidth]{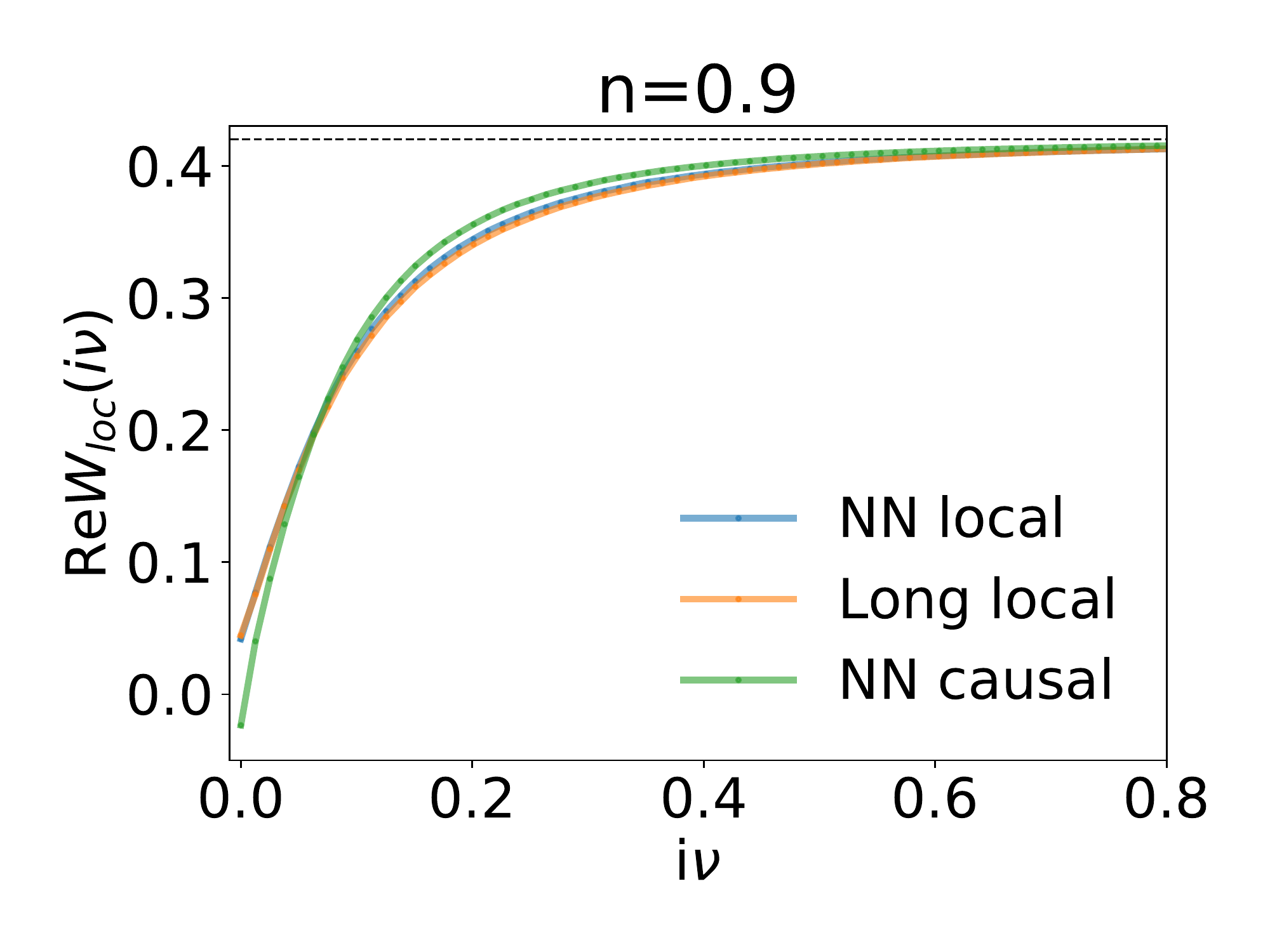}
	\includegraphics[width=0.475\linewidth]{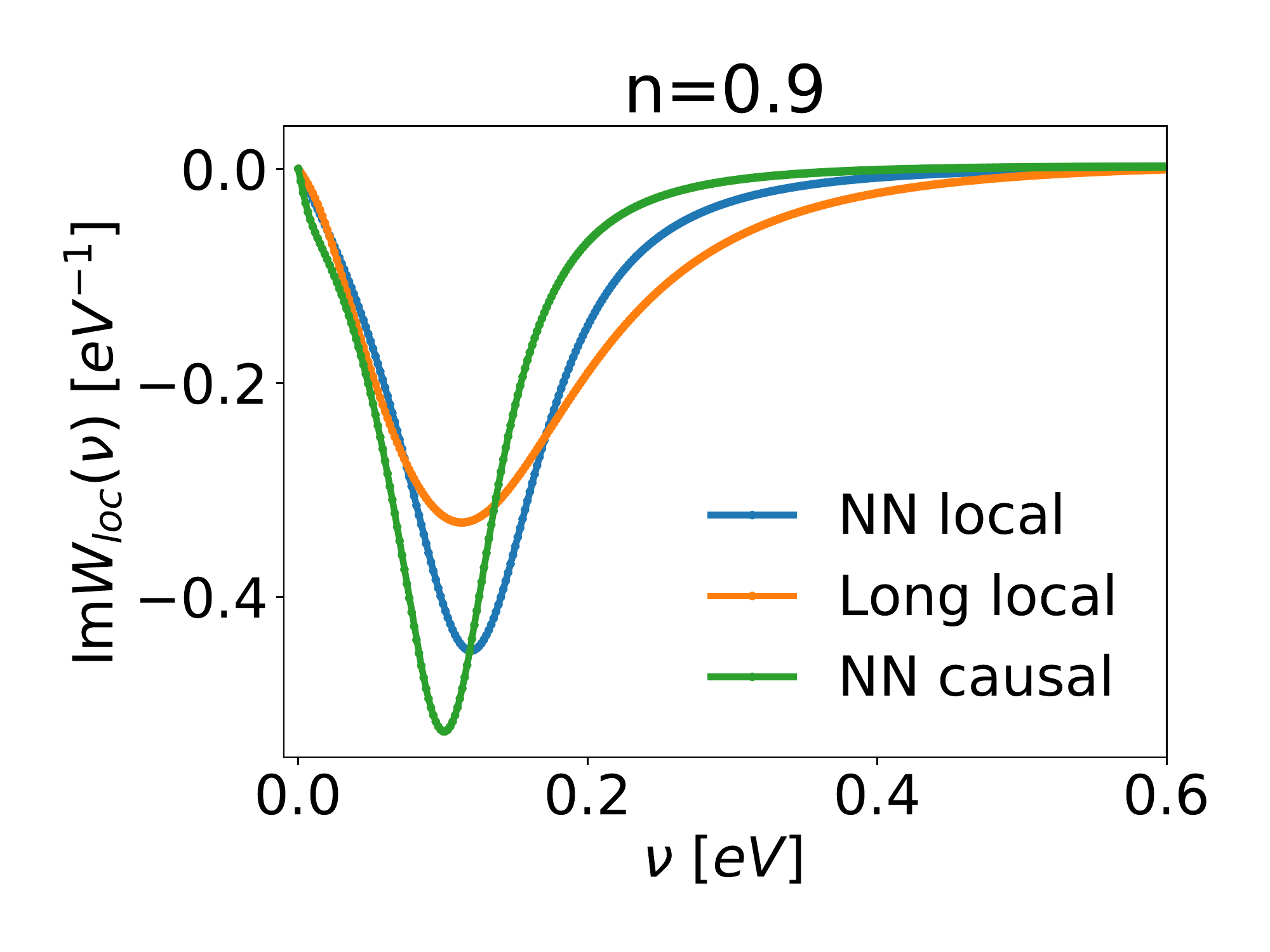}
	\caption{The local fully screened interaction $W_{\text{loc}}$ in the Matsubara frequency domain and in the real frequency domain after Pad\'e analytical continuation. In the top panels, the model is half-filled ($n=1$). The middle panels are for $5\%$ hole doping ($n=0.95$), and the bottom panels for $10\%$ hole doping ($n=0.9$). The interaction parameters are $U=0.42$, $V=0.08$.
	}\label{fig:doping_W}
\end{figure}

Figure~\ref{fig:doping_W} shows how the screening properties are modified by the hole doping. The left panels plot the results for $W_\text{loc}$ on the Matsubara frequency axis, and the middle and right panels the real and imaginary parts after maximum entropy analytical continuation. 
Again, we show results for the local scheme with NN interactions (blue), the local scheme with long-ranged interactions (orange) and the causal scheme with NN interactions (green). By comparing the static values, one notices that while the local screened interaction in the half-filled insulator is slightly larger in the causal scheme than in the local scheme, the opposite is true in the hole-doped systems. Hence, the effect of the correction terms $\Delta_\text{cor}$ and $\mathcal{U}_\text{cor}$ (Eq.~\eqref{eq:EDMFT_Generalised_correction}) on the correlation strength depends on doping, and more generally on the choice of parameters. Overall, however, we find a good agreement of the physical observables obtained by the two self-consistency schemes and also between the models with NN and long-ranged interactions. 

The imaginary part of the real-frequency $W_\text{loc}(\omega)$ reveals the relevant screening modes in the system. 
For the undoped (half-filled) system in the first row, the dominant screening mode, both in the fully screened local interaction Im$W_\text{loc}(\omega)$ and the retarded impurity interaction Im$\mathcal{U}(\omega)$ (not shown) approximately matches the 0.4 eV energy separation between the LHB and UHB. This suggests that screening in the Mott insulator is associated with charge excitations across the Mott gap.\cite{huang2014} The screening effect is however small, as one can deduce from the small reduction of the real part, compared to the bare $U$. For the doped systems, the peak appears around $~0.1$ eV, which approximately corresponds to the gap between the LHB and the quasi-particle peak in the spectral function. Hence, these types of charge excitations contribute primarily to the additional screening in the doped Mott system. Because of this additional screening, the static value of $W_\text{loc}$ is now substantially smaller than the bare $U$, and in the case of $n=0.9$ close to zero (in $\mathcal U$, the reduction is only about 15-20\%).

The effect of doping on the imaginary part of the local self-energy is plotted in the left panels of Fig.~\ref{fig:doping_SGA}. These results confirm the transition from Mott insulating to metallic behavior, and the previous observation that in the doped systems (especially for 5\% doping), the results from the causal scheme are less correlated. They also confirm that the differences between the local and causal scheme are rather small, as far as the correlation strength is concerned, and the same is true for the difference between the models with NN and long-ranged interactions.  

\subsubsection{Temperature dependence}

\begin{figure}[!]
	\centering
	\includegraphics[width=0.475\linewidth]{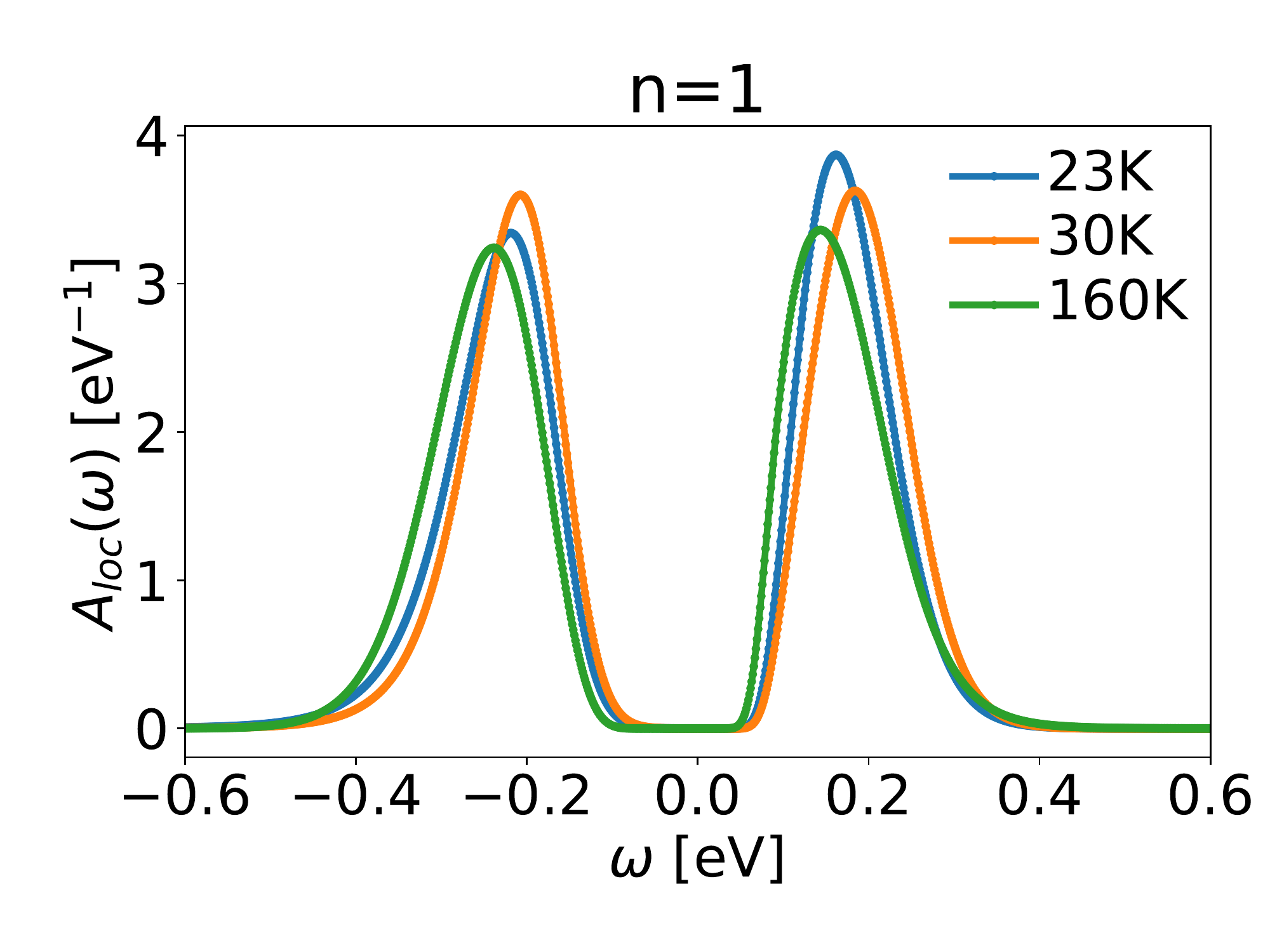}
	\includegraphics[width=0.475\linewidth]{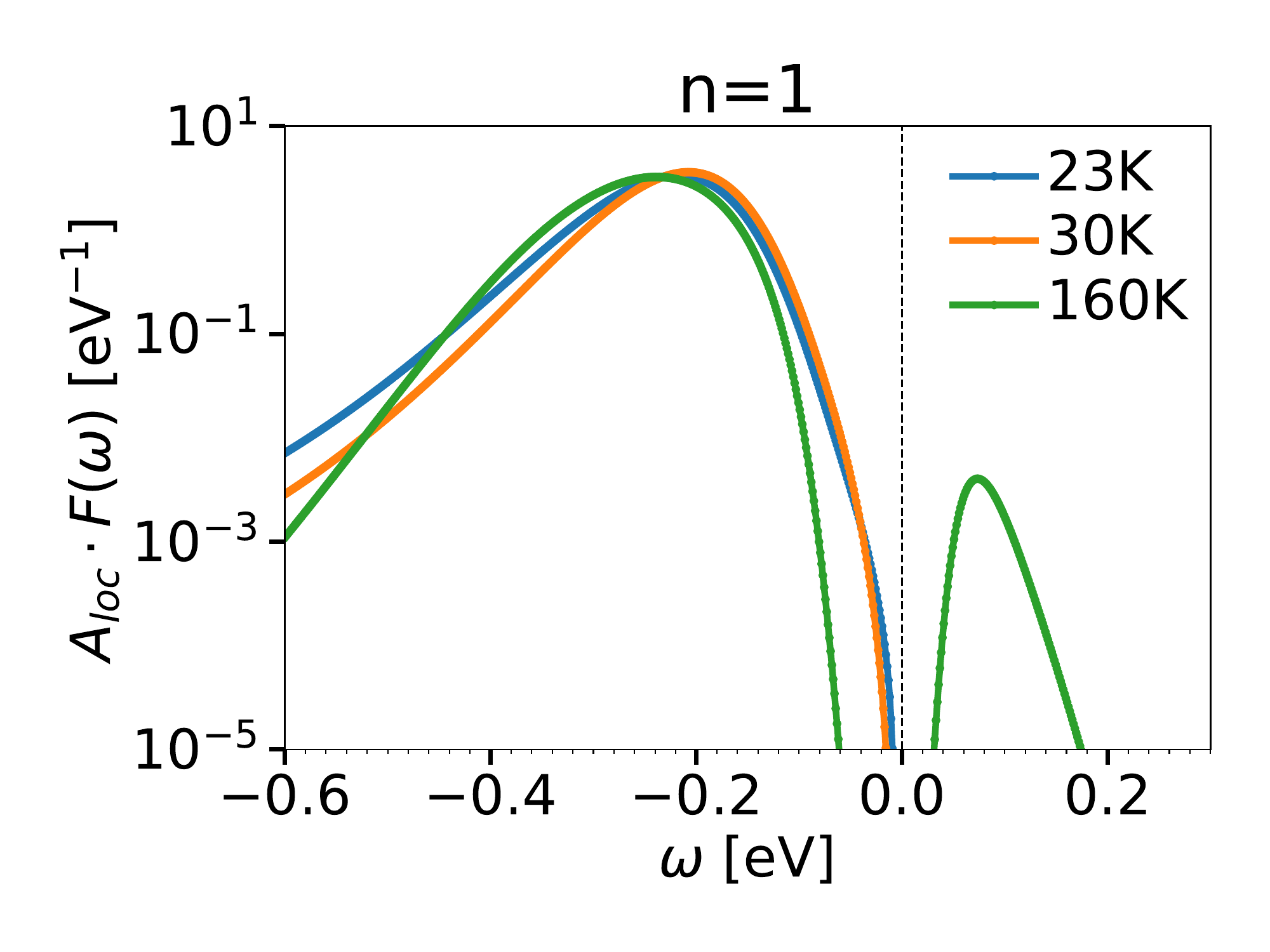}
	\includegraphics[width=0.475\linewidth]{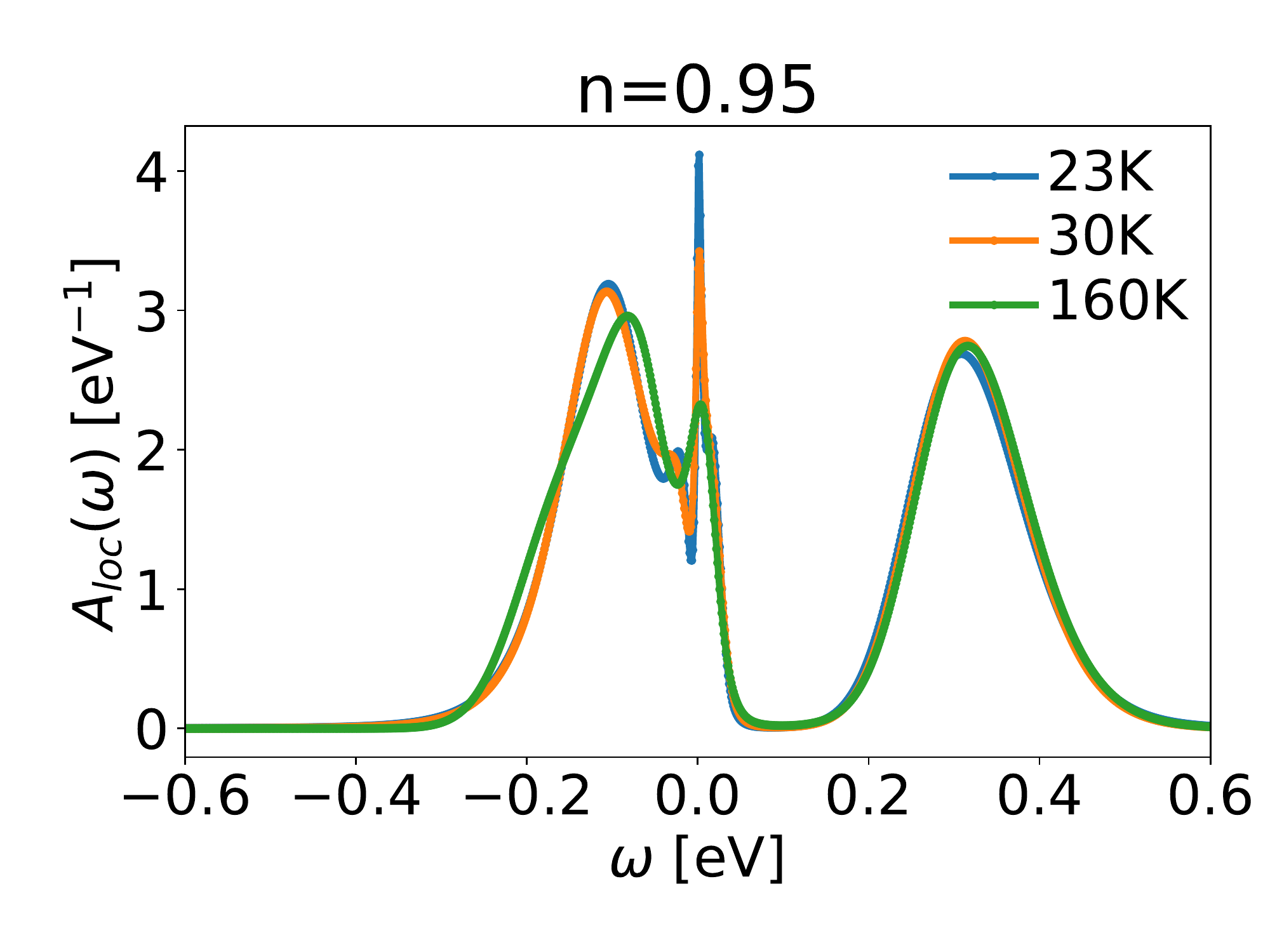}
	\includegraphics[width=0.475\linewidth]{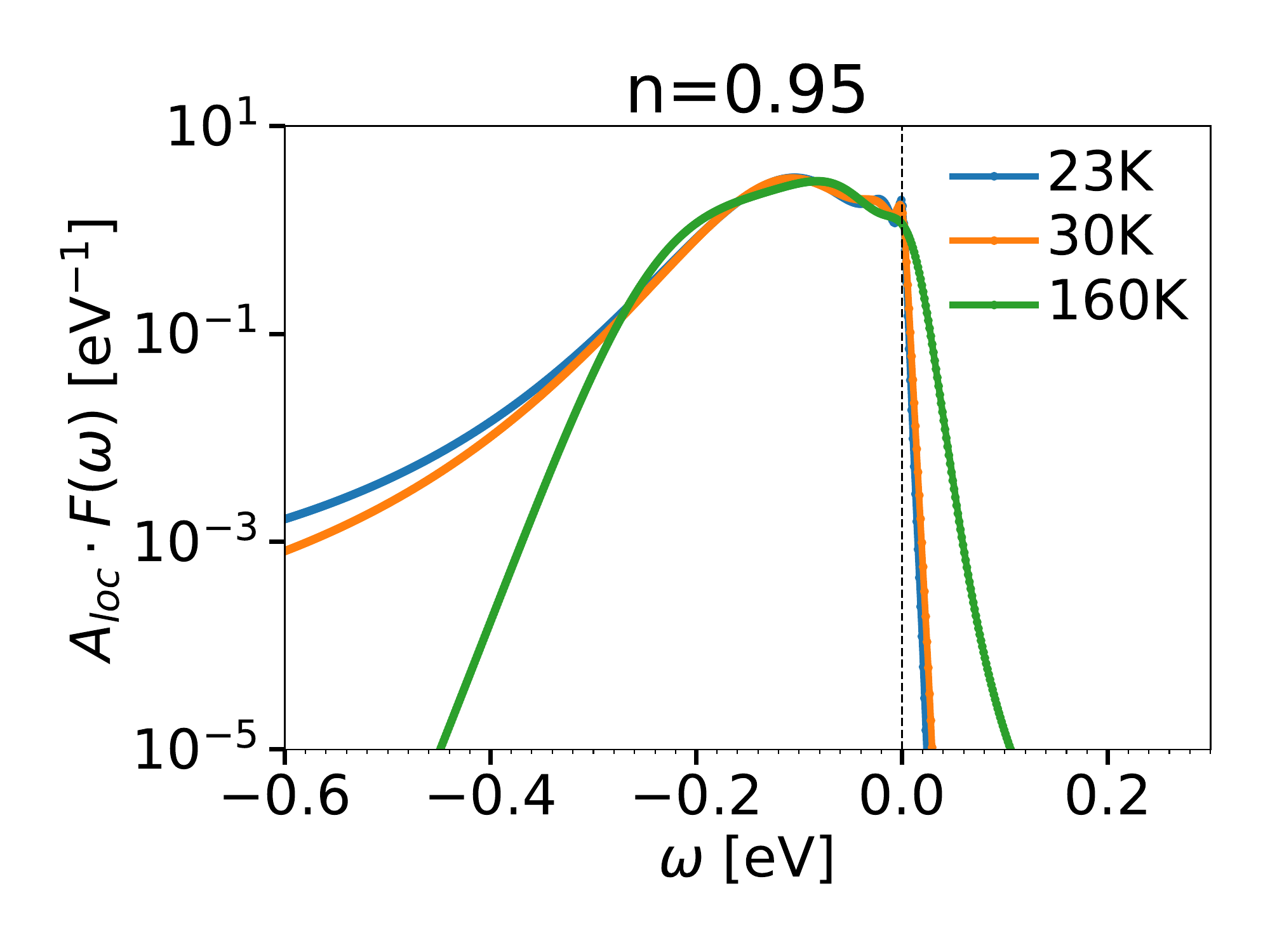}
	\includegraphics[width=0.475\linewidth]{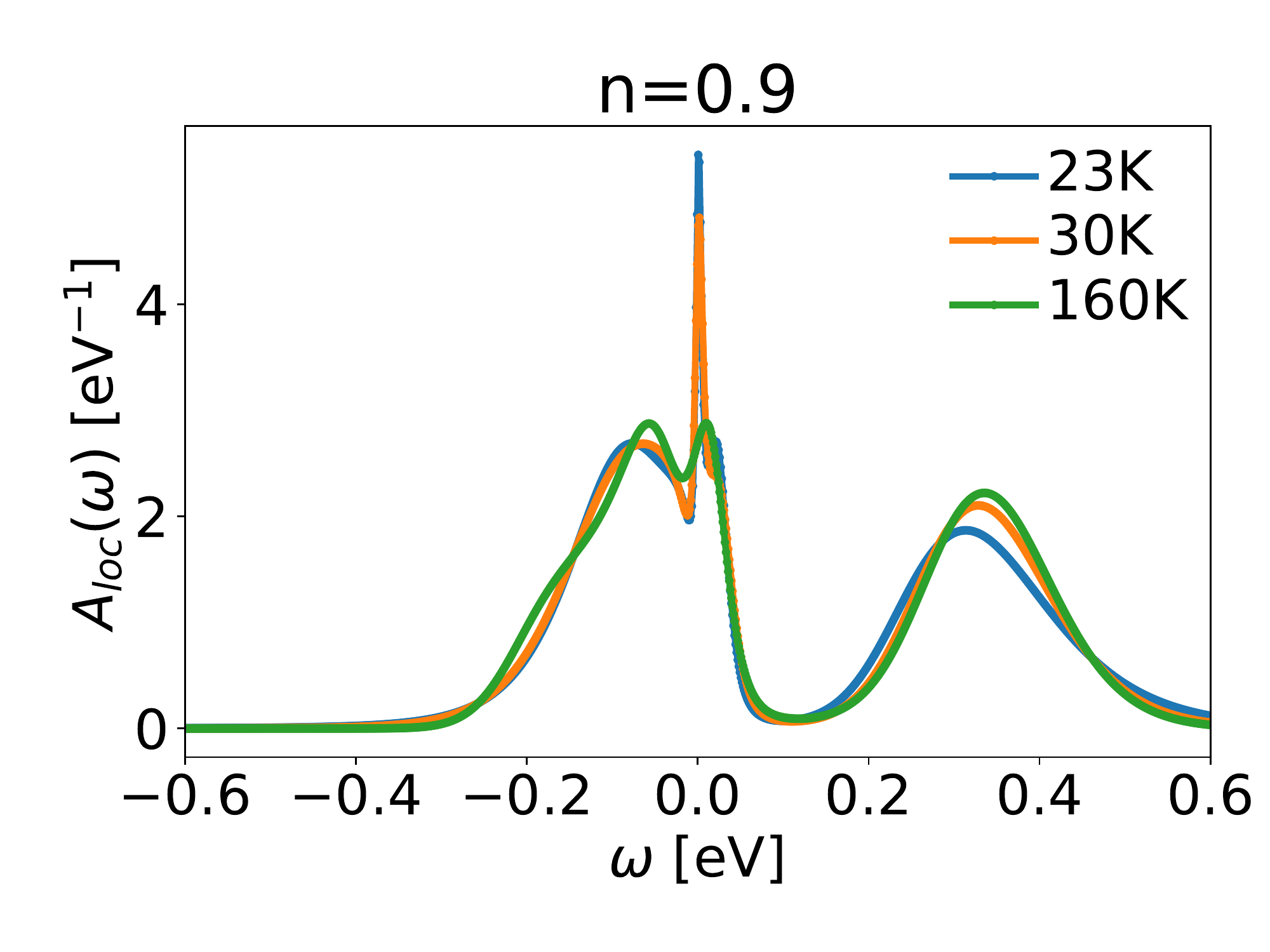}
	\includegraphics[width=0.475\linewidth]{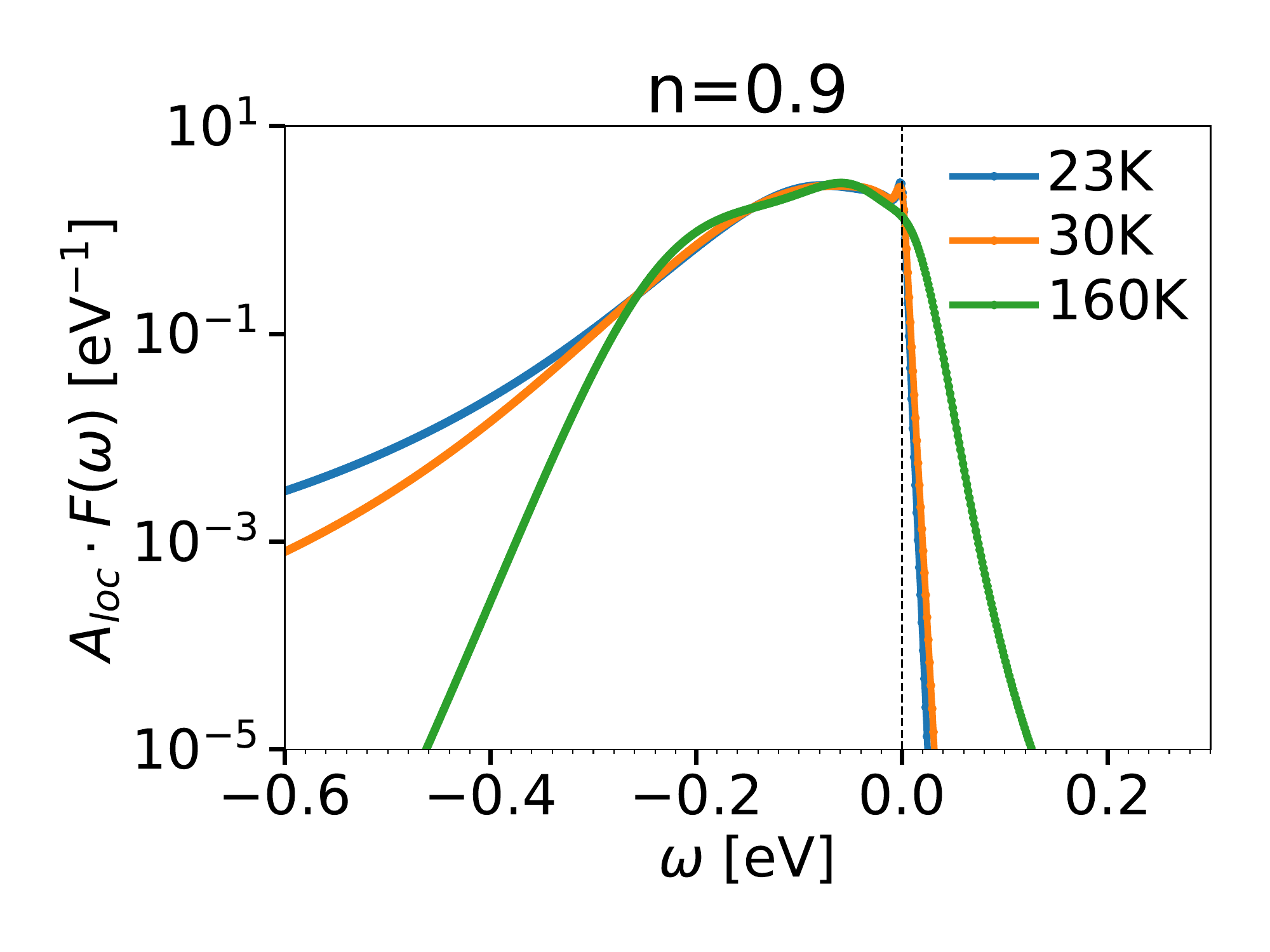}
	\caption{Spectral function $A_\text{loc}(\omega)$ and occupation function $A_{\text{loc}}(\omega) F(\omega)$ for the half-filled, 5\% hole-doped and 10\% hole-doped system, respectively, for the indicated values of temperature ($U=0.42$, $V=0.08$).
	}\label{fig:T_compare1}
\end{figure}

In Fig.~\ref{fig:T_compare1} we show how a temperature increase changes the local spectral function $A_\text{loc}(\omega)$ and the electron distribution function $A_\text{loc}(\omega) F(\omega)$ for different hole concentrations. The three rows are for half-filling, 5\% hole-doping, and 10\% hole-doping, respectively, while $U=0.42$, $V=0.08$ are fixed. In each panel, we plot the results for $T=23$~K, $30$~K and $160$~K. 
Up to $160$~K, i.e. inside the C-CDW phase, temperature does not significantly alter the spectra nor any of the physical fermionic and bosonic fields. Also the gap size is merely affected by thermal broadening. The main effect of increasing temperature is a partial melting of the quasi-particle peaks in the doped systems. In the occupation functions (right panels), the higher temperature leads to a slower decay of the occupation near the Fermi edge in the doped system, because of the broader $F(\omega)$. In the Mott insulator, at the highest temperature, one finds a more prominent thermal (doublon) population of the upper Hubbard band, and a small shift of the whole spectrum relative to the chemical potential. 

\subsubsection{Comparison with experiment}
\label{sec:exp}

\begin{figure}[t]
	\centering
	\includegraphics[width=0.99\linewidth]{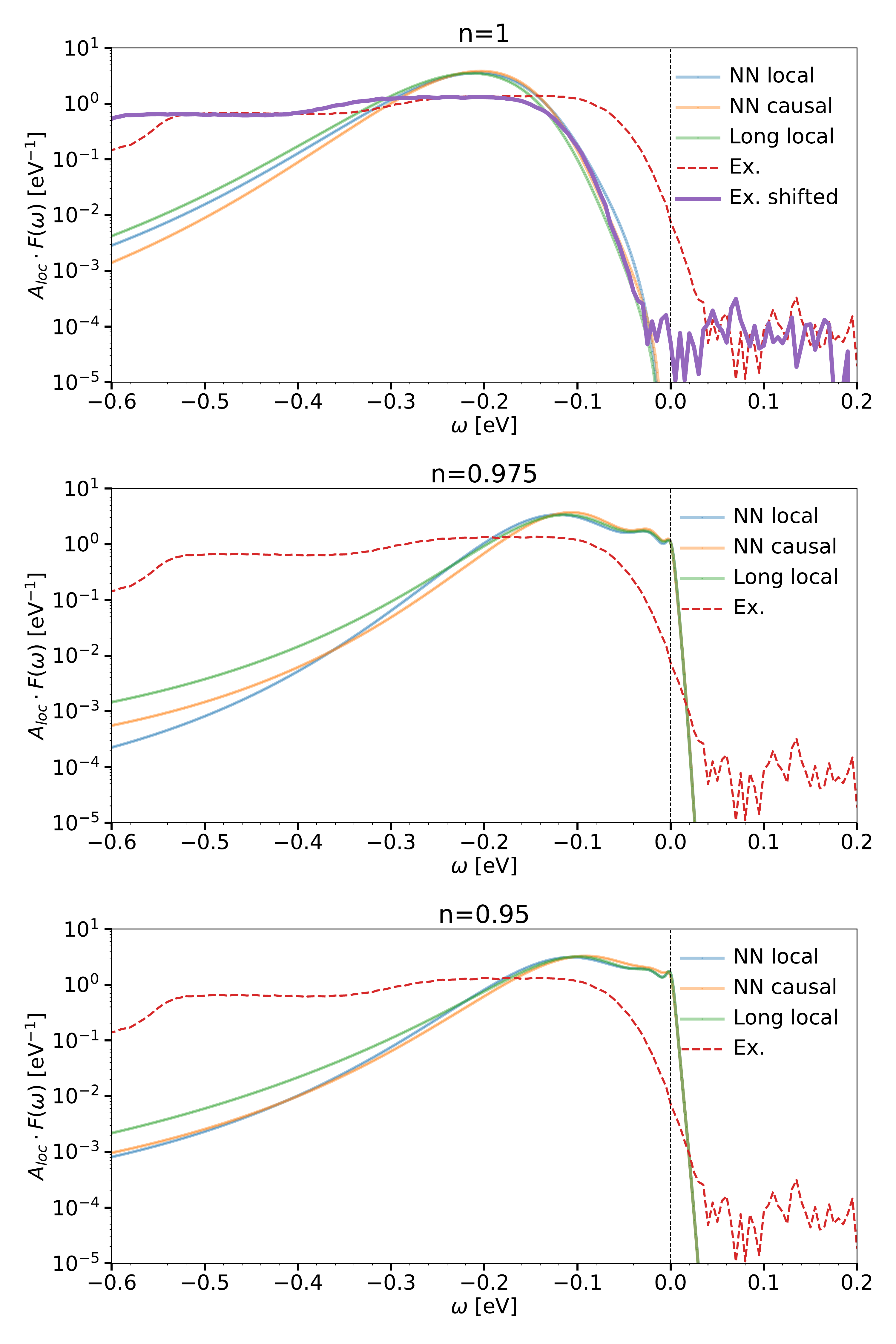}
	\caption{Occupation function $A_{\text{loc}}(\omega) F(\omega)$ for the half-filled, 2.5\% and 5\% hole-doped system (from top to bottom). The dashed red line corresponds to the equilibrium data plotted in Fig.~2(a) of Ref.~\onlinecite{ligges2018}. To match the experiment, the $GW$+EDMFT calculations are carried out here for $T=30$~K ($U=0.42$, $V=0.08$). The purple line is the experimental distribution shifted to match the calculated results (the chemical potential in the experiment is not precisely known). 
	}\label{fig:30K_compare}
\end{figure}

In Fig.~\ref{fig:30K_compare} we compare the distribution functions for the half-filled, 2.5\% and 5\% hole-doped system to the equilibrium photoemission spectrum reported in Fig.~2a of Ref.~\onlinecite{ligges2018}. To enable a direct comparison, we use the same temperature, $T=30$~K, as in the experiment. The red dashed line plots the raw experimental data, while the purple curve in the top panel plots the experimental result shifted by $\Delta \omega=0.065$~eV  on the energy axis. Experimentally, it is difficult to determine the chemical potential in an insulating system, and it is thus more meaningful in the gapped case to match the experimental curve with the upper edge of the lower Hubbard band in the simulation data. By doing so we notice that the measured distribution function reproduces very well the shape of the Hubbard band near the gap edge, and that the decay over three orders of magnitude is much slower than in a metallic system (middle and bottom panel), where it is controlled by the Fermi function cutoff, even for a small hole doping concentration. In addition, while the metallic quasi-particle peak in the distribution function and on a log-scale plot is not very prominent, no hint of such a feature is evident in the experimental data. These results suggest that the 1$T$-TaS$_2$ sample used in the experiment was not significantly hole-doped and that the doublon population dynamics measured in Ref.~\onlinecite{ligges2018} might be controlled by phenomena that 
their modeling did not capture. 

Apart from the slope near the chemical potential, there are two notable differences between the measured and calculated distributions. (i) The measured spectrum shows two plateau-like structures, with the lower one extending down to energies of approximately $-0.6$ eV. This latter feature comes from lower-lying bands, including sulphur bands, and are thus absent in our simulations. (ii) Even the higher-energy hump, which can be interpreted as the lower Hubbard band, is flatter and broader than in the simulations. This appears to be a consequence of the three-dimensional nature of  1$T$-TaS$_2$, which consists of a stacking of triangular-lattice-type layers. The hopping in the stacking direction is actually large, compared to the in-plane hopping, \cite{lee2019,pasquier2021} so that one may expect a broader, more 1-dimensional DOS than what comes out of our triangular-lattice simulation. The latter is in rough agreement with the DOS measured in STM experiments,\cite{cho2016,wu2021} but these experiments are less sensitive to the stacking direction than photoemission measurements.

\section{Conclusions}
\label{sec:conclusions}

In this work, we studied the phase diagram and correlation functions of the single-band extended Hubbard model on the two-dimensional triangular lattice using the standard local implementation of the self-consistency equations\cite{biermann2003,ayral2013} and a recently proposed causal variant.\cite{backes2020} The Ewald lattice summation method has also been implemented in order to investigate the role of longer-ranged nonlocal interactions.  Our test calculations in different regions of the phase diagram confirmed that the modified $GW$+EDMFT self-consistency loop of Backes, Sun and Biermann\cite{backes2020} removes the non-causal features in the effective impurity interaction while the hybridization function turns out to be causal in both schemes (for the model parameters considered). The causal variant results in slightly stronger correlations effects in the half-filled system, while in hole-doped systems, the correlations -- as measured by the imaginary part of the self-energy -- can be weakened.  We showed that this is the result of two opposing effects: while the effective interaction $\mathcal{U}$ produced by the causal scheme tends to be smaller than in the case of the local implementation, the hybridization function $\Delta$ is also smaller, which suppresses the kinetic energy. Overall, this results in physical observables (Green's functions and screened interactions) which do not significantly differ between the two schemes. This is comforting in the sense that recent {\it ab-initio} studies based on the local implementation of $GW$+DMFT have produced results which are in good agreement with experiments.\cite{petocchi2020,petocchi2021} 

While noncausal structures in auxiliary quantities like $\mathcal{U}$ and $\Delta$ are, conceptually, not a problem,\cite{nilsson2017} the causal dynamical mean fields produced by the scheme of Backes {\it et al.} can be advantageous from a numerical point of view. In particular, we expect that the real-time propagation in nonequilibrium implementations of $GW$+EDMFT\cite{golez2017,golez2019} becomes more stable. Causality also enables the use of maximum entropy analytical continuation for the analysis of the spectral content of the auxiliary fields.  

We presented simulation results for parameter values appropriate for the layered dichalchogenide 1$T$-TaS$_2$. In the low-temperature C-phase, this material is a polaronic Mott insulator that can be described by an effective single-band Hubbard model on the triangular lattice.\cite{wilson1975,fazekas1979a,sipos2008} Based on a rough estimate of the size of the molecular orbitals, we determined the ratio $V/U=0.19$ and, by comparing the width and separation of the Hubbard bands measured by photoemission and STM to the $GW$+EDMFT spectra, fixed the values of $U$ and $t$. This yields the parameters $U=0.42$~eV, $V=0.08$~eV, and $t=-0.02$~eV for our low energy description. Models with NN and longer-ranged nonlocal interactions treated with the Ewald summation were found to produce very similar results. Here we should note that the bandwidth for a single layer of TaS$_2$, as estimated by {\it ab-initio} calculations, is much smaller than 0.2 eV.~\cite{pasquier2021} This means that the above-estimated parameters implicitly take into account the three-dimensional nature of 1$T$-TaS$_2$, which has larger hoppings in the stacking direction than in the in-plane direction. The triangular-lattice simulations do however not explicitly capture the anisotropy of the lattice and this may be the reason why the Hubbard bands in the simulations are more peaked than in the (bulk sensitive) photoemission experiments. 

Motivated by the results of Ref.~\onlinecite{ligges2018}, which suggested a significant hole-doping in a nominally undoped sample of 1$T$-TaS$_2$, we investigated the hole-doping effect on the spectral function, self-energy, screened interaction and effective local interaction of the model with realistic parameters. These results show that in the absence of disorder or other effects not captured by $GW$+EDMFT, a hole doping of a few percent should result in a prominent quasi-particle peak at the edge of the lower Hubbard band at $T=30$~K. This peak was not seen in the photo-emission data.  Also, the direct comparison of the experimental spectrum to the simulated occupation functions for half-filling, 2.5\% and 5\% hole doping showed a very good agreement with the undoped result (near the gap edge, and after an appropriate energy shift of the experimental data), while there were significant differences to the distribution functions of the hole doped systems. In particular, the latter exhibit a much more rapid decay of the occupation, which is controlled by the temperature, rather than by the shape of the lower Hubbard band. These findings suggest that the 1$T$-TaS$_2$ crystal used in the experiments of Ref.~\onlinecite{ligges2018} was not significantly hole-doped, as was concluded based on the short doublon life-time.
An improved nonequilibrium DMFT description of 1$T$-TaS$_2$ (compared to Ref.~\onlinecite{ligges2018}) might be obtained by employing more reliable impurity solvers and by  considering photo-excitations from deeper-lying bands. It would also be worthwhile to study the stacking effect in nonequilibrium simulations, as well as different cooling mechanisms.\cite{murakami2015,werner2019}

We also considered the temperature effect in equilibrium systems, but apart from the weight of the quasi-particle peak, the physical properties (correlation strength, screening) are not much temperature dependent in the range ($T\lesssim 160$ K) relevant for the C-phase of 1$T$-TaS$_2$.

\begin{acknowledgments}
We thank S. Biermann for helfpul discussions, and U.~Bovensiepen for providing the raw data for the equilibrium spectrum in Fig. 2a of Ref.~\onlinecite{ligges2018}. This work was supported by  
the Swiss National Science Foundation via NCCR Marvel and Grant No. 200021-196966.
The calculations have been performed on the Beo05 cluster at the University of Fribourg.  
\end{acknowledgments}

\appendix

\section{Ewald summation}\label{app_ewald}
We consider a Coulomb repulsion $V(\mathbf{r})=\frac{1}{r}$ in a two-dimensional system.
Applying the Fourier transformation,  one finds
\begin{equation}
	V(\mathbf{k})= \int d\mathbf{r} e^{-i\mathbf{k}\cdot \mathbf{r}}\frac{1}{r}=\frac{2\pi}{k}.
\end{equation}
With the help of the the identity
\begin{equation}
	\frac{1}{k}=\frac{2}{\sqrt{\pi}}\int_0^\infty dt e^{-(kt)^2}
\end{equation}
one can express $V(\mathbf{k})$ as an integral. We separate the integral into the short-range contribution from $[0, \eta]$ and the long-range contribution from $[\eta, \infty]$, which yields
\begin{equation}
	\begin{aligned}
		V^S(\mathbf{k})=\int_0^\eta dt\frac{2\pi}{k}=4\sqrt{\pi}\int_0^\eta dt e^{-(kt)^2},\\ 
		V^L(\mathbf{k})=\int_\eta^\infty dt\frac{2\pi}{k}=4\sqrt{\pi}\int_\eta^\infty dt e^{-(kt)^2}.
	\end{aligned}
\end{equation}
As shown in the main text and in Fig.~\ref{fig:Vk}, the parameter $\eta$ plays a similar role as the $\alpha$ parameter in the Yukawa potential. 

Now we treat the two parts separately. For the short range part, the inverse Fourier transformation to real space gives
\begin{equation}
	\begin{aligned}
		V^S(\mathbf{r}) 
		&=\frac{1}{N_k}\sum_\mathbf{k}e^{i\mathbf{k}\cdot\mathbf{r}}4\sqrt{\pi}\int_0^\eta dt e^{-(kt)^2}\\
		&=\frac{1}{\sqrt{\pi}}\int_0^\eta dt \frac{1}{t^2} e^{-\frac{r^2}{4t^2}},
	\end{aligned}
\end{equation}
which, using the variable change $x\equiv \frac{r}{2t}$, becomes
\begin{equation}\label{eq:ewald_short}
	\begin{aligned}
		V^S(\mathbf{r})&
		=\frac{1}{r}\text{erfc}\Big(\frac{r}{2\eta}\Big).
	\end{aligned}
\end{equation}

Here, $\text{erfc}(x)$ is the complementary error function, defined as $\text{erfc}(x)=\frac{2}{\sqrt{\pi}}\int_x^\infty dx e^{-x^2}$.
For the long-range part, using the variable change $x\equiv kt$, one finds
\begin{equation}\label{eq:ewald_long}
	\begin{aligned}
		V^L(\mathbf{r}) 
		&=\frac{1}{N_\mathbf{k}}\sum_\mathbf{k}e^{i\mathbf{k}\cdot\mathbf{r}}4\sqrt{\pi}\int_\eta^\infty dt e^{-(kt)^2}\\
		&=\frac{1}{N_{\mathbf{k}}}\sum_\mathbf{k}e^{i\mathbf{k}\cdot\mathbf{r}}\frac{2\pi}{k}\text{erfc}(k\eta).
	\end{aligned}
\end{equation}
Due to charge neutrality, the $\mathbf{k}=0$ term, corresponding to the average potential,  
should be removed from the sum. Adding the expressions for the long-range and short-range potentials in Eq.~\eqref{eq:ewald_short} and Eq.~\eqref{eq:ewald_long} one obtains Eq.~\eqref{eq:ewald_potential} in the main text. 

\section{Derivation of the (causal) EDMFT self-consistency equations}\label{app_causal}

In this appendix, we derive, following Ref.~\onlinecite{backes2020}, the causal EDMFT self-consistency equations (Eqs.~\eqref{eq:EDMFT_Generalised_sc} and ~\eqref{eq:EDMFT_Generalised_correction}) using a generalized cavity method.\cite{georges1996} The basic idea of the cavity method is to separate a lattice into two parts, a single site denoted as ``site 0" and the remaining part of the lattice, i.e. the lattice with cavity, denoted by a superscript ``$^{(0)}$". After separating the action of the lattice model into terms related to site 0, the rest, and terms connecting the two parts, we integrate out all the degrees of freedom from the lattice with cavity, which leaves only those of the single site, denoted by the subscript ``$_0$". Starting from the lattice action, one thus obtains an effective action $S_{\text{eff}}$ of an impurity model. Formally, one can express the relationship between the lattice action and the action of the effective single-site model (impurity model) as
\begin{equation}\label{eq:s_eff}
	\frac{1}{Z} \int \prod_{i \neq 0, \sigma} \mathcal{D}[c^*_{i\sigma}, c_{i \sigma}] e^{-S[c_{i \sigma}^*, c_{i \sigma}]} =\frac{1}{Z_{\text {eff }}} e^{-S_{\text {eff }}\left[c_{0 \sigma}^*, c_{0 \sigma}\right]}.
\end{equation}

In the action of the full lattice model, the terms contributed by the degrees of freedom with no connection to the impurity site represent the action of the lattice with cavity, $S^{(0)}$. The degrees of freedom coupled to the impurity site are viewed as external sources, as the generators of the cavity Green's functions. Together with terms involving only the impurity site, contained in $S_0$, we can thus split the full lattice action defined in Eq.~\eqref{eq:hubbard_action2} into the three parts $S=S_{0}+S^{(0)}+\Delta S$, 
with the explicit form
\begin{equation}\label{eq:hubbard_partion2}
	\begin{aligned} 
		S_{0}=& \int_{0}^{\beta} d \tau\left\{\sum_{\sigma} c_{0 \sigma}^{*}(\tau)\left(\partial_{\tau}-\mu\right) c_{0 \sigma}(\tau)+\frac{1}{2} n_{0}v_{00} n_0\right\}, \\
		S^{(0)}=& \int_{0}^{\beta} d \tau\left\{\sum_{i j \neq 0, \sigma} c_{i \sigma}^{*}\left(\partial_{\tau}-\mu-t_{i j}\right) c_{j \sigma}+\frac{1}{2} \sum_{i j, \neq 0} n_{i}v_{i j} n_{j}\right\}, \\
		\Delta S=& \int_{0}^{\beta} d \tau\left\{-\sum_{i \neq 0, \sigma} t_{i 0}\left(c_{0 \sigma}^{*} c_{i \sigma}+c_{i \sigma}^{*} c_{0 \sigma}\right)+\sum_{i \neq 0} n_{i}v_{i 0} n_{0}\right\}.
	\end{aligned}    
\end{equation}

Defining $\eta_{i} \equiv t_{i 0} c_{0 \sigma}$ and $j_{i} \equiv v_{i 0} n_{0}$ we can rewrite the last term as
\begin{equation}
	\Delta S[\eta_i,\eta^*_i,j_i] =\int^\beta_0 d\tau\left\{-\sum_{i\neq 0,\sigma}(\eta^*_i c_{i\sigma}+c^*_{i\sigma}\eta_i)+\sum_{i\neq 0}j_i n_i\right\}.
\end{equation}
Now, the effective action $S_{\operatorname{eff}}$ defined in Eq.~\eqref{eq:s_eff} can be explicitly written as
\begin{align}
	&\frac{e^{-S_{\operatorname{eff}}}}{Z_{\operatorname{eff}}}=\frac{e^{-S_0}}{Z}\int \mathcal{D}_{i\neq 0}[c^*_{i\sigma},c_{i\sigma}]e^{-(S^{(0)}+\Delta S)}\\
	&\Longrightarrow S_{\operatorname{eff}}=S_0-\Omega+ \text{const},
\end{align}
where $\Omega[c^*_{0\sigma},c_{0\sigma}]=\ln \int \mathcal{D}_{i \neq 0}\left[c_{i}^{*}, c_{i}\right] e^{-\left(S^{(0)}+\Delta S\right)}$ is the generating functional of the cavity lattice Green's function,\cite{georges1996}
\begin{equation}
\begin{aligned}
	&G_{i_{1} \ldots i_{n} j_{n} \ldots j_{1}}^{(0)}\left(\tau_{1} \ldots \tau_{n}, \tau_{1}^{\prime} \ldots \tau_{n}^{\prime}\right)\\
	&~~~~=(-1)^{n} \frac{\delta^{2 n} \Omega}{\delta \eta_{i_{1}}^{*}\left(\tau_{1}\right) \ldots \delta \eta_{i_{n}}^{*}\left(\tau_{n}\right) \delta \eta_{j_{n}}\left(\tau_{n}^{\prime}\right) \ldots \delta \eta_{j_{1}}\left(\tau_{1}^{\prime}\right)}, \\
	&\chi_{i_{1} \ldots i_{n} j_{n} \ldots j_{1}}^{(0)}\left(\tau_{1} \ldots \tau_{n}, \tau_{1}^{\prime} \ldots \tau_{n}^{\prime}\right) \\
	&~~~~= \frac{\delta^{2 n} \Omega}{\delta j_{i_{1}}\left(\tau_{1}\right) \ldots \delta j_{i_{n}}\left(\tau_{n}\right) \delta j_{j_{n}}\left(\tau_{n}^{\prime}\right) \ldots \delta j_{j_{1}}\left(\tau_{1}^{\prime}\right)}.
\end{aligned}
\end{equation}
Thus, we obtain the series expansion of the effective action $S_{\operatorname{eff}}$,
\begin{equation}
	\begin{aligned}
		S_{\operatorname{eff}}&\left[\eta_{i}^{*}, \eta_{i}, j_{i}\right]=
		\\& -\sum_{n=1}^{\infty} \sum_{i_{1} \ldots j_{n}} \int d \tau_{1} \ldots d \tau_{n}^{\prime} \eta_{i_{1}}^{*}\left(\tau_{1}\right) \ldots \eta_{j_{n}}\left(\tau_{n}^{\prime}\right)\\&~\times(-1)^{n} G_{i_{1} \ldots i_{n} j_{n} \ldots j_{1}}^{(0)}\left(\tau_{1} \ldots \tau_{n}^{\prime}\right)\\
		& -\sum_{n=1}^{\infty} \sum_{i_{1} \ldots j_{n}} \int d \tau_{1} \ldots d \tau_{n}^{\prime} j_{i_{1}}\left(\tau_{1}\right) \ldots j_{j_{n}}\left(\tau_{n}^{\prime}\right) \\&~\times \chi_{i_{1} \ldots i_{n} j_{n} \ldots j_{1}}^{(0)}\left(\tau_{1} \ldots \tau_{n}^{\prime}\right)+S_0+\text{const}.
	\end{aligned}
\end{equation}

Thus far, every step is exact. Now, we 
truncate the series expansion of $\Omega$, which is only a good approximation in a system with large connectivity. Assuming that each lattice site has $z\to \infty$ neighbors, we have to rescale $t\to t/\sqrt{z}$ and $V\to V/z$ in order to keep the balance between kinetic and potential energy.\footnote{From the local perspective, only the case when one election hops in while one hops out with amplitude $t_{j0}t_{0i}$ contributes to the local Hamiltonian. Thus, $(zt)^2$ has to be an $\mathcal{O}(1)$ quantity as $z$ goes to infinity.\cite{muller-hartmann1989}}
 Given this rescaling, it has been proven \cite{metzner1989,georges1996} that the large connectivity limit will dramatically simplify the expansion of the effective action, because the $n^\text{th}$ terms scale as $(1/z)^{n-1}$ and only the $n=1$ terms survive, 
\begin{equation}
\begin{aligned}
	S_{\operatorname{eff}}=&\int d\tau d\tau' \left\{c^*_{i\sigma}(\tau) t_{0i}t_{j0}G^{(0)}_{ij}(\tau-\tau')c_{j\sigma}(\tau')\right.
	\\&\left.-n_0(\tau) v_{0i}v_{j0}\chi^{(0)}_{ij}(\tau-\tau')n_0(\tau')\right\}+S_0 +\mathcal{O}\Big(\frac{1}{z}\Big)\\
	=&-\int^\beta_0 d\tau d\tau'\sum_{\sigma}c_\sigma^*(\tau)\mathcal{G}(\tau-\tau')c_\sigma(\tau')\\
	&+\frac{1}{2}\int^\beta_0 d \tau d\tau' n(\tau)\mathcal{U}(\tau-\tau')n(\tau')+\mathcal{O}\Big(\frac{1}{z}\Big),
\end{aligned}
\end{equation}
where the Weiss fields $\mathcal{G}$ and $\mathcal{U}$ are defined as, 
\begin{equation}
	\begin{aligned}
		\mathcal{G}^{-1}\left(\tau_1-\tau_2\right) =& \ \delta(\tau_1-\tau_2)
		(-\partial_{\tau_1}+\mu)\\& -\sum_{i j\neq 0} t_{0 i} t_{j 0} G_{i j}^{(0)}\left(\tau_1-\tau_2\right), \\
		\mathcal{U}\left(\tau_1-\tau_2\right) =&  \ \delta(\tau_1-\tau_2)v_{00}\\&-\sum_{ij\neq 0} v_{ 0i} v_{j 0} \chi_{i j}^{(0)}\left(\tau_1-\tau_2\right).
	\end{aligned}
\end{equation}
Here, $G^{(0)}$ and $\chi^{(0)}$ denote the Green's function and charge susceptibility of the lattice with cavity,
\begin{equation}
	\begin{aligned}
		{{G}^{(0)}_{ij\sigma}}\left(\tau_1-\tau_2\right) &\equiv -\langle\timeorder c_{i\sigma}(\tau_1)c^*_{j\sigma}(\tau_2) \rangle^{(0)}, \\
		\chi_{ij}^{(0)}\left(\tau_1-\tau_2\right) &\equiv\langle n_i(\tau_1)n_j(\tau_2)\rangle^{(0)}.
	\end{aligned}
\end{equation}
In the Matsubara frequency $i\omega_n$ $(i\nu_n)$ domain, the Weiss fields thus become
\begin{equation}\label{eq:Weiss_matsu1}
	\begin{aligned}
		\mathcal{G}^{-1}\left(i\omega_n\right) &= i\omega_n+\mu-\sum_{i j} t_{0i} t_{j 0} G_{i j}^{(0)}\left(i\omega_n\right), \\
		\mathcal{U}\left(i\nu_n\right) &= v_{00}-\sum_{ij} v_{0i} v_{j 0} \chi_{i j}^{(0)}\left(i\nu_n\right).
	\end{aligned}
\end{equation}
In some previous works,\cite{sun2002,biermann2003,ayral2013,huang2014} an alternative form of the interaction part of the action has been considered by applying the Hubbard-Stratonovich transformation\cite{hubbard1959} to the interaction terms in Eq.~\eqref{eq:hubbard_action2}. This allows to decouple the density-density interactions and to connect the formalism to the LW functional $\Psi$. The resulting alternative form of the action is
\begin{equation}\label{eq:hubbard_action}
	\begin{aligned}
		&S\left[c^{*}, c, \phi\right] = \!\int_{0}^{\beta} \!\!d \tau\left\{\!-\sum_{i j \sigma} \! c_{i \sigma}^{*}(\tau)[(-\partial_\tau+\tilde{\mu})\delta_{ij}-t_{ij}] c_{j \sigma}(\tau)\!\right\}\\&+\int_{0}^{\beta} d \tau 
		\left\{\frac{1}{2} \sum_{i j} \phi_{i}(\tau)\left[v^{-1}\right]_{i j} \phi_{j}(\tau)+i \sum_{i} \phi_{i}(\tau) n_{i}(\tau)\right\},
	\end{aligned}
\end{equation}
with $\phi$ the bosonic Hubbard-Stratonovich field and $ \tilde \mu =\mu+U/2$. Following the same steps as before, one obtains 
\begin{equation}\label{eq:Weiss_matsu1_alternative}
	\begin{aligned}
		\mathcal{G}^{-1}\left(i\omega_n\right) &= i\omega_n+\mu-\sum_{i j} t_{0i} t_{j 0} G_{i j}^{(0)}\left(i\omega_n\right), \\
		\mathcal{U}^{-1}\left(i\nu_n\right) &= v^{-1}_{00}-\sum_{ij} v^{-1}_{0i} v^{-1}_{j 0} W_{i j}^{(0)}\left(i\nu_n\right),
	\end{aligned}
\end{equation}
where $W_{ij}^{(0)}=\langle\phi_i(0)\phi_j(\tau)\rangle^{(0)}$ is the bosonic Green's function of the lattice with cavity. Eqs.~\eqref{eq:Weiss_matsu1} and \eqref{eq:Weiss_matsu1_alternative} show two different definitions of the Weiss field $\mathcal{U}$. As we will see, they become equivalent if one uses the standard (local) self-consistency scheme, while in the causal scheme, they are no longer obviously equivalent. 

The second approximation (for lattices with finite coordination number) concerns the relation between the correlation functions $G^{(0)}_{ij}$, $\chi^{(0)}_{ij}$ and $W^{(0)}_{ij}$ of the lattice with cavity, and the full lattice correlation functions $G_{ij}$, $\chi_{ij}$ and $W_{ij}$. For a general lattice, the relations are
\begin{equation}\label{eq:cavity_function}
	\begin{aligned}
		G_{i j}^{(0)} &=G_{i j}-G_{i 0} G_{0 j} / G_{00}, \\ 
		\chi_{i j}^{(0)} &=\chi_{i j}-\chi_{i 0} \chi_{0 j} / \chi_{00},\\
		W_{i j}^{(0)} &=W_{i j}-W_{i 0} W_{0 j} / W_{00},
	\end{aligned}
\end{equation}
which become exact in the large connectivity limit. The proof is done by expressing the Green's functions in terms of an inverse Hamiltonian matrix in the basis of lattice sites, as described in Sec.~III.C of Ref.~\onlinecite{georges1996}, and derived originally by Hubbard.\cite{hubbard1964} We can understand the first relation by counting the paths contributing to the Green's function. In the large connectivity limit, the difference between $G^{(0)}_{ij}$ and $G_{ij}$ lies in the contribution from the paths connecting the sites $i$ and $j$ through the site $0$ \textit{only once}. Symbolically, this contribution is proportional to $G_{i0}G_{0j}$ but has to be normalized by $G_{00}$, the probability of leaving and returning to site 0, in order to cancel paths passing through site 0 more than once. The treatment of the susceptibility $\chi$ and the screened interaction $W$ is analogous.

Now, inserting Eq.~(\ref{eq:cavity_function}) into Eq.~\eqref{eq:Weiss_matsu1} or Eq.~\eqref{eq:Weiss_matsu1_alternative}, we find
\begin{equation}\label{eq:Weiss_matsu2}
	\begin{aligned}
		\mathcal{G}^{-1}\left(i\omega_n\right) =& \  i\omega_n+\mu-\sum_{i j} t_{0i} t_{j 0} G_{i j}\left(i\omega_n\right)\\&+\sum_{i j} t_{0i}G_{i 0}\left(i\omega_n\right) G_{0j}\left(i\omega_n\right)t_{j 0}/ G_{0 0}\left(i\omega_n\right), \\
		\mathcal{U}\left(i\nu_n\right) =&\ v_{00}-\sum_{ij} v_{0i} v_{j 0} \chi_{i j}\left(i\nu_n\right)\\&+\sum_{ij} v_{ 0i} \chi_{i 0}\left(i\nu_n\right) \chi_{ 0j}\left(i\nu_n\right)v_{j 0}/\chi_{0 0}\left(i\nu_n\right),
	\end{aligned}
\end{equation}
and in the alternative form,
\begin{equation}\label{eq:Weiss_matsu2_alternative}
	\begin{aligned}
		\mathcal{U}^{-1}\left(i\nu_n\right) =& \ v^{-1}_{00}-\sum_{ij} v^{-1}_{0i} v^{-1}_{j 0} W_{i j}\left(i\nu_n\right)\\&+\sum_{ij} v^{-1}_{i 0} W_{ 0i}\left(i\nu_n\right) W_{0j}\left(i\nu_n\right)v^{-1}_{j 0}/W_{0 0}\left(i\nu_n\right).
	\end{aligned}
\end{equation}
Fourier transforming to momentum space and introducing the notation  $\langle\ldots\rangle_k\equiv\frac{1}{N_k} \sum_k$, one finds\cite{backes2020}
\begin{equation}\label{eq:Weiss_matsu3}
	\begin{aligned}
		\mathcal{G}^{-1}\left(i\omega_n\right) =& \ i\omega_n+\mu-\left(\langle \epsilon G \epsilon\rangle_k-\langle \epsilon G\rangle_k\langle G\rangle_k^{-1}\langle G \epsilon\rangle_k\right), \\
		\mathcal{U}(i\nu_n) =& \ \langle v\rangle_k-\left(\langle v \chi v\rangle_k-\langle v \chi\rangle_k\langle\chi\rangle_k^{-1}\langle\chi v\rangle_k\right),
	\end{aligned}
\end{equation}
or alternatively,
\begin{equation}\label{eq:Weiss_matsu3_alternative}
	\begin{aligned}
		\mathcal{U}^{-1}\left(i\omega_n\right) =& \ \langle v^{-1}\rangle_k-(\langle v^{-1} W v^{-1}\rangle_k\\&\quad \quad \quad \quad \quad -\langle v^{-1} W\rangle_k\langle W\rangle_k^{-1}\langle W v^{-1}\rangle_k).
	\end{aligned}
\end{equation}
 
We can further replace the band dispersion $\epsilon_k$ used in Eq.~\eqref{eq:Weiss_matsu3} by the lattice Dyson equation $\epsilon_k\to-G^{-1}(k,i\omega_n)+i\omega_n+\mu-\Sigma(k,i\omega_n)$. In the case of the screened interaction, we replace the Coulomb interaction $v_k$ used in Eq.~\eqref{eq:Weiss_matsu3_alternative} by the bosonic Dyson equation $v_k\to W^{-1}(k,i\omega_n)+\Pi(k,i\omega_n)$ and replace the charge susceptibility $\chi$ by the screened interaction $W$ using the identity
$W=v-v\chi v$. After some manipulations, we obtain the generalized EDMFT self-consistency equations\cite{backes2020}
\begin{equation}\label{eq:EDMFT_Generalised_sc_app}
	\begin{aligned}
		\mathcal{G}^{-1}\left(i\omega_n\right) &=\quad\langle G\rangle^{-1}+\langle\Sigma\rangle+\Delta_\text{cor},\\
		\mathcal{U}(i \nu_n)&=\left[\langle W\rangle^{-1}+\langle \Pi\rangle\right]^{-1}+\mathcal{U}_\text{cor},
	\end{aligned}
\end{equation}
and the alternative form
\begin{equation}\label{eq:EDMFT_Generalised_sc_alternative}
	\begin{aligned}
		\mathcal{U}^{-1}(i \nu_n)&=\left[\langle W\rangle^{-1}+\langle \Pi\rangle\right]+{\tilde{\mathcal{U}}}^{-1}_\text{cor},
	\end{aligned}
\end{equation}
where the correction terms are given by
\begin{widetext}   
\begin{equation}\label{eq:EDMFT_Generalised_correction_app}
	\begin{aligned}
		\Delta_\text{cor} &= -\langle\Sigma G \Sigma\rangle_k+\langle\Sigma G\rangle_k\langle G\rangle_k^{-1}\langle G \Sigma\rangle_k-2\langle\Sigma\rangle_k+\langle\Sigma G\rangle_k\langle G\rangle_k^{-1}+\langle G\rangle_k^{-1}\langle G \Sigma\rangle_k , \\
        \mathcal{U}_\text{cor} &=-\langle \Pi W\rangle_k[\langle \Pi\rangle_k+\langle \Pi W \Pi\rangle_k]^{-1}\langle W \Pi\rangle_k+\langle \Pi\rangle_k\langle W\rangle_k[\langle \Pi\rangle_k+\langle \Pi\rangle_k\langle W\rangle_k\langle \Pi\rangle_k]^{-1}\langle W\rangle_k\langle \Pi\rangle_k , \\
        {\tilde{\mathcal{U}}}^{-1}_\text{cor} &=-\langle\Pi W \Pi\rangle_k+\langle\Pi W\rangle_k\langle W\rangle^{-1}_k\langle W \Pi\rangle_k-2\langle\Pi\rangle_k+\langle\Pi W\rangle_k\langle W\rangle_k^{-1}+\langle W\rangle_k^{-1}\langle W \Pi\rangle_k.
	\end{aligned}
\end{equation}
\end{widetext}

In the standard DMFT \cite{georges1996} or EDMFT \cite{sun2002} treatment, the self-energy $\Sigma$ and the polarization $\Pi$ are momentum independent and calculated by the impurity solver: $\Sigma=\Sigma_\text{imp}$ and $\Pi=\Pi_{\text{imp}}$. In this case, we have $\langle G\Sigma\rangle=\langle G\rangle \Sigma_{\text{imp}}$ and $\langle W\Pi\rangle=\langle W\rangle \Pi_\text{imp}$ and the correction terms defined by Eq.~\eqref{eq:EDMFT_Generalised_correction_app} vanish. In this case the generalized schemes reduce to the standard (local) self-consistency loop and there is no difference between the two alternative procedures (Eqs.~\eqref{eq:EDMFT_Generalised_sc_app} and \eqref{eq:EDMFT_Generalised_sc_alternative}). In the $GW$+EDMFT scheme, however, one combines the local impurity contributions $(\Sigma_{\text{imp}},\Pi_{\text{imp}})$ with the nonlocal parts of the $GW$ contributions $\Sigma^{GW}(k)$ and $\Pi^{GW}(k)$. In this case, the correction terms defined in Eq.~\eqref{eq:EDMFT_Generalised_correction_app} will not cancel. 

The standard (local) treatment of $GW$+(E)DMFT used in previous works \cite{sun2002,biermann2003,ayral2013,nilsson2017} 
considers only the local part of the $k$-dependent self-energy and polarization 
in the self-consistency equations (Eqs.~\eqref{eq:EDMFT_Generalised_sc_app} and \eqref{eq:EDMFT_Generalised_sc_alternative}) which fix the dynamical mean fields. This amounts to neglecting the correction terms. While the local procedure becomes exact in the infinite-connectivity limit, and thus is consistent with the first two approximations, it can produce non-causal Weiss fields (especially $\mathcal{U}$), as discussed in Refs.~\onlinecite{nilsson2017,backes2020} and the main text. 

To summarize, for the EDMFT procedure or the standard (local) $GW$+EDMFT procedure, the correction terms defined by Eq.~\eqref{eq:EDMFT_Generalised_correction_app} cancel, and one obtains the local EDMFT-type self-consistency equations,
\begin{equation}\label{eq:EDMFT_sc}
	\begin{aligned}
		\mathcal{G}^{-1}=G_{\mathrm{loc}}^{-1}+\Sigma_{\mathrm{loc}}, \\
		\mathcal{U}^{-1}=W_{\mathrm{loc}}^{-1}+\Pi_{\mathrm{loc}}.
	\end{aligned}
\end{equation}
The $GW$+EDMFT variant with the correction terms in the self-consistency equations,  Eqs.~\eqref{eq:EDMFT_Generalised_sc_app} or \eqref{eq:EDMFT_Generalised_sc_alternative}, will be referred to as the \textit{causal} self-consistency scheme, since as shown in Ref.~\onlinecite{backes2020}, it results in causal dynamical mean fields. This causal scheme is relevant only in the case of $k$-dependent self-energies and polarizations, i.e. if EDFMT is combined with the $GW$ approximation (or some other scheme which can provide non-local components). 

In this work, we only implement and test the first of the two alternative forms, i.e. the causal scheme defined by Eq.~\eqref{eq:EDMFT_Generalised_sc_app}.

\bibliography{ref}

\begin{thebibliography}{49}
\expandafter\ifx\csname natexlab\endcsname\relax\def\natexlab#1{#1}\fi
\expandafter\ifx\csname bibnamefont\endcsname\relax
  \def\bibnamefont#1{#1}\fi
\expandafter\ifx\csname bibfnamefont\endcsname\relax
  \def\bibfnamefont#1{#1}\fi
\expandafter\ifx\csname citenamefont\endcsname\relax
  \def\citenamefont#1{#1}\fi
\expandafter\ifx\csname url\endcsname\relax
  \def\url#1{\texttt{#1}}\fi
\expandafter\ifx\csname urlprefix\endcsname\relax\def\urlprefix{URL }\fi
\providecommand{\bibinfo}[2]{#2}
\providecommand{\eprint}[2][]{\url{#2}}

\bibitem[{\citenamefont{Mott}(1968)}]{mott1968}
\bibinfo{author}{\bibfnamefont{N.~F.} \bibnamefont{Mott}},
  \bibinfo{journal}{Rev. Mod. Phys.} \textbf{\bibinfo{volume}{40}},
  \bibinfo{pages}{677} (\bibinfo{year}{1968}),
  \urlprefix\url{https://link.aps.org/doi/10.1103/RevModPhys.40.677}.

\bibitem[{\citenamefont{Imada et~al.}(1998)\citenamefont{Imada, Fujimori, and
  Tokura}}]{imada1998}
\bibinfo{author}{\bibfnamefont{M.}~\bibnamefont{Imada}},
  \bibinfo{author}{\bibfnamefont{A.}~\bibnamefont{Fujimori}}, \bibnamefont{and}
  \bibinfo{author}{\bibfnamefont{Y.}~\bibnamefont{Tokura}},
  \bibinfo{journal}{Rev. Mod. Phys.} \textbf{\bibinfo{volume}{70}},
  \bibinfo{pages}{1039} (\bibinfo{year}{1998}),
  \urlprefix\url{https://link.aps.org/doi/10.1103/RevModPhys.70.1039}.

\bibitem[{\citenamefont{Hedin}(1999)}]{hedin1999}
\bibinfo{author}{\bibfnamefont{L.}~\bibnamefont{Hedin}}, \bibinfo{journal}{J.
  Phys.: Condens. Matter} \textbf{\bibinfo{volume}{11}}, \bibinfo{pages}{R489}
  (\bibinfo{year}{1999}),
  \urlprefix\url{https://iopscience.iop.org/article/10.1088/0953-8984/11/42/201}.

\bibitem[{\citenamefont{Werner and Casula}(2016)}]{werner2016}
\bibinfo{author}{\bibfnamefont{P.}~\bibnamefont{Werner}} \bibnamefont{and}
  \bibinfo{author}{\bibfnamefont{M.}~\bibnamefont{Casula}},
  \bibinfo{journal}{J. Phys.: Condens. Matter} \textbf{\bibinfo{volume}{28}},
  \bibinfo{pages}{383001} (\bibinfo{year}{2016}), \eprint{1602.00584},
  \urlprefix\url{http://arxiv.org/abs/1602.00584}.

\bibitem[{\citenamefont{Hansmann et~al.}(2013)\citenamefont{Hansmann, Ayral,
  Vaugier, Werner, and Biermann}}]{hansmann2013}
\bibinfo{author}{\bibfnamefont{P.}~\bibnamefont{Hansmann}},
  \bibinfo{author}{\bibfnamefont{T.}~\bibnamefont{Ayral}},
  \bibinfo{author}{\bibfnamefont{L.}~\bibnamefont{Vaugier}},
  \bibinfo{author}{\bibfnamefont{P.}~\bibnamefont{Werner}}, \bibnamefont{and}
  \bibinfo{author}{\bibfnamefont{S.}~\bibnamefont{Biermann}},
  \bibinfo{journal}{Phys. Rev. Lett.} \textbf{\bibinfo{volume}{110}},
  \bibinfo{pages}{166401} (\bibinfo{year}{2013}),
  \urlprefix\url{https://link.aps.org/doi/10.1103/PhysRevLett.110.166401}.

\bibitem[{\citenamefont{Ayral et~al.}(2012)\citenamefont{Ayral, Werner, and
  Biermann}}]{ayral2012}
\bibinfo{author}{\bibfnamefont{T.}~\bibnamefont{Ayral}},
  \bibinfo{author}{\bibfnamefont{P.}~\bibnamefont{Werner}}, \bibnamefont{and}
  \bibinfo{author}{\bibfnamefont{S.}~\bibnamefont{Biermann}},
  \bibinfo{journal}{Phys. Rev. Lett.} \textbf{\bibinfo{volume}{109}},
  \bibinfo{pages}{226401} (\bibinfo{year}{2012}),
  \urlprefix\url{https://link.aps.org/doi/10.1103/PhysRevLett.109.226401}.

\bibitem[{\citenamefont{Huang et~al.}(2014)\citenamefont{Huang, Ayral,
  Biermann, and Werner}}]{huang2014}
\bibinfo{author}{\bibfnamefont{L.}~\bibnamefont{Huang}},
  \bibinfo{author}{\bibfnamefont{T.}~\bibnamefont{Ayral}},
  \bibinfo{author}{\bibfnamefont{S.}~\bibnamefont{Biermann}}, \bibnamefont{and}
  \bibinfo{author}{\bibfnamefont{P.}~\bibnamefont{Werner}},
  \bibinfo{journal}{Phys. Rev. B} \textbf{\bibinfo{volume}{90}},
  \bibinfo{pages}{195114} (\bibinfo{year}{2014}),
  \urlprefix\url{https://link.aps.org/doi/10.1103/PhysRevB.90.195114}.

\bibitem[{\citenamefont{Sun and Kotliar}(2002)}]{sun2002}
\bibinfo{author}{\bibfnamefont{P.}~\bibnamefont{Sun}} \bibnamefont{and}
  \bibinfo{author}{\bibfnamefont{G.}~\bibnamefont{Kotliar}},
  \bibinfo{journal}{Phys. Rev. B} \textbf{\bibinfo{volume}{66}},
  \bibinfo{pages}{085120} (\bibinfo{year}{2002}),
  \urlprefix\url{https://link.aps.org/doi/10.1103/PhysRevB.66.085120}.

\bibitem[{\citenamefont{Biermann et~al.}(2003)\citenamefont{Biermann,
  Aryasetiawan, and Georges}}]{biermann2003}
\bibinfo{author}{\bibfnamefont{S.}~\bibnamefont{Biermann}},
  \bibinfo{author}{\bibfnamefont{F.}~\bibnamefont{Aryasetiawan}},
  \bibnamefont{and} \bibinfo{author}{\bibfnamefont{A.}~\bibnamefont{Georges}},
  \bibinfo{journal}{Phys. Rev. Lett.} \textbf{\bibinfo{volume}{90}},
  \bibinfo{pages}{086402} (\bibinfo{year}{2003}),
  \urlprefix\url{https://link.aps.org/doi/10.1103/PhysRevLett.90.086402}.

\bibitem[{\citenamefont{Aryasetiawan et~al.}(2004)\citenamefont{Aryasetiawan,
  Biermann, and Georges}}]{aryasetiawan2004}
\bibinfo{author}{\bibfnamefont{F.}~\bibnamefont{Aryasetiawan}},
  \bibinfo{author}{\bibfnamefont{S.}~\bibnamefont{Biermann}}, \bibnamefont{and}
  \bibinfo{author}{\bibfnamefont{A.}~\bibnamefont{Georges}},
  \bibinfo{journal}{arXiv:cond-mat/0401626}  (\bibinfo{year}{2004}),
  \eprint{cond-mat/0401626},
  \urlprefix\url{http://arxiv.org/abs/cond-mat/0401626}.

\bibitem[{\citenamefont{Biermann et~al.}(2004)\citenamefont{Biermann,
  Aryasetiawan, and Georges}}]{biermann2004}
\bibinfo{author}{\bibfnamefont{S.}~\bibnamefont{Biermann}},
  \bibinfo{author}{\bibfnamefont{F.}~\bibnamefont{Aryasetiawan}},
  \bibnamefont{and} \bibinfo{author}{\bibfnamefont{A.}~\bibnamefont{Georges}},
  \bibinfo{journal}{arXiv:cond-mat/0401653}  (\bibinfo{year}{2004}),
  \eprint{cond-mat/0401653},
  \urlprefix\url{http://arxiv.org/abs/cond-mat/0401653}.

\bibitem[{\citenamefont{Petocchi et~al.}(2021)\citenamefont{Petocchi,
  Christiansson, and Werner}}]{petocchi2021}
\bibinfo{author}{\bibfnamefont{F.}~\bibnamefont{Petocchi}},
  \bibinfo{author}{\bibfnamefont{V.}~\bibnamefont{Christiansson}},
  \bibnamefont{and} \bibinfo{author}{\bibfnamefont{P.}~\bibnamefont{Werner}},
  \bibinfo{journal}{arXiv:2106.03689 [cond-mat]}  (\bibinfo{year}{2021}),
  \eprint{2106.03689}, \urlprefix\url{http://arxiv.org/abs/2106.03689}.

\bibitem[{\citenamefont{Ewald}(1921)}]{ewald1921}
\bibinfo{author}{\bibfnamefont{P.~P.} \bibnamefont{Ewald}},
  \bibinfo{journal}{Annalen der Physik} \textbf{\bibinfo{volume}{369}},
  \bibinfo{pages}{253} (\bibinfo{year}{1921}),
  \urlprefix\url{https://onlinelibrary.wiley.com/doi/abs/10.1002/andp.19213690304}.

\bibitem[{\citenamefont{Harris}(1998)}]{harris1998}
\bibinfo{author}{\bibfnamefont{F.~E.} \bibnamefont{Harris}},
  \bibinfo{journal}{International Journal of Quantum Chemistry}
  \textbf{\bibinfo{volume}{68}}, \bibinfo{pages}{385} (\bibinfo{year}{1998}),
  \urlprefix\url{https://onlinelibrary.wiley.com/doi/abs/10.1002/%28SICI%291097-461X%281998%2968%3A6%3C385%3A%3AAID-QUA2%3E3.0.CO%3B2-R}.

\bibitem[{\citenamefont{Backes et~al.}(2020)\citenamefont{Backes, Sim, and
  Biermann}}]{backes2020}
\bibinfo{author}{\bibfnamefont{S.}~\bibnamefont{Backes}},
  \bibinfo{author}{\bibfnamefont{J.-H.} \bibnamefont{Sim}}, \bibnamefont{and}
  \bibinfo{author}{\bibfnamefont{S.}~\bibnamefont{Biermann}},
  \bibinfo{journal}{arXiv:2011.05311 [cond-mat]}  (\bibinfo{year}{2020}),
  \eprint{2011.05311}, \urlprefix\url{http://arxiv.org/abs/2011.05311}.

\bibitem[{\citenamefont{Nilsson et~al.}(2017)\citenamefont{Nilsson, Boehnke,
  Werner, and Aryasetiawan}}]{nilsson2017}
\bibinfo{author}{\bibfnamefont{F.}~\bibnamefont{Nilsson}},
  \bibinfo{author}{\bibfnamefont{L.}~\bibnamefont{Boehnke}},
  \bibinfo{author}{\bibfnamefont{P.}~\bibnamefont{Werner}}, \bibnamefont{and}
  \bibinfo{author}{\bibfnamefont{F.}~\bibnamefont{Aryasetiawan}},
  \bibinfo{journal}{Phys. Rev. Materials} \textbf{\bibinfo{volume}{1}},
  \bibinfo{pages}{043803} (\bibinfo{year}{2017}),
  \urlprefix\url{https://link.aps.org/doi/10.1103/PhysRevMaterials.1.043803}.

\bibitem[{\citenamefont{Wang et~al.}(2020)\citenamefont{Wang, Yao, Xin, Han,
  Wang, Chen, Cai, Li, and Zhang}}]{wang2020}
\bibinfo{author}{\bibfnamefont{Y.~D.} \bibnamefont{Wang}},
  \bibinfo{author}{\bibfnamefont{W.~L.} \bibnamefont{Yao}},
  \bibinfo{author}{\bibfnamefont{Z.~M.} \bibnamefont{Xin}},
  \bibinfo{author}{\bibfnamefont{T.~T.} \bibnamefont{Han}},
  \bibinfo{author}{\bibfnamefont{Z.~G.} \bibnamefont{Wang}},
  \bibinfo{author}{\bibfnamefont{L.}~\bibnamefont{Chen}},
  \bibinfo{author}{\bibfnamefont{C.}~\bibnamefont{Cai}},
  \bibinfo{author}{\bibfnamefont{Y.}~\bibnamefont{Li}}, \bibnamefont{and}
  \bibinfo{author}{\bibfnamefont{Y.}~\bibnamefont{Zhang}},
  \bibinfo{journal}{Nat Commun} \textbf{\bibinfo{volume}{11}},
  \bibinfo{pages}{4215} (\bibinfo{year}{2020}),
  \urlprefix\url{https://www.nature.com/articles/s41467-020-18040-4}.

\bibitem[{\citenamefont{Ligges et~al.}(2018)\citenamefont{Ligges, Avigo,
  Gole{\v z}, Strand, Beyazit, Hanff, Diekmann, Stojchevska, Kall{\"a}ne, Zhou
  et~al.}}]{ligges2018}
\bibinfo{author}{\bibfnamefont{M.}~\bibnamefont{Ligges}},
  \bibinfo{author}{\bibfnamefont{I.}~\bibnamefont{Avigo}},
  \bibinfo{author}{\bibfnamefont{D.}~\bibnamefont{Gole{\v z}}},
  \bibinfo{author}{\bibfnamefont{H.~U.~R.} \bibnamefont{Strand}},
  \bibinfo{author}{\bibfnamefont{Y.}~\bibnamefont{Beyazit}},
  \bibinfo{author}{\bibfnamefont{K.}~\bibnamefont{Hanff}},
  \bibinfo{author}{\bibfnamefont{F.}~\bibnamefont{Diekmann}},
  \bibinfo{author}{\bibfnamefont{L.}~\bibnamefont{Stojchevska}},
  \bibinfo{author}{\bibfnamefont{M.}~\bibnamefont{Kall{\"a}ne}},
  \bibinfo{author}{\bibfnamefont{P.}~\bibnamefont{Zhou}}, \bibnamefont{et~al.},
  \bibinfo{journal}{Phys. Rev. Lett.} \textbf{\bibinfo{volume}{120}},
  \bibinfo{pages}{166401} (\bibinfo{year}{2018}),
  \urlprefix\url{https://link.aps.org/doi/10.1103/PhysRevLett.120.166401}.

\bibitem[{\citenamefont{Negele and Orland}(1988)}]{negele1988}
\bibinfo{author}{\bibfnamefont{J.~W.} \bibnamefont{Negele}} \bibnamefont{and}
  \bibinfo{author}{\bibfnamefont{H.}~\bibnamefont{Orland}},
  \emph{\bibinfo{title}{Quantum {{Many}}-Particle {{Systems}}}}
  (\bibinfo{publisher}{{Basic Books}}, \bibinfo{year}{1988}), ISBN
  \bibinfo{isbn}{978-0-201-12593-1}.

\bibitem[{\citenamefont{Ayral et~al.}(2013)\citenamefont{Ayral, Biermann, and
  Werner}}]{ayral2013}
\bibinfo{author}{\bibfnamefont{T.}~\bibnamefont{Ayral}},
  \bibinfo{author}{\bibfnamefont{S.}~\bibnamefont{Biermann}}, \bibnamefont{and}
  \bibinfo{author}{\bibfnamefont{P.}~\bibnamefont{Werner}},
  \bibinfo{journal}{Phys. Rev. B} \textbf{\bibinfo{volume}{87}},
  \bibinfo{pages}{125149} (\bibinfo{year}{2013}),
  \urlprefix\url{https://link.aps.org/doi/10.1103/PhysRevB.87.125149}.

\bibitem[{\citenamefont{Hedin}(1965)}]{hedin1965}
\bibinfo{author}{\bibfnamefont{L.}~\bibnamefont{Hedin}},
  \bibinfo{journal}{Phys. Rev.} \textbf{\bibinfo{volume}{139}},
  \bibinfo{pages}{A796} (\bibinfo{year}{1965}),
  \urlprefix\url{https://link.aps.org/doi/10.1103/PhysRev.139.A796}.

\bibitem[{\citenamefont{Petocchi
  et~al.}(2020{\natexlab{a}})\citenamefont{Petocchi, Nilsson, Aryasetiawan, and
  Werner}}]{petocchi2020}
\bibinfo{author}{\bibfnamefont{F.}~\bibnamefont{Petocchi}},
  \bibinfo{author}{\bibfnamefont{F.}~\bibnamefont{Nilsson}},
  \bibinfo{author}{\bibfnamefont{F.}~\bibnamefont{Aryasetiawan}},
  \bibnamefont{and} \bibinfo{author}{\bibfnamefont{P.}~\bibnamefont{Werner}},
  \bibinfo{journal}{Phys. Rev. Research} \textbf{\bibinfo{volume}{2}},
  \bibinfo{pages}{013191} (\bibinfo{year}{2020}{\natexlab{a}}),
  \urlprefix\url{https://link.aps.org/doi/10.1103/PhysRevResearch.2.013191}.

\bibitem[{\citenamefont{Georges et~al.}(1996)\citenamefont{Georges, Kotliar,
  Krauth, and Rozenberg}}]{georges1996}
\bibinfo{author}{\bibfnamefont{A.}~\bibnamefont{Georges}},
  \bibinfo{author}{\bibfnamefont{G.}~\bibnamefont{Kotliar}},
  \bibinfo{author}{\bibfnamefont{W.}~\bibnamefont{Krauth}}, \bibnamefont{and}
  \bibinfo{author}{\bibfnamefont{M.~J.} \bibnamefont{Rozenberg}},
  \bibinfo{journal}{Rev. Mod. Phys.} \textbf{\bibinfo{volume}{68}},
  \bibinfo{pages}{13} (\bibinfo{year}{1996}),
  \urlprefix\url{https://link.aps.org/doi/10.1103/RevModPhys.68.13}.

\bibitem[{\citenamefont{Si and Smith}(1996)}]{si1996}
\bibinfo{author}{\bibfnamefont{Q.}~\bibnamefont{Si}} \bibnamefont{and}
  \bibinfo{author}{\bibfnamefont{J.~L.} \bibnamefont{Smith}},
  \bibinfo{journal}{Phys. Rev. Lett.} \textbf{\bibinfo{volume}{77}},
  \bibinfo{pages}{3391} (\bibinfo{year}{1996}),
  \urlprefix\url{https://link.aps.org/doi/10.1103/PhysRevLett.77.3391}.

\bibitem[{\citenamefont{Aryasetiawan and Gunnarsson}(1998)}]{aryasetiawan1998}
\bibinfo{author}{\bibfnamefont{F.}~\bibnamefont{Aryasetiawan}}
  \bibnamefont{and}
  \bibinfo{author}{\bibfnamefont{O.}~\bibnamefont{Gunnarsson}},
  \bibinfo{journal}{Rep. Prog. Phys.} \textbf{\bibinfo{volume}{61}},
  \bibinfo{pages}{237} (\bibinfo{year}{1998}), \eprint{cond-mat/9712013},
  \urlprefix\url{http://arxiv.org/abs/cond-mat/9712013}.

\bibitem[{\citenamefont{Pavarini et~al.}(2011)\citenamefont{Pavarini, Koch,
  Vollhardt, and Lichtenstein}}]{pavarini2011a}
\bibinfo{editor}{\bibfnamefont{E.}~\bibnamefont{Pavarini}},
  \bibinfo{editor}{\bibfnamefont{E.}~\bibnamefont{Koch}},
  \bibinfo{editor}{\bibfnamefont{D.}~\bibnamefont{Vollhardt}},
  \bibnamefont{and}
  \bibinfo{editor}{\bibfnamefont{A.}~\bibnamefont{Lichtenstein}}, eds.,
  \emph{\bibinfo{title}{The {{LDA}}+{{DMFT}} Approach to Strongly Correlated
  Materials}}, vol.~\bibinfo{volume}{1} of \emph{\bibinfo{series}{Schriften Des
  {{Forschungszentrums J\"ulich Reihe Modeling}} and {{Simulation}}}}
  (\bibinfo{publisher}{{Forschungszentrum J\"ulich}},
  \bibinfo{address}{{J\"ulich}}, \bibinfo{year}{2011}), ISBN
  \bibinfo{isbn}{978-3-89336-734-4},
  \urlprefix\url{http://www.cond-mat.de/events/correl11}.

\bibitem[{\citenamefont{Luttinger and Ward}(1960)}]{luttinger1960}
\bibinfo{author}{\bibfnamefont{J.~M.} \bibnamefont{Luttinger}}
  \bibnamefont{and} \bibinfo{author}{\bibfnamefont{J.~C.} \bibnamefont{Ward}},
  \bibinfo{journal}{Phys. Rev.} \textbf{\bibinfo{volume}{118}},
  \bibinfo{pages}{1417} (\bibinfo{year}{1960}),
  \urlprefix\url{https://link.aps.org/doi/10.1103/PhysRev.118.1417}.

\bibitem[{\citenamefont{Almbladh et~al.}(1999)\citenamefont{Almbladh, Barth,
  and Leeuwen}}]{almbladh1999}
\bibinfo{author}{\bibfnamefont{C.-O.} \bibnamefont{Almbladh}},
  \bibinfo{author}{\bibfnamefont{U.~V.} \bibnamefont{Barth}}, \bibnamefont{and}
  \bibinfo{author}{\bibfnamefont{R.~V.} \bibnamefont{Leeuwen}},
  \bibinfo{journal}{Int. J. Mod. Phys. B} \textbf{\bibinfo{volume}{13}},
  \bibinfo{pages}{535} (\bibinfo{year}{1999}),
  \urlprefix\url{https://www.worldscientific.com/doi/abs/10.1142/S0217979299000436}.

\bibitem[{\citenamefont{Werner and Millis}(2006)}]{werner2006}
\bibinfo{author}{\bibfnamefont{P.}~\bibnamefont{Werner}} \bibnamefont{and}
  \bibinfo{author}{\bibfnamefont{A.~J.} \bibnamefont{Millis}},
  \bibinfo{journal}{Phys. Rev. B} \textbf{\bibinfo{volume}{74}},
  \bibinfo{pages}{155107} (\bibinfo{year}{2006}),
  \urlprefix\url{https://link.aps.org/doi/10.1103/PhysRevB.74.155107}.

\bibitem[{\citenamefont{Werner and Millis}(2010)}]{werner2010}
\bibinfo{author}{\bibfnamefont{P.}~\bibnamefont{Werner}} \bibnamefont{and}
  \bibinfo{author}{\bibfnamefont{A.~J.} \bibnamefont{Millis}},
  \bibinfo{journal}{Phys. Rev. Lett.} \textbf{\bibinfo{volume}{104}},
  \bibinfo{pages}{146401} (\bibinfo{year}{2010}),
  \urlprefix\url{https://link.aps.org/doi/10.1103/PhysRevLett.104.146401}.

\bibitem[{\citenamefont{Wu et~al.}(2021)\citenamefont{Wu, Bu, Zhang, Fei,
  Zheng, Gao, Luo, Liu, Sun, and Yin}}]{wu2021}
\bibinfo{author}{\bibfnamefont{Z.}~\bibnamefont{Wu}},
  \bibinfo{author}{\bibfnamefont{K.}~\bibnamefont{Bu}},
  \bibinfo{author}{\bibfnamefont{W.}~\bibnamefont{Zhang}},
  \bibinfo{author}{\bibfnamefont{Y.}~\bibnamefont{Fei}},
  \bibinfo{author}{\bibfnamefont{Y.}~\bibnamefont{Zheng}},
  \bibinfo{author}{\bibfnamefont{J.}~\bibnamefont{Gao}},
  \bibinfo{author}{\bibfnamefont{X.}~\bibnamefont{Luo}},
  \bibinfo{author}{\bibfnamefont{Z.}~\bibnamefont{Liu}},
  \bibinfo{author}{\bibfnamefont{Y.-p.} \bibnamefont{Sun}}, \bibnamefont{and}
  \bibinfo{author}{\bibfnamefont{Y.}~\bibnamefont{Yin}},
  \bibinfo{journal}{arXiv:2105.08663 [cond-mat]}  (\bibinfo{year}{2021}),
  \eprint{2105.08663}, \urlprefix\url{http://arxiv.org/abs/2105.08663}.

\bibitem[{\citenamefont{Vidberg and Serene}(1977)}]{vidberg1977}
\bibinfo{author}{\bibfnamefont{H.~J.} \bibnamefont{Vidberg}} \bibnamefont{and}
  \bibinfo{author}{\bibfnamefont{J.~W.} \bibnamefont{Serene}},
  \bibinfo{journal}{J Low Temp Phys} \textbf{\bibinfo{volume}{29}},
  \bibinfo{pages}{179} (\bibinfo{year}{1977}),
  \urlprefix\url{https://doi.org/10.1007/BF00655090}.

\bibitem[{\citenamefont{Boehnke et~al.}(2016)\citenamefont{Boehnke, Nilsson,
  Aryasetiawan, and Werner}}]{boehnke2016}
\bibinfo{author}{\bibfnamefont{L.}~\bibnamefont{Boehnke}},
  \bibinfo{author}{\bibfnamefont{F.}~\bibnamefont{Nilsson}},
  \bibinfo{author}{\bibfnamefont{F.}~\bibnamefont{Aryasetiawan}},
  \bibnamefont{and} \bibinfo{author}{\bibfnamefont{P.}~\bibnamefont{Werner}},
  \bibinfo{journal}{Phys. Rev. B} \textbf{\bibinfo{volume}{94}},
  \bibinfo{pages}{201106} (\bibinfo{year}{2016}),
  \urlprefix\url{https://link.aps.org/doi/10.1103/PhysRevB.94.201106}.

\bibitem[{\citenamefont{Petocchi
  et~al.}(2020{\natexlab{b}})\citenamefont{Petocchi, Christiansson, Nilsson,
  Aryasetiawan, and Werner}}]{petocchi2020a}
\bibinfo{author}{\bibfnamefont{F.}~\bibnamefont{Petocchi}},
  \bibinfo{author}{\bibfnamefont{V.}~\bibnamefont{Christiansson}},
  \bibinfo{author}{\bibfnamefont{F.}~\bibnamefont{Nilsson}},
  \bibinfo{author}{\bibfnamefont{F.}~\bibnamefont{Aryasetiawan}},
  \bibnamefont{and} \bibinfo{author}{\bibfnamefont{P.}~\bibnamefont{Werner}},
  \bibinfo{journal}{Phys. Rev. X} \textbf{\bibinfo{volume}{10}},
  \bibinfo{pages}{041047} (\bibinfo{year}{2020}{\natexlab{b}}),
  \urlprefix\url{https://link.aps.org/doi/10.1103/PhysRevX.10.041047}.

\bibitem[{\citenamefont{Cho et~al.}(2016)\citenamefont{Cho, Cheon, Kim, Lee,
  Cho, Cheong, and Yeom}}]{cho2016}
\bibinfo{author}{\bibfnamefont{D.}~\bibnamefont{Cho}},
  \bibinfo{author}{\bibfnamefont{S.}~\bibnamefont{Cheon}},
  \bibinfo{author}{\bibfnamefont{K.-S.} \bibnamefont{Kim}},
  \bibinfo{author}{\bibfnamefont{S.-H.} \bibnamefont{Lee}},
  \bibinfo{author}{\bibfnamefont{Y.-H.} \bibnamefont{Cho}},
  \bibinfo{author}{\bibfnamefont{S.-W.} \bibnamefont{Cheong}},
  \bibnamefont{and} \bibinfo{author}{\bibfnamefont{H.~W.} \bibnamefont{Yeom}},
  \bibinfo{journal}{Nat Commun} \textbf{\bibinfo{volume}{7}},
  \bibinfo{pages}{10453} (\bibinfo{year}{2016}),
  \urlprefix\url{https://www.nature.com/articles/ncomms10453}.

\bibitem[{\citenamefont{Eckstein and Werner}(2011)}]{eckstein2011}
\bibinfo{author}{\bibfnamefont{M.}~\bibnamefont{Eckstein}} \bibnamefont{and}
  \bibinfo{author}{\bibfnamefont{P.}~\bibnamefont{Werner}},
  \bibinfo{journal}{Phys. Rev. B} \textbf{\bibinfo{volume}{84}},
  \bibinfo{pages}{035122} (\bibinfo{year}{2011}),
  \urlprefix\url{https://link.aps.org/doi/10.1103/PhysRevB.84.035122}.

\bibitem[{\citenamefont{Lee et~al.}(2019)\citenamefont{Lee, Goh, and
  Cho}}]{lee2019}
\bibinfo{author}{\bibfnamefont{S.-H.} \bibnamefont{Lee}},
  \bibinfo{author}{\bibfnamefont{J.~S.} \bibnamefont{Goh}}, \bibnamefont{and}
  \bibinfo{author}{\bibfnamefont{D.}~\bibnamefont{Cho}},
  \bibinfo{journal}{Phys. Rev. Lett.} \textbf{\bibinfo{volume}{122}},
  \bibinfo{pages}{106404} (\bibinfo{year}{2019}),
  \urlprefix\url{https://link.aps.org/doi/10.1103/PhysRevLett.122.106404}.

\bibitem[{\citenamefont{Pasquier and Yazyev}(2021)}]{pasquier2021}
\bibinfo{author}{\bibfnamefont{D.}~\bibnamefont{Pasquier}} \bibnamefont{and}
  \bibinfo{author}{\bibfnamefont{O.~V.} \bibnamefont{Yazyev}},
  \bibinfo{journal}{arXiv:2108.11277 [cond-mat]}  (\bibinfo{year}{2021}),
  \eprint{2108.11277}, \urlprefix\url{http://arxiv.org/abs/2108.11277}.

\bibitem[{\citenamefont{Gole{\v z} et~al.}(2017)\citenamefont{Gole{\v z},
  Boehnke, Strand, Eckstein, and Werner}}]{golez2017}
\bibinfo{author}{\bibfnamefont{D.}~\bibnamefont{Gole{\v z}}},
  \bibinfo{author}{\bibfnamefont{L.}~\bibnamefont{Boehnke}},
  \bibinfo{author}{\bibfnamefont{H.~U.~R.} \bibnamefont{Strand}},
  \bibinfo{author}{\bibfnamefont{M.}~\bibnamefont{Eckstein}}, \bibnamefont{and}
  \bibinfo{author}{\bibfnamefont{P.}~\bibnamefont{Werner}},
  \bibinfo{journal}{Phys. Rev. Lett.} \textbf{\bibinfo{volume}{118}},
  \bibinfo{pages}{246402} (\bibinfo{year}{2017}),
  \urlprefix\url{https://link.aps.org/doi/10.1103/PhysRevLett.118.246402}.

\bibitem[{\citenamefont{Gole{\v z} et~al.}(2019)\citenamefont{Gole{\v z},
  Eckstein, and Werner}}]{golez2019}
\bibinfo{author}{\bibfnamefont{D.}~\bibnamefont{Gole{\v z}}},
  \bibinfo{author}{\bibfnamefont{M.}~\bibnamefont{Eckstein}}, \bibnamefont{and}
  \bibinfo{author}{\bibfnamefont{P.}~\bibnamefont{Werner}},
  \bibinfo{journal}{Phys. Rev. B} \textbf{\bibinfo{volume}{100}},
  \bibinfo{pages}{235117} (\bibinfo{year}{2019}),
  \urlprefix\url{https://link.aps.org/doi/10.1103/PhysRevB.100.235117}.

\bibitem[{\citenamefont{Wilson et~al.}(1975)\citenamefont{Wilson, Di~Salvo, and
  Mahajan}}]{wilson1975}
\bibinfo{author}{\bibfnamefont{J.}~\bibnamefont{Wilson}},
  \bibinfo{author}{\bibfnamefont{F.}~\bibnamefont{Di~Salvo}}, \bibnamefont{and}
  \bibinfo{author}{\bibfnamefont{S.}~\bibnamefont{Mahajan}},
  \bibinfo{journal}{Advances in Physics} \textbf{\bibinfo{volume}{24}},
  \bibinfo{pages}{117} (\bibinfo{year}{1975}),
  \urlprefix\url{https://doi.org/10.1080/00018737500101391}.

\bibitem[{\citenamefont{Fazekas and Tosatti}(1979)}]{fazekas1979a}
\bibinfo{author}{\bibfnamefont{P.}~\bibnamefont{Fazekas}} \bibnamefont{and}
  \bibinfo{author}{\bibfnamefont{E.}~\bibnamefont{Tosatti}},
  \bibinfo{journal}{Philosophical Magazine B} \textbf{\bibinfo{volume}{39}},
  \bibinfo{pages}{229} (\bibinfo{year}{1979}),
  \urlprefix\url{https://doi.org/10.1080/13642817908245359}.

\bibitem[{\citenamefont{Sipos et~al.}(2008)\citenamefont{Sipos, Kusmartseva,
  Akrap, Berger, Forr{\'o}, and Tuti{\v s}}}]{sipos2008}
\bibinfo{author}{\bibfnamefont{B.}~\bibnamefont{Sipos}},
  \bibinfo{author}{\bibfnamefont{A.~F.} \bibnamefont{Kusmartseva}},
  \bibinfo{author}{\bibfnamefont{A.}~\bibnamefont{Akrap}},
  \bibinfo{author}{\bibfnamefont{H.}~\bibnamefont{Berger}},
  \bibinfo{author}{\bibfnamefont{L.}~\bibnamefont{Forr{\'o}}},
  \bibnamefont{and} \bibinfo{author}{\bibfnamefont{E.}~\bibnamefont{Tuti{\v
  s}}}, \bibinfo{journal}{Nature Mater} \textbf{\bibinfo{volume}{7}},
  \bibinfo{pages}{960} (\bibinfo{year}{2008}),
  \urlprefix\url{http://www.nature.com/articles/nmat2318}.

\bibitem[{\citenamefont{Murakami et~al.}(2015)\citenamefont{Murakami, Werner,
  Tsuji, and Aoki}}]{murakami2015}
\bibinfo{author}{\bibfnamefont{Y.}~\bibnamefont{Murakami}},
  \bibinfo{author}{\bibfnamefont{P.}~\bibnamefont{Werner}},
  \bibinfo{author}{\bibfnamefont{N.}~\bibnamefont{Tsuji}}, \bibnamefont{and}
  \bibinfo{author}{\bibfnamefont{H.}~\bibnamefont{Aoki}},
  \bibinfo{journal}{Phys. Rev. B} \textbf{\bibinfo{volume}{91}},
  \bibinfo{pages}{045128} (\bibinfo{year}{2015}),
  \urlprefix\url{https://link.aps.org/doi/10.1103/PhysRevB.91.045128}.

\bibitem[{\citenamefont{Werner et~al.}(2019)\citenamefont{Werner, Eckstein,
  M{\"u}ller, and Refael}}]{werner2019}
\bibinfo{author}{\bibfnamefont{P.}~\bibnamefont{Werner}},
  \bibinfo{author}{\bibfnamefont{M.}~\bibnamefont{Eckstein}},
  \bibinfo{author}{\bibfnamefont{M.}~\bibnamefont{M{\"u}ller}},
  \bibnamefont{and} \bibinfo{author}{\bibfnamefont{G.}~\bibnamefont{Refael}},
  \bibinfo{journal}{Nat Commun} \textbf{\bibinfo{volume}{10}},
  \bibinfo{pages}{5556} (\bibinfo{year}{2019}),
  \urlprefix\url{https://www.nature.com/articles/s41467-019-13557-9}.

\bibitem[{\citenamefont{Metzner and Vollhardt}(1989)}]{metzner1989}
\bibinfo{author}{\bibfnamefont{W.}~\bibnamefont{Metzner}} \bibnamefont{and}
  \bibinfo{author}{\bibfnamefont{D.}~\bibnamefont{Vollhardt}},
  \bibinfo{journal}{Phys. Rev. Lett.} \textbf{\bibinfo{volume}{62}},
  \bibinfo{pages}{324} (\bibinfo{year}{1989}),
  \urlprefix\url{https://link.aps.org/doi/10.1103/PhysRevLett.62.324}.

\bibitem[{\citenamefont{Hubbard}(1959)}]{hubbard1959}
\bibinfo{author}{\bibfnamefont{J.}~\bibnamefont{Hubbard}},
  \bibinfo{journal}{Phys. Rev. Lett.} \textbf{\bibinfo{volume}{3}},
  \bibinfo{pages}{77} (\bibinfo{year}{1959}),
  \urlprefix\url{https://link.aps.org/doi/10.1103/PhysRevLett.3.77}.

\bibitem[{\citenamefont{Hubbard and Flowers}(1964)}]{hubbard1964}
\bibinfo{author}{\bibfnamefont{J.}~\bibnamefont{Hubbard}} \bibnamefont{and}
  \bibinfo{author}{\bibfnamefont{B.~H.} \bibnamefont{Flowers}},
  \bibinfo{journal}{Proceedings of the Royal Society of London. Series A.
  Mathematical and Physical Sciences} \textbf{\bibinfo{volume}{281}},
  \bibinfo{pages}{401} (\bibinfo{year}{1964}),
  \urlprefix\url{https://royalsocietypublishing.org/doi/abs/10.1098/rspa.1964.0190}.

\bibitem[{\citenamefont{{M{\"u}ller-Hartmann}}(1989)}]{muller-hartmann1989}
\bibinfo{author}{\bibfnamefont{E.}~\bibnamefont{{M{\"u}ller-Hartmann}}},
  \bibinfo{journal}{Z. Physik B - Condensed Matter}
  \textbf{\bibinfo{volume}{74}}, \bibinfo{pages}{507} (\bibinfo{year}{1989}),
  \urlprefix\url{https://doi.org/10.1007/BF01311397}.

\end{thebibliography}
\end{document}